\documentclass{aa}
\usepackage[pdftex]{graphicx}
\usepackage{epsfig}
\usepackage{amsmath}
\usepackage{wasysym}
\usepackage{txfonts}
\pdfoutput=1
\usepackage[pdftex]{color,xcolor}
\usepackage[pdftex]{graphicx}
\usepackage[backref]{hyperref}
\hypersetup{pdfauthor=Paul Molli\`{e}re}
\hypersetup{backref=true, pagebackref=true, hyperindex=true, breaklinks=true,colorlinks=true,urlcolor=blue, linkcolor=blue,  citecolor=blue,pagecolor=red, bookmarks=true, bookmarksopen=true}
\usepackage{epstopdf}
\hypersetup{backref=true}

\def\mearth{M_\oplus}

\def\msun{M_\odot}

\def\f1{f_{\rm I}}
\def\mj{M_{\rm J}}

\def\mstar{M_*}

\def\beq{\begin{equation}}
\def\eeq{\end{equation}}

\def\fopa{f_{\rm opa}}
\def\t2{\tau_{\rm II}}

\def\sigmas0{\Sigma_{\rm s,0}}

\newcommand{\lsun}{L_{\odot}}

\def\npa{Nuc. Phys. A} 
\def\rmp{Rev. Mod. Phys.} 
\def\araa{ARA\&A}             
\def\apj{ApJ}                 
\def\apjs{ApJS}               
\def\aap{A\&A}                
\def\mnras{MNRAS}             

\def\({\left(}
\def\){\right)}
\def\<{\left<}
\def\>{\right>}

\begin{document}

\title{Deuterium burning in objects forming \\ via the core accretion scenario}
\subtitle{Brown dwarfs or planets?}

\author{P. Molli\`{e}re\inst{1} \and  C. Mordasini\inst{1}}

\institute{Max-Planck-Institut f\"ur Astronomie, K\"onigstuhl 17, D-69117 Heidelberg, Germany}

\offprints{Paul  MOLLIERE, \email{molliere@mpia.de}}

\date{Received 19 June 2012 / Accepted 30 August 2012 }

\abstract
{}
{{Our aim is to study deuterium burning in objects forming according to the core accretion scenario in the hot and cold start assumption and what minimum deuterium burning mass limit is found for these objects. We also study how the burning process influences the structure and luminosity of the objects. Furthermore we want to test and verify our results by comparing them to already existing hot start simulations which did not consider, however, the formation process.}
}
{We present a new method to calculate deuterium burning of objects in a self-consistently coupled model of planet formation and evolution. We discuss which theory is used to describe the process of deuterium burning and how it was implemented.
}
{{We find that the objects forming according to a hot start scenario behave approximately in the same way as found in previous works of evolutionary calculations, which did not consider the formation.
However, for cold start objects one finds that the objects expand during deuterium burning instead of being partially stabilized against contraction.} In both cases, hot and cold start, the mass of the solid core has an influence on the minimum mass limit of deuterium burning. The general position of the mass limit, 13 $\mj$, stays however approximately the same.
None of the investigated parameters was able to change this mass limit by more than $0.8$ $\mj$. {Due to deuterium burning,} the luminosity of hot and cold start objects becomes comparable after $\sim200 $ Myrs.
}
{}

\titlerunning{Deuterium burning in objects forming via core accretion}
\authorrunning{P. Molli\`{e}re and C. Mordasini}

\maketitle

\section{Introduction}
It is generally accepted that compact gaseous objects with a mass greater than 12-13 $\mj$, where $\mj$ is the mass of Jupiter, will start deuterium burning via the reaction $ \mathrm{p} + \mathrm{d} \rightarrow \gamma + ~^3\mathrm{He}$ (Saumon et al. \cite{saumonetal1996}, Chabrier \& Baraffe \cite{chabrieretal2000}, Burrows et al. \cite{burrowsetal2001}). In masses greater than approximately 63 $\mj$, lithium burning will set in via the two reactions $\mathrm{p} + ~^7\mathrm{Li} \rightarrow 2\alpha$ and $\mathrm{p} + ~^6\mathrm{Li} \rightarrow \alpha + ^3\mathrm{He}$ (see e. g. Burrows et al. \cite{burrowsetal2001}). As long as the object's mass does not exceed a value of approximately 80 $\mj$, which is the lower mass limit for hydrogen burning (Burrows et al. \cite{burrowsetal2001}), it belongs to the class of so-called Brown Dwarfs. Objects with masses below 13 $\mj$, i.e. objects which are not able to burn deuterium, are called planets (Boss et al. \cite{bossetal}), if they are in an orbit around a 
star 
or a stellar remnant and fulfill the properties which are demanded for planets in the solar system as well (independent of the actual formation mode). Observationally, for companions in orbit around roughly solar like stars, it seems to be difficult to discriminate between planets and low-mass Brown Dwarfs: The observed frequencies of objects in the mass range around 13 $\mj$ behave smoothly and exhibit no special pattern which might be expected if two different formation processes (one for planets and one for Brown Dwarfs) would be at work (S{\'e}gransan et al \cite{segransanetal2010}). Thus, high-mass planets and low-mass Brown Dwarfs might form through the same processes. Two different ways of forming (giant) planets are being discussed. The first is the so-called disk instability or direct collapse model, which explains the formation of planets by a gravitational instability in the protostellar disk, whereas the second model is the core accretion model, which explains the formation of giant planets with 
building up a solid core by planetesimal accretion which, at some point, becomes heavy enough to accrete large amounts of gas from the protostellar disk (for a detailed review of both methods see e.g. D'Angelo et al. (\cite{dangeloetal})). In this paper we present a new model that combines the effect of deuterium burning with the theory of core accretion planet formation (based on the implementation by Alibert et al. (\cite{alibertmordasini2004}) and Mordasini et al. \cite{mordasinietal2011}). A difference to the investigation conducted so far on deuterium burning objects is the presence of a solid core in the center of the planet. It is interesting to study how the presence of the core affects the burning process, as previous work has shown that deuterium burning in objects harboring a solid core is possible (Baraffe et al. \cite{baraffeetal}). An advantage of the combined study of deuterium 
burning and planet formation is the the possibility to investigate the reaction of deuterium burning to the variation of the parameters that constrain the planetary formation process. Such parameters can be the dust-to-gas ratio in the protostellar disk, the maximum allowed gas accretion rate or the mass fraction of helium in the gas etc. As the existence of solid cores in giant planets has been subject to a discussion (e.g. Guillot et al. (\cite{guillotetal2003}) and Wilson \& Militzer (\cite{wilsonetal}) suggest that the cores might dissolve), our results are for the limiting assumption that the core does not dissolve. \\ \\
 The results of Lubow et al. (\cite{lubowetal}) suggest that objects forming by core accretion might not get massive enough to burn deuterium. The reason for this is the formation of a gap in the disk for sufficiently massive objects which prevents them from further significant mass accretion. This would limit the maximal masses to $\sim 10 \ \mj$. However, Kley \& Dirksen (\cite{kleyetal}) 
find that as an object reaches a 
certain mass threshold, the gap gets wide enough to clear the locations of the eccentricity damping resonances, such that the resonances exciting eccentricity become important. The system is then prone to an eccentric instability, resulting in again high accretion rates of the object. This eccentric instability is the motivation for considering objects forming via the core accretion scenario with masses high enough to fuse deuterium. As the exact temporal behavior of the mass accretion rate in these phases is still unknown, we used the simplification of assuming a constant maximum allowed gas accretion rate $\dot{M}_{\rm max}$ in the runaway accretion phase.
\subsection{Organization of the paper}
The paper is organized as follows: In Section \ref{sect:burningtheory} we explain the theory which was used to describe the deuterium burning process. In Section \ref{sect:impmethod} we describe how deuterium burning was implemented into the already existing code. In Section \ref{sect:hotstartcomp} we compare our results obtained for the deuterium burning assuming a hot start to already published work (in which also a hot start was assumed, but the formation process was left out). In Section \ref{subsect:coldstart} we look at the deuterium burning process in objects forming via the cold start assumption. In Section \ref{sect:minmass} we investigate the implications of changes in quantities such as the dust-to-gas ratio in the protoplanetary disk, the initial deuterium abundance, the maximum allowed mass accretion rate or the helium mass fraction of the gas for the minimum mass limit for deuterium burning. In Section \ref{sec:distinguish} we show how deuterium burning can shorten the timescale after which the 
luminosity of hot and cold start objects becomes comparable. Our results are summarized in Section \ref{sect:summary}, followed by the conclusion in Section
\ref{sect:conclusion}. {Finally, analytical approximations concerning the cooling phase of the objects can be found in Appendix \ref{appendix:A} and \ref{appendix:B}.}

\section{Theory of deuterium burning}\label{sect:burningtheory}
In order to describe the deuterium fusion process one has to calculate its contribution to the luminosity. This means that we have to know the energy generation rate $\epsilon$, i.e. energy per unit mass and unit time. We use the rate as it is defined in Kippenhahn \& Weigert (\cite{kippenhahnetal}), assuming fully ionized, non-degenerate nuclei.
The value assumed for the energy released in every fusion process was taken from Fowler et al. (\cite{fowleretal}), whereas the value for the cross-section factor was taken from Angulo et al. (\cite{anguloetal}).
\subsection{Screening}
In order to calculate the nuclear reaction rates of the deuterium fusion process correctly, screening theory had to be applied, i.e. the effect that enhances the nuclear fusion rate if the repelling positive charges of the nuclei are shielded from each other by surrounding electrons.
We implemented screening following the papers of Dewitt et al. (\cite{dewittetal}) and Graboske et al. (\cite{graboskeetal}).
\subsection{Electron degeneracy}
As we treat objects with high central densities, electron degeneracy will become important. {The electron degeneracy will affect the screening behavior. The factor $\Theta_e$ accounts for this effect and measures the strength of degeneracy.} It is defined as (Dewitt et al. (\cite{dewittetal})):
\beq
\Theta_e = \frac{1}{2}\frac{F_{-1/2}(\psi)}{F_{1/2}(\psi)}
\label{equ:thetae}
\eeq
where $F_{\nu}(\psi)$ are the Fermi-Dirac integrals
\beq
F_{\nu}(\psi) = \int_0^{\infty}\frac{x^\nu}{e^{x-\psi}+1}dx
\eeq
and $\psi$ is the so-called degeneracy parameter. For $\psi \rightarrow \infty$ the electrons are fully degenerate, whereas for $\psi \rightarrow -\infty$ they are fully non-degenerate. As $\psi$ goes from $-\infty$ to $\infty$ $\Theta_e$ goes from 1 to 0. 
{To obtain $\psi$ we tabulated $F_{1/2}(\psi)$ in the $\psi$-range of interest and used that it holds that} 
{(see e.g. Kippenhahn \& Weigert (\cite{kippenhahnetal}))}
\beq
F_{1/2}(\psi)=\frac{n_eh^3}{4\pi\left(2m_ek_{\mathrm{B}}T\right)^{3/2}} \ \ .
\eeq
\section{Model}\label{sect:impmethod}
A detailed description of the simulation method used in this paper can be found in Mordasini et al. (\cite{mordasinietal2011}). Here we only give a brief outline of its functioning, concentrating on changes to the already existing method. The following four equations needed to calculate the object structure hold for spherical symmetric and hydrostatic conditions. {They are the standard equations for calculating stellar structures.} The assumption of hydrostatic equilibrium can be justified as the planet is nearly hydrostatic even in the most critical phase of the radial collapse of the envelope. This happens at the transition from the attached to the detached phase, where one finds the kinetic energy of the gas to be many orders of magnitude smaller than the total energy of the object.
\begin{align}
   \frac{\partial m}{\partial r} &= 4 \pi r^2 \rho &
   \label{equ:mass}
   \frac{\partial P}{\partial r} &= - \frac{Gm}{r^2} \rho \\
   \frac{\partial T}{\partial r} &= \frac{T}{P}\frac{\partial P}{\partial r}\nabla & 
   \label{equ:temp}
   \frac{\partial l}{\partial r} &=- 4 \pi r^2 \rho \( T\frac{\partial S}{\partial t} + \epsilon \)
\end{align}
However, instead of Eq. \ref{equ:temp}, we use the following equation
\beq
   \frac{\partial l}{\partial r} = 0
  \label{equ:lumiuse}
\eeq
and calculate the luminostiy as described in Sect. \ref{subsec:lumcalc}. The reason why it is justified to choose $\partial l/\partial r = 0$ is outlined in Sect. \ref{sect:lumsim}. \\ \\
Eq. \ref{equ:temp} describes the radial evolution of the temperature, where the temperature gradient $\nabla$ is defined as
\beq
\nabla = \frac{d~\mathrm{ln}T}{d~\mathrm{ln}P}.
\eeq
Following the Schwarzschild criterion and assuming that the object is either only convective or radiative in a shell, we set
\beq
\nabla = \mathrm{min}\(\nabla_{\mathrm{ad}},\nabla_{\mathrm{rad}}\)
\label{equ:nablacond}
\eeq
where $\nabla_{\mathrm{ad}}$, the adiabatic temperature gradient, is given directly by the EOS. The radiative temperature gradient $\nabla_{\mathrm{rad}}$ is defined as
\beq
\nabla_{\mathrm{rad}}=\frac{3}{16 \pi a c G}\frac{\kappa l P}{m T^4}
\label{equ:nablarad}
\eeq
In Eq. \ref{equ:nablarad}, $a$ denotes the radiation energy density constant, $c$ the speed of light, $\kappa$ the opacity and $l=l(r)$ the luminosity.
Finally, the right part of Eq. \ref{equ:temp} yields the luminosity change per unit radius, where $S$ denotes the entropy per unit mass.
$\epsilon$ denotes the thermonuclear energy generation rate per unit mass and unit time.
\subsection{Calculation of the luminosity}
\label{subsec:lumcalc}
Starting from the detached phase\footnote{We start the consideration from the detached phase here as only after the collapse and the subsequent runaway accretion of mass one reaches temperatures and densities high enough to fuse deuterium.}, the model, as described in Mordasini et al. (\cite{mordasinietal2011}), uses a shooting method to find the structure of the object, iterating on the outer radius. As one needs an surface boundary condition for the luminosity, the following derivation is chosen: The total energy of the object is
\beq
E_{\mathrm{tot}}= E_{\mathrm{grav}}+E_{\mathrm{int}}=-\int_0^M{\frac{Gm}{r}dm}+\int_0^M{udm}
\label{equ:energyintegral}
\eeq
where $u$ is the internal energy per unit mass.
Motivated by the virial theorem and the fact that our solutions are found assuming a hydrostatic structure, one can express the total energy of the object as
\beq
E_{\mathrm{tot}}= - \xi \frac{GM^2}{2R}
\label{equ:etot}
\eeq
The factor $\xi$ depends on the radial mass distribution inside the object and its internal energy content\footnote{For an ideal gas with an EOS in the form $P=\(\gamma-1\)\rho u$ the dependence of $\xi$ on the internal energy content would be expressed as an dependence on $\gamma$, the adiabatic index of the gas.}. In order to obtain the luminosity one can differentiate Eq. \ref{equ:etot} with respect to time which yields
\beq
L = \xi\frac{GM}{R}\dot{M}-\xi\frac{GM^2}{2R^2}\dot{R}+\frac{GM^2}{2R}\dot{\xi}
\eeq
This is only correct, of course, if we do not have the effect of deuterium burning, which will be included in the luminosity consideration in the next subsection. As we never know $\xi$ in advance of the structure calculation at a certain timestep (and thus also $\dot{\xi}$ is unknown), we use the $\xi$ of the previous timestep and approximate
\beq
L \approx C\cdot\(\xi\frac{GM}{R}\dot{M}-\xi\frac{GM^2}{2R^2}\dot{R}\)
\label{equ:lumapprox}
\eeq
The correction factor $C$ accounts for the fact that we excluded the $\dot{\xi}$-term from our calculation (which we assume to be small, this can be obtained by choosing the timestep sufficiently short) and that we used $\xi$ from the previous timestep.
An estimate of the correction factor for the next timestep to $t+dt$ is then found by
\beq
C = \frac{E_{\mathrm{tot}}(t)-E_{\mathrm{tot}}(t-dt)}{-L(t)dt}
\eeq
where the total energies are obtained via the integrals defined in Eq. \ref{equ:energyintegral}.
When considering a hot start, the internal luminosity $L_{\mathrm{int}}$ (i.e. the luminosity supposed to originate from within the planet) used to calculate the internal structure of the planet is simply the luminosity as defined in Eq. \ref{equ:lumapprox}
\beq
L_{\mathrm{int}} = L.
\label{equ:internallumhot}
\eeq
In the cold start case we set
\beq
L_{\mathrm{int}} = L - \frac{GM\dot{M}_{\mathrm{gas}}}{R}
\label{equ:internallumcold}
\eeq
assuming that the gravitational potential energy released by the gas freely falling onto the planet is radiated away very efficiently in a shock at the objects surface.
This causes the gas to lose its initially quite high entropy and it is incorporated into the object at a low specific entropy.
One must add, however, that it is still unclear whether objects forming via the core accretion scenario really form ''cold'', as this is depending completely on the structure of the shock through which the gas is accreted in the runaway accretion phase (Marley et al. \cite{marleyetal}, Stahler et al. \cite{stahleretal1980}). It might be possible that the gas accretion luminosity should, at least partially, be treated as an internal energy source (commonly denoted as a ''warm start''), i.e. the gas is accreted with a higher specific entropy. Motivated by the fact that the shock structure and thus the initial entropy of the freshly accreted gas is not really understood{,} the post-formation evolution of planets or objects with initial conditions which lie in the warm start regime were studied recently (Spiegel \& Burrows (\cite{spiegeletal2012})).
\subsubsection{Further simplifications}
\label{sect:lumsim}
As one finds the objects to be fully convective in the detached and post-formation evolution phase (except for a thin radiative layer at the outer boundary) we do not calculate the radial structure of the luminosity, as it is not necessary. The radial structure of the luminosity only enters the objects structure calculation via Eq. \ref{equ:temp} and only if the radial shells under consideration are radiative (see Eq. \ref{equ:nablacond}). As the objects are basically fully convective, we set $d~l/d~r=0$ (for simplicity), which is not affecting the simulation. Furthermore it has the advantage to also cover the objects evolution in the attached phase, were the main luminosity is due to the solid accretion of the core, thus the luminosity is approximately constant in the gaseous envelope.
As shown in Mordasini et al. (\cite{mordasinietal2011}) the calculation of the structure as explained here leads to evolutionary sequences which are in excellent agreement with the conventional entropy method (see e.g. Burrows et al. \cite{burrowsetal1997}, Burrows et al. \cite{burrowsetal2001} and Chabrier \& Baraffe \cite{chabrieretal2000}).
However, a possible caveat might be that we assume the objects to have very effective, large scale convection (i.e. the radiative losses of rising convective blobs are small compared to the energy they transport). This is equal to treating the objects as adiabatic in the convective regions. This, however, must not necessarily be the case (Leconte \& Chabrier \cite{leconte}), given that less effective convection is superadiabatic.
\subsection{Adopting the calculation of the correction factor $C$ to deuterium burning}
As outlined in the previous section, the boundary luminosity $L_{\mathrm{int}}$ needed for the structure calculation of the objects can be approximated by using the virial theorem.  In this method, however, the luminosity contribution of deuterium burning cannot be included directly. To overcome this problem we utilized the simplification of using the deuterium burning luminosity $L_{\rm D}$ calculated at the previous timestep, i.e. at $t-dt$, and set (for the cold start):
\beq
L_{\mathrm{int}}(t) = L(t) - \frac{GM(t)\dot{M}_{\mathrm{gas}}(t)}{R(t)} + L_{\rm D}(t-dt)
\eeq
The error of this approximation, similarly to the error we make for the approximation of the correction factor $C$, gets smaller the smaller we choose the timesteps. In order to get the correct value of the correction factor $C$ one has to take into account that the total energy definition in Eq. \ref{equ:energyintegral} does not include any term considering the deuterium fusion process. Thus, we set
\beq
C = \frac{E_{\mathrm{tot}}(t)-E_{\mathrm{tot}}(t-dt)}{-\(L(t)-L_{\rm D}(t-dt)\)dt}
\eeq
in order to correct for this. The fact that $L_{\rm D}(t-dt)$ appears here instead of $L_{\rm D}(t)$ comes from the just mentioned problem that we cannot ''predict'' the deuterium burning luminosity with the virial theorem ansatz and thus use $L_{\rm D}$ from the previous timestep. $C$ is thus still the ratio of the actual difference of potential and internal energy and the estimated energy radiated away due to contraction, cooling and accretion processes in the timestep of length $dt$.
\subsection{Calculation of the deuterium abundance}
In order to calculate the decrease of the deuterium abundance we assumed that the deuterium decreases homogeneously in the gaseous envelope. The justification for this is that convective mixing is found to be very effective. Using the mixing length theory and assuming a mixing length equal to the pressure scale height, we found that the convective mixing timescale is always at least 2 or 3 orders of magnitudes smaller than the timescale of deuterium burning. 
\section{Results of the hot start accreting model and comparison to hot start simulations}\label{sect:hotstartcomp}
In order to test the results we obtain for deuterium burning with our simulation, especially when varying parameters such as the object's helium abundance, deuterium abundance etc., we will compare with the results found by Spiegel et al. (\cite{spiegeletal2011}) (hereafter called SBM11). As SBM11 considered hot start initial conditions for their evolutionary models we will also use the hot start condition for the outer boundary condition of the luminosity as described in subsection \ref{subsec:lumcalc}. However, it remains an important difference that in our model the formation of the objects is taken into account whereas in the model of SBM11 only the evolution of the objects is considered.
The other outer boundary conditions for the evolutionary phase after the formation are
\begin{align}
    P&=\frac{2 g}{3 \kappa} & T_{\rm int}^{4}&=\frac{ L_{\rm int}}{4 \pi \sigma R^2}\\
  T_{\rm equi}&=280\,{\rm K}  \left(\frac{a}{1 {\rm AU}}\right)^{-\frac{1}{2}}\left(\frac{\mstar}{\msun}\right) &  T^{4}&=(1-A) T_{\rm equi}^{4}+T_{\rm int}^{4}
\end{align}
This are the gray photospheric boundary conditions in the so-called Eddington approximation. $A$ denotes the albedo of the object. The simulation of SBM11 uses a more sophisticated atmospheric model,
assuming proper non-gray atmospheres. One finds, however, that the results one obtains with our simple treatment agree well with the results obtained from
the non-gray atmospheric model (Bodenheimer et al. (\cite{bodenheimeretal2000})). In Tab. \ref{tab:standart} the definition of our fiducial model is given.
\begin{table}
\caption{Fiducial model for the in-situ formation and evolution calculation}\label{tab:standart}
\begin{center}
\begin{tabular}{lc}
\hline\hline
Quantity / Property & Value / Status \\
\hline
a (AU) & 5.2 \\
$\sigmas0$ (g/cm$^{2}$)                         & 10          \\                                              
$\dot{M}_{\rm gas,max}$ ($\mearth$/yr)& 0.01 \\
$T_{\rm neb}$ (K)           &150\\
$P_{\rm neb}$ (dyn/cm$^{2}$)           &   0.275     \\ 
Dust-to-gas ratio  &0.04 \\
Initial embryo mass ($\mearth$) & 0.1\\
Migration & not included\\
Disk evolution & not included\\
Planetesimal ejection & included\\
Simulation duration & $10^{10}$ yrs\\
Grain opacity red. factor $\fopa$ &0.003\\
Helium mass fraction Y &0.25\\
Deuterium number fraction [D/H] &$2 \times 10^{-5}$ \\
$\mstar$ & 1 $\msun$ \\ \hline
\end{tabular}
\end{center}
\end{table}
Thus, all other runs of the simulation will only be specified by the parameters in which they deviate from this fiducial model. In Tab. \ref{tab:standart},
$a$ is the distance to the star at which the planet is supposed to form (migration is disabled), $\sigmas0$ denotes the initial local surface density of solids,
$T_{\rm neb}$ and $P_{\rm neb}$ stand for the temperature and the pressure in the disk at the location of the planet (the disk evolution is neglected) and $f_{\rm opa}$ stands for the
grain opacity reduction factor, i. e. the factor by which we assume the grain opacity of the interstellar medium to be reduced (for further details, see
Mordasini et al. (\cite{mordasiniklahr2011})). The deuterium number fraction [D/H] is defined as the ratio of the deuterium and hydrogen atoms.
In Tab. \ref{tab:compruns} the definition of the runs carried out in order to compare to SBM11 is given.
\begin{figure*}
\begin{minipage}{0.5\textwidth}
	      \centering
       \includegraphics[width=0.95\textwidth]{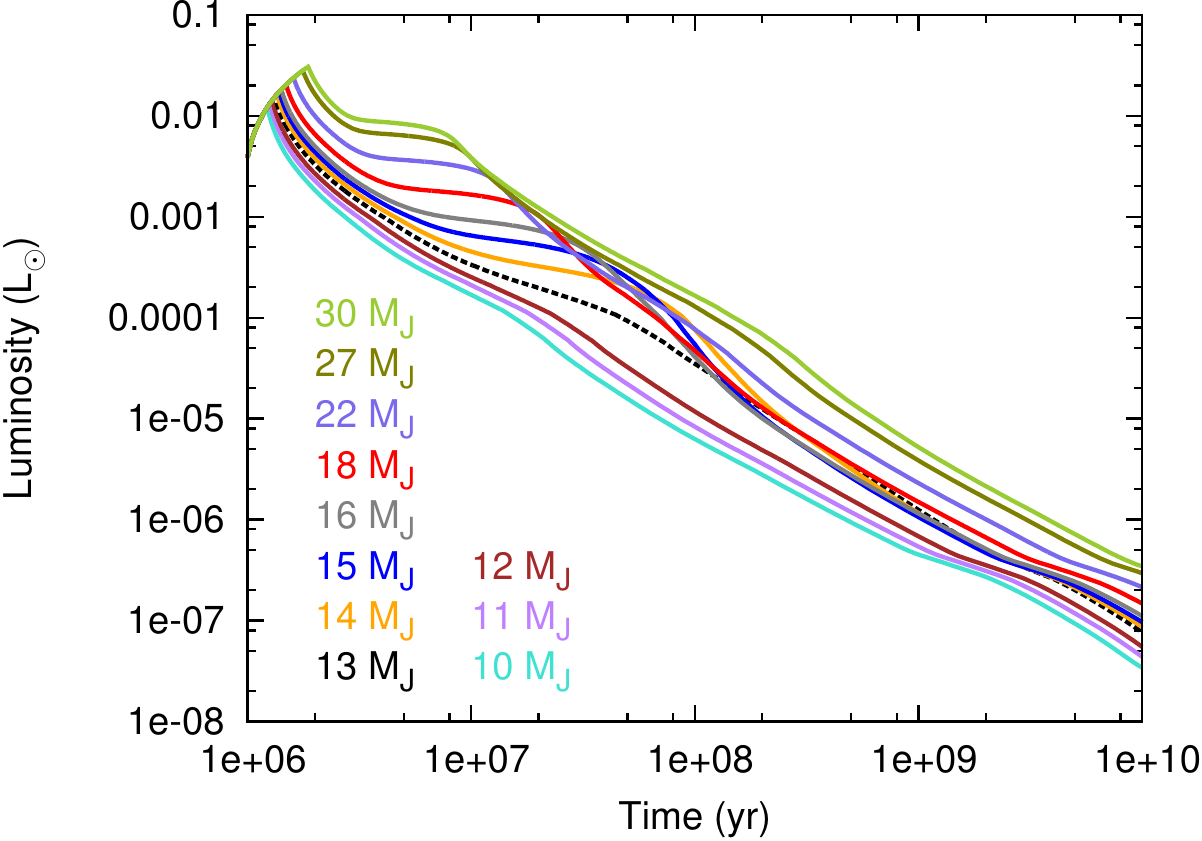}
    \end{minipage}\hfill
      \begin{minipage}{0.5\textwidth}
	      \centering
       \includegraphics[width=0.95\textwidth]{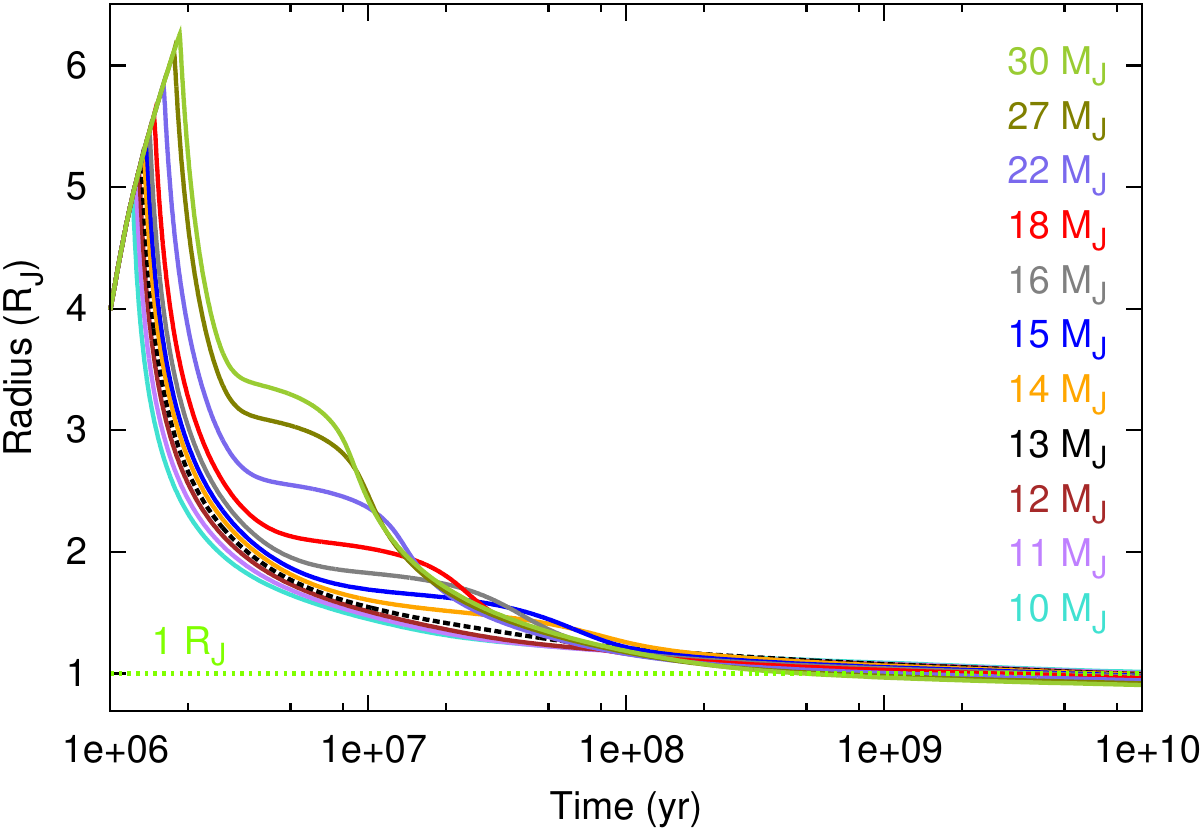}
     \end{minipage}\hfill
     \begin{minipage}{0.5\textwidth}
      \centering
       \includegraphics[width=0.95\textwidth]{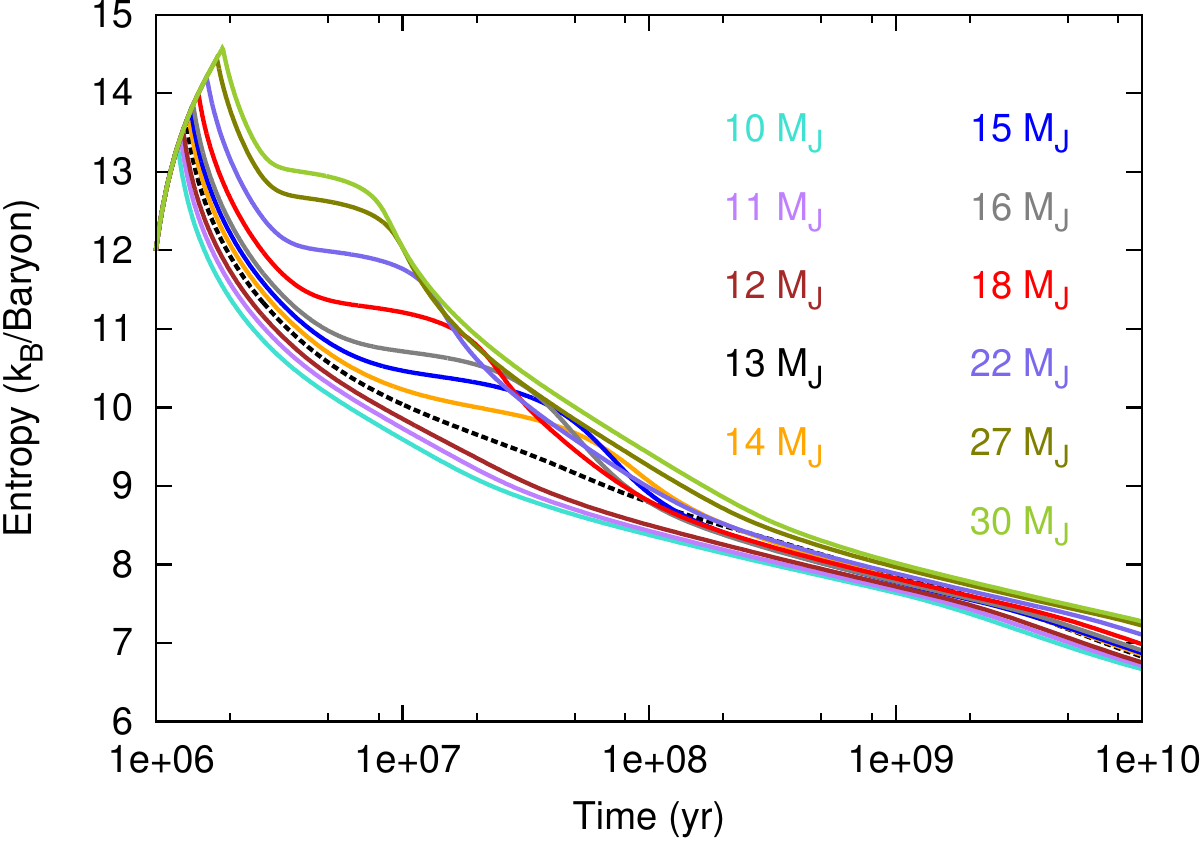}
     \end{minipage}
     \begin{minipage}{0.5\textwidth}
      \centering
       \includegraphics[width=0.95\textwidth]{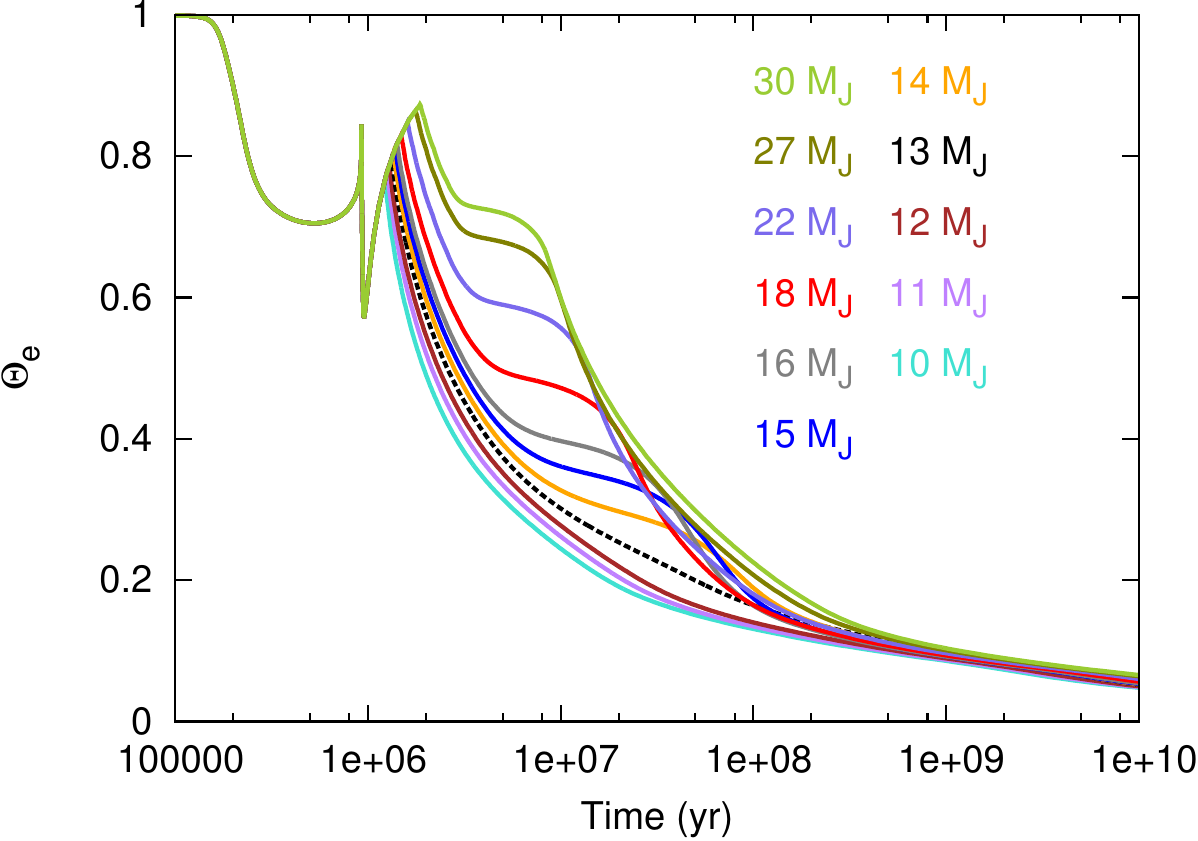}
     \end{minipage}
     \caption{Temporal evolution of the {luminosity} (upper left), the radius (upper right), the entropy above the core (lower left) and the mass averaged $\Theta_e$-factor in the envelope (lower right) for \textbf{hot start} objects with masses varying from 10 $\mj$
              to 30 $\mj$ using our fiducial model as defined in Tab. \ref{tab:standart}. The black dashed line in all plots corresponds to the 13 $\mj$ object. The masses corresponding to the individual lines are indicated in the plot. The horizontal green line in the plot of the radius indicates the position of 1 $R_{\rm J}$.}
     \label{fig:overall}
\end{figure*}
\begin{table}
\caption{Hot start comparison runs}\label{tab:compruns}
\begin{center}
\begin{tabular}{lc}
\hline\hline
Run name & Property \\ \hline
Y22 & $Y=0.22$ \\
Y24 & $Y=0.24$ \\
Y25 & $Y=0.25$ \\
Y26 & $Y=0.26$ \\
Y28 & $Y=0.28$ \\
Y30 & $Y=0.3$ \\ \hline
D1 & ${\rm [D/H]}=1\times 10^{-5}$ \\
D2 & ${\rm [D/H]}=2\times 10^{-5}$ \\
D4 & ${\rm [D/H]}=4\times 10^{-5}$ \\ \hline
fpg1 & Dust to gas ratio = 0.04 $\hat{=} \ {\rm [Fe/H]} = 0$\\
fpg2 & Dust to gas ratio = 0.09 $\hat{=} \ {\rm [Fe/H]} = 0.36$\\
fpg3 & Dust to gas ratio = 0.12 $\hat{=} \ {\rm [Fe/H]} = 0.48$\\ \hline
A1 & $\dot{M}_{\rm gas,max}$ = $10^{-1}$ $\mearth$/yr\\
A2 & $\dot{M}_{\rm gas,max}$ = $10^{-2}$ $\mearth$/yr\\
A3 & $\dot{M}_{\rm gas,max}$ = $10^{-3}$ $\mearth$/yr\\
\hline
\end{tabular}
\end{center}
\end{table}
The definition of the metallicity used in Tab. \ref{tab:compruns} is the following (we assume scaled solar compositions)
\beq
{\rm [Fe/H]} = {\rm log}_{10}\(\frac{{\rm Dust-to-gas \ ratio}}{0.04}\)
\label{equ:metall}
\eeq
This deviates from the usual definition of the metallicity as the protosolar mass fraction is found to be 1.49 \% (Lodders \cite{lodders}). In the inner parts of the disk, however, one should have a higher dust-to-gas
ratio: Due to the different orbital velocities of gas and solids the solids will start to drift inwards such that one finds
a dust-to-gas ratio of approximately 0.04 at 5.2 AU if one considers a solar like star ({R{\'o}zyczka et al. \cite{rozyczkaetal}).
The choice of our fiducial model is mainly motivated by the initial conditions Pollack et al. (\cite{pollacketal}) chose and represents a model which
lies in the regime of conditions one might expect for an object forming via the core accretion scenario.
The reason for taking the initial deuterium abundance to be $2\times 10^{-5}$ is motivated by Prodanovi{\'c} et al. (\cite{prodanovicetal}), as this
is the result they found for the local interstellar medium. This deuterium abundance of $2\times 10^{-5}$ has more or less become the standard value when investigating deuterium burning in Brown Dwarfs. It is also the value used in the fiducial model of SBM11. The choice we made for the initial conditions of the comparison runs
where we vary the helium or the deuterium abundance is based on SBM11 as we want to compare to their results.
Our choice for the metallicity (or dust-to-gas ratio), however, are made due to a different reason. In SBM11
the metallicity has an impact on the opacity of the object (as a higher metallicity means more molecules and thus a higher opacity). In our model, however,
we assume that all the solids accreted by the objects are eventually accreted on the core. Thus a higher metallicity (or higher dust-to-gas
ratio) will not change the opacity, but the core mass of the object. For the opacities we assumed a solar composition and used the opacities of Freedman et al. (\cite{freedmanetal}). The dust-to-gas ratios we chose correspond to the solar case (i.e. [Fe/H] = 0)
and to higher metallicities. However, they still lie in the feasible regime (Santos et al. \cite{santosetal}, Fischer \& Valenti \cite{fischeretal}).
Varying the maximum allowed gas accretion rate is, of course, also something we cannot compare to SBM11, as they
do not consider the formation process.
The motivation for studying different maximum allowed gas accretion rates ($\dot{M}_{\rm max}$) comes from the fact that different circumstellar disks will have different properties at different stages during their evolution. The disk properties such as the viscosity and the surface density determine $\dot{M}_{\rm max}$ (Mordasini et al. \cite{mordasiniklahr2011}). The maximum allowed gas accretion rates we chose also
lie in the expected regime (Mordasini et al. \cite{mordasiniklahr2011}).
Finally and of importance to qualitatively understand the results, is an approximative expression for the nuclear energy generation rate (Stahler \cite{stahler1988}),
\beq
\label{equ:stahler}
\epsilon \propto {\rm [D/H]} \cdot \rho \cdot T^{11.8}
\eeq
{and the following equation, which approximates the central temperature (see, e. g., Kippenhahn \& Weigert \cite{kippenhahnetal}):}
\beq
T_{\rm cent} \propto \frac{M}{R}
\label{equ:centtempapprox}
\eeq
{This equation is obtained from the equation of hydrostatic equilibrium using the EOS of an ideal gas. We found that this expression approximates the qualitative behavior of the central temperature quite well in terms of predicting where it will increase during the phase of collapse and runaway accretion. However, an important caveat of this approximation is that it assumes an ideal gas. This is not true as the objects will become more and more degenerate as they contract and their mass increases (see Section \ref{sect:minmass} and Appendix \ref{appendix:B}).} 

\subsection{Results of the hot start accreting model: Overall behavior}  
First we study and compare the overall behavior of the objects, forming {via} hot start {core} accretion, employing our fiducial model.
In Fig \ref{fig:overall} one can see the {luminosity}, the radius, {the specific entropy above the core and the mass averaged degeneracy related $\Theta_e$ factor (see Equ. \ref{equ:thetae}).}
We show the evolution starting at $10^6$ years,
as the foregoing growth of the solid core to the isolation mass and the subsequent attached phase of the object are equal in the cold start case of the
core accretion scenario. \\ \\
As one can see in the radius plot of Fig. \ref{fig:overall}, as the planets are still growing in mass via runaway mass accretion,
they will also grow in size (after having collapsed due to the transition from the attached to the detached phase) until they have reached their final mass. The end of the mass accretion
produces as sharp bend at approximately $3 \times 10^6$ years in the radius evolution for the 30 $\mj$ object (and earlier for the lower mass objects) and
inverts the radius growth phase into a contraction phase. This is a difference to the cold start model of the core accretion scenario, where one observes a gradual contraction of the planet
once it has collapsed when entering the detached phase (see e.g. Mordasini et al. \cite{mordasinietal2011}, Marley et al. \cite{marleyetal}). This difference is
due to the inclusion of accretion luminosity into the internal luminosity. Thus, the gas is not accreted onto the already contracting object as in a cold start, but it is accreted injecting all its released gravitational potential energy into the object, which prevents it from contraction. This means that the gas is accreted at high entropy.
The radii one finds directly at the end of formation can be regarded as possible initial radii for hot start evolution models if one assumes that the objects have formed according to the core accretion scenario. However, there is one caveat which must be addressed: As the accreted gas first deposits it's liberated gravitational energy in the upper regions of the object in the runaway accretion phase of a hot start, the simplifiying assumptions made for the luminosity in Sect. \ref{sect:lumsim} might break down. During a ''hot'' accretion radiative zones well inside the object could develop (see e.g. Mercer-Smith et al. 
\cite{mercersmith}), so that our assumption that ${\rm 
d} \ l / {\rm d} \ r = 0$ becomes invalid in these regions. Thus by forcing the object to have an adiabatic structure our results could depart quite significantly from the correct values in this phase. However, as the Kelvin-Helmholtz timescale of an object at the end of formation is very short the objects would quickly forget about their thermodynamic state directly after the formation, evolving back on the correct evolutionary track (provided that the latter is not characterized by a much lower entropy). Thus one should be cautious when interpreting our results directly after formation as initial conditions for a hot start evolutionary model.\\ \\
Looking at the specific entropy above the core in Fig. \ref{fig:overall} one can see how the most massive object (30 $\mj$) reaches a specific entropy as high as 14.5 $k_{\rm B}/{\rm Baryon}$ due to the accretion of high entropy material (which is characteristic for a hot start). We note here as well that these values found for the specific entropy at the end of formation might be realistic initial values for evolution calculations if one assumes that the objects have formed via core accretion. However one should keep in mind the caveat outlined above. As the objects are fully convective in their interior (and thus have an adiabatic temperature gradient) the specific entropy is constant throughout the envelope. Again one sees that also the decrease of the entropy is partially stabilized by deuterium burning. \\ \\
In the $\Theta_e$ plot of Fig. \ref{fig:overall} one sees how the degeneracy related factor $\Theta_e$ (see Eq. \ref{equ:thetae}) evolves. For this plot $\Theta_e$ was mass averaged in the gaseous envelope. As one can see the envelope starts fully non-degenerate ($\Theta_e = 1$) and reaches a first minimum at approximately $\Theta_e = 0.7$. Instead of starting at $10^6$ years as all other plots of Fig. \ref{fig:overall}, the $\Theta_e$ plot starts at $10^5$ years in order to show the initial value of $\Theta_e=1$. After having reached a value of 0.7, $\Theta_e$ then starts to increase again at the beginning of the runaway phase and then drops down to 0.6 at the moment where the objects envelope collapses (at $\approx 10^6$ years). One can see that deuterium burning will partially stabilize $\Theta_e$ as well. After this stabilization phase, however, $\Theta_e$ will start do decrease stronger again, approaching small values between 0 and 0.1 at the end of the simulation, corresponding to stronger degeneracy.
 \\ \\
\subsubsection{Comparison of the stabilization radii to the results of Burrows et al. (\cite{burrowsetal1997})}
As the SBM11 hot start calculations (to which we compare our work in the next section) are based on the model of Burrows et al. (\cite{burrowsetal1997}),
it seemed worthwhile to compare to those results, especially as the data of the Burrows et al. (\cite{burrowsetal1997}) is available online. In Fig. \ref{fig:burrcomprad} one can see the radius evolution obtained by Burrows et al. (\cite{burrowsetal1997}) (dashed green lines) and our own results
(red solid lines). The results of Burrows et al. (\cite{burrowsetal1997}) have been shifted in time in order to make the phases of deuterium burning coincide.
This is justified as the Burrows et al. (\cite{burrowsetal1997}) results do not consider the formation of the object, thus the definition of $t=0$ is not equal. 
As one can see in Fig. \ref{fig:burrcomprad}, the radius at which the partial stabilization of the objects occurs is the same in both models and also the subsequent
radial evolutions are nearly equal. This again underlines the reliability of our model assumptions, i.e. the way how we calculate the luminosity and the way how deuterium burning
was included into the simulation.
\begin{figure*}
\begin{minipage}{0.5\textwidth}
	      \centering
       \includegraphics[width=0.95\textwidth]{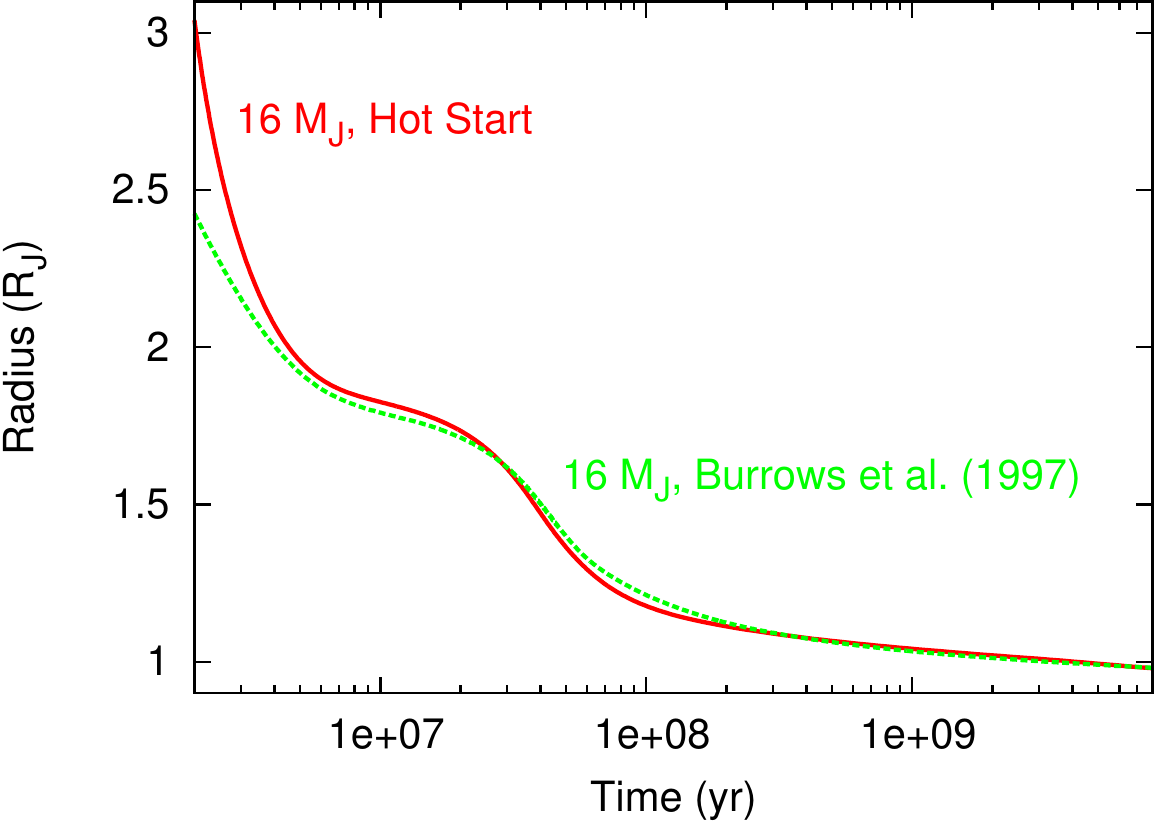}
     \end{minipage}\hfill
     \begin{minipage}{0.5\textwidth}
	      \centering
       \includegraphics[width=0.95\textwidth]{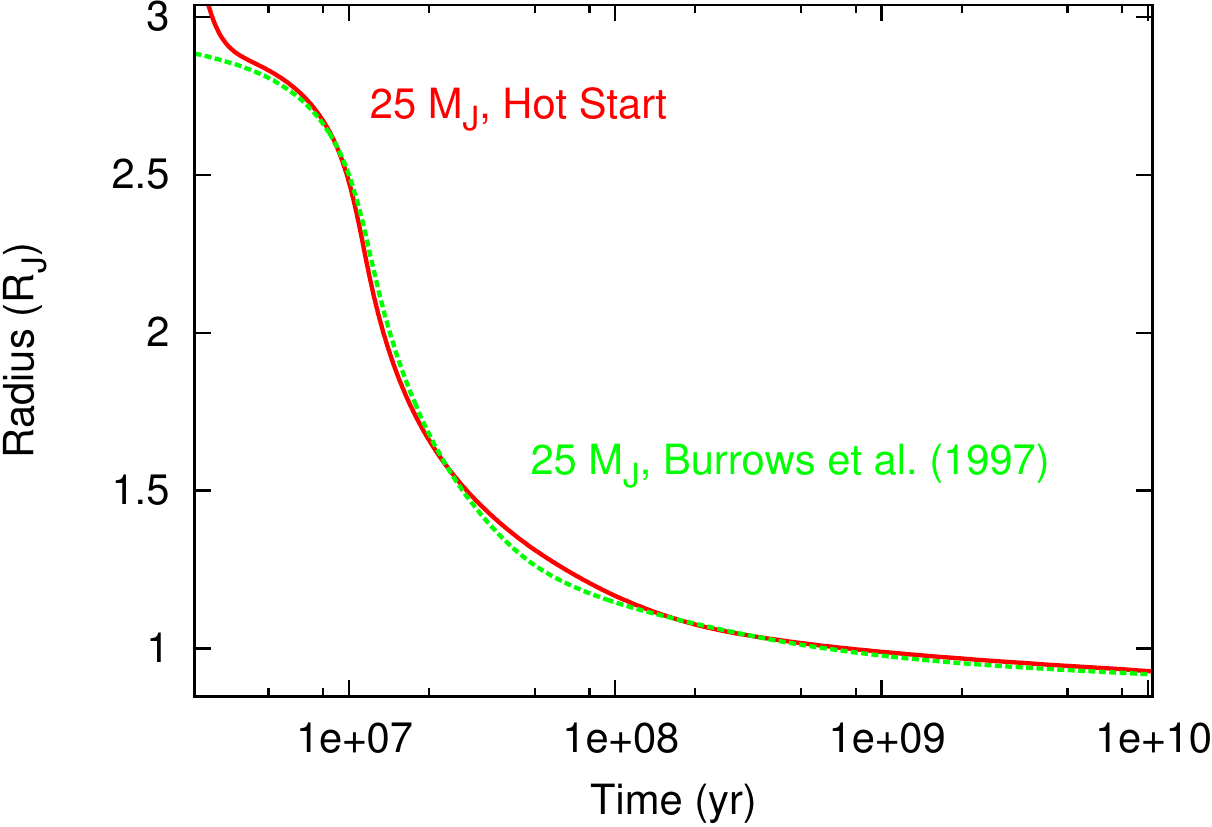}
     \end{minipage}\hfill
     \caption{Radial evolution of deuterium burning objects (\textbf{hot start} accretion). The solid red lines show our own results, whereas the dashed green lines show the results
	      of Burrows et al. (\cite{burrowsetal1997}). As the $t=0$ definition is not the same, the Burrows et al. (\cite{burrowsetal1997}) results where shifted
	      in time in order to make the phases of deuterium burning coincide. The mass of the objects is indicated in the plots.} 
      \label{fig:burrcomprad}
\end{figure*}
\subsubsection{Comparison to the SBM11 results}
Comparison with SBM11 (SBM11 also assumed a Helium mass fraction of 0.25, along with
a deuterium number fraction with respect to hydrogen of $2 \times 10^{-5}$ as their fiducial model)	
shows that the radii at which the planets become partially supported against contraction by
deuterium burning roughly coincide with our results. A 20 $\mj$ object will be in this phase at a radius between 2 and 2.5 Jupiter radii in the SBM11
results, which agrees with our findings. For the lower mass objects in the 14 to 15 $\mj$ range this phase lies around 1.5 Jupiter radii which is also true
for our results. \\ \\
Furthermore, when looking at the evolution of the ratio $L_{\rm D}/L_{\rm tot}$ we find that for a 13 $\mj$ object it peaks at approximately
0.5 which is the same result SBM11 obtain. Also, we and SBM11 both find that for $M\gtrsim$ 14 $\mj$
the peak of $L_{\rm D}/L_{\rm tot}$ is above 0.8, reaching more than 0.9 for the heavier objects $\gtrsim$ 20 $\mj$. The results for $L_{\rm D}/L_{\rm tot}$
for objects below 12 $\mj$ are also similar to the SBM11 results, the only deviation is observed for the 12 $\mj$ object,
where our peak value for $L_{\rm D}/L_{\rm tot}$ is twice as high. One must note, however, that $L_{\rm D}/L_{\rm tot}$ is a very steep function of the mass in this range, thus slight differences in the models might be seen quite strongly. \\ \\
Also, when looking at the luminosities
at which the objects are partially stabilized due to deuterium burning we find a very good agreement between our results and those of
SBM11. \\ \\
In our model and in the model of SBM11 the metallicity enters in two different ways: It sets the opacity in the SBM11 case and the core mass in our case. As both SBM11 and we ourselves find that the metallicity has a significant influence on the deuterium burning process (see Section \ref{subsubsec:paramstudy}), it is rather a coincidence
that we straightaway found the metallicity for our fiducial model such that it yields approximately the same deuterium burning behavior as in the
SBM11 case. Thus even though they also assumed a solar metallicity
in their fiducial model it was per se not clear that we would get a good agreement. It shows, however, that once one has found this metallicity value which seems to correct for the effect of the different implications
of the metallicities in both models, the results seem to be approximately the same in all considered quantities.
\subsection{Parameter study}
\label{subsubsec:paramstudy}
\begin{figure*}
\begin{minipage}{0.5\textwidth}
	      \centering
       \includegraphics[width=0.95\textwidth]{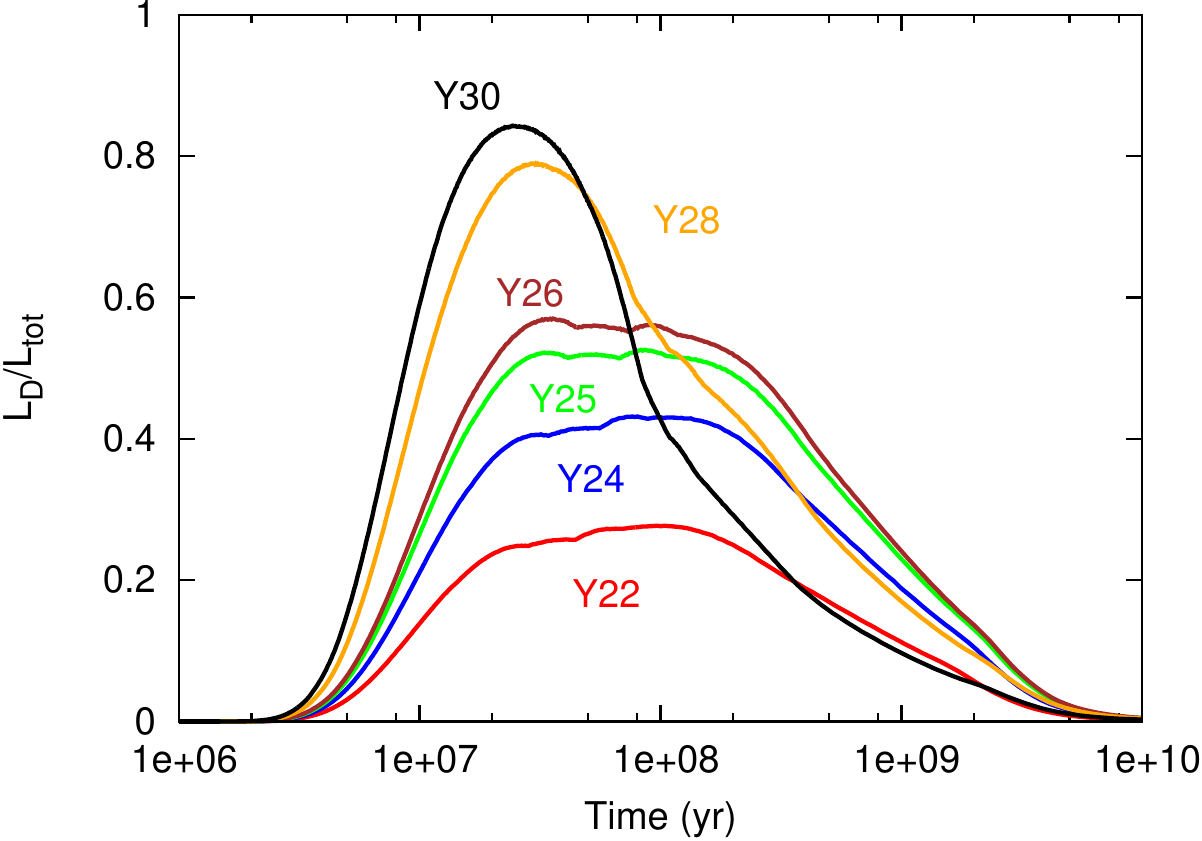}
     \end{minipage}\hfill
     \begin{minipage}{0.5\textwidth}
      \centering
       \includegraphics[width=0.95\textwidth]{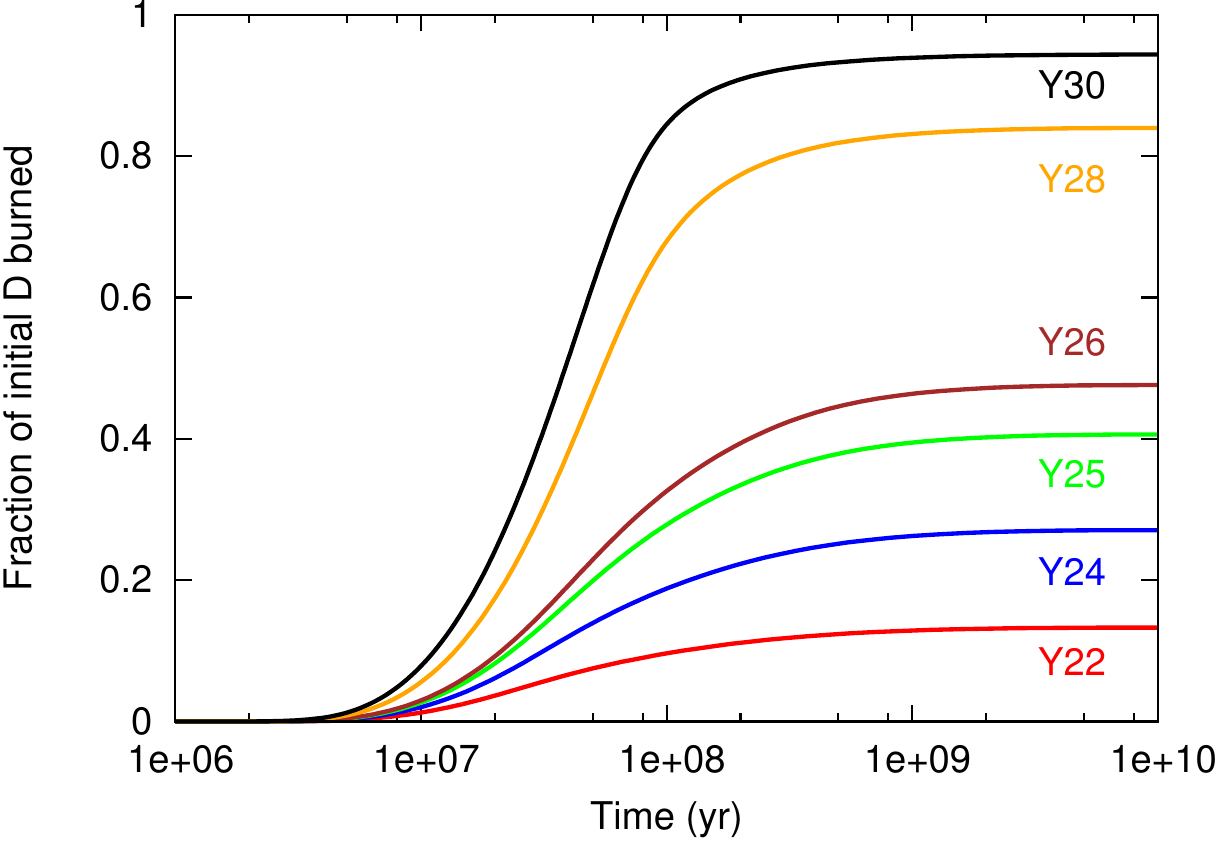}
     \end{minipage}
     \begin{minipage}{0.5\textwidth}
	      \centering
       \includegraphics[width=0.95\textwidth]{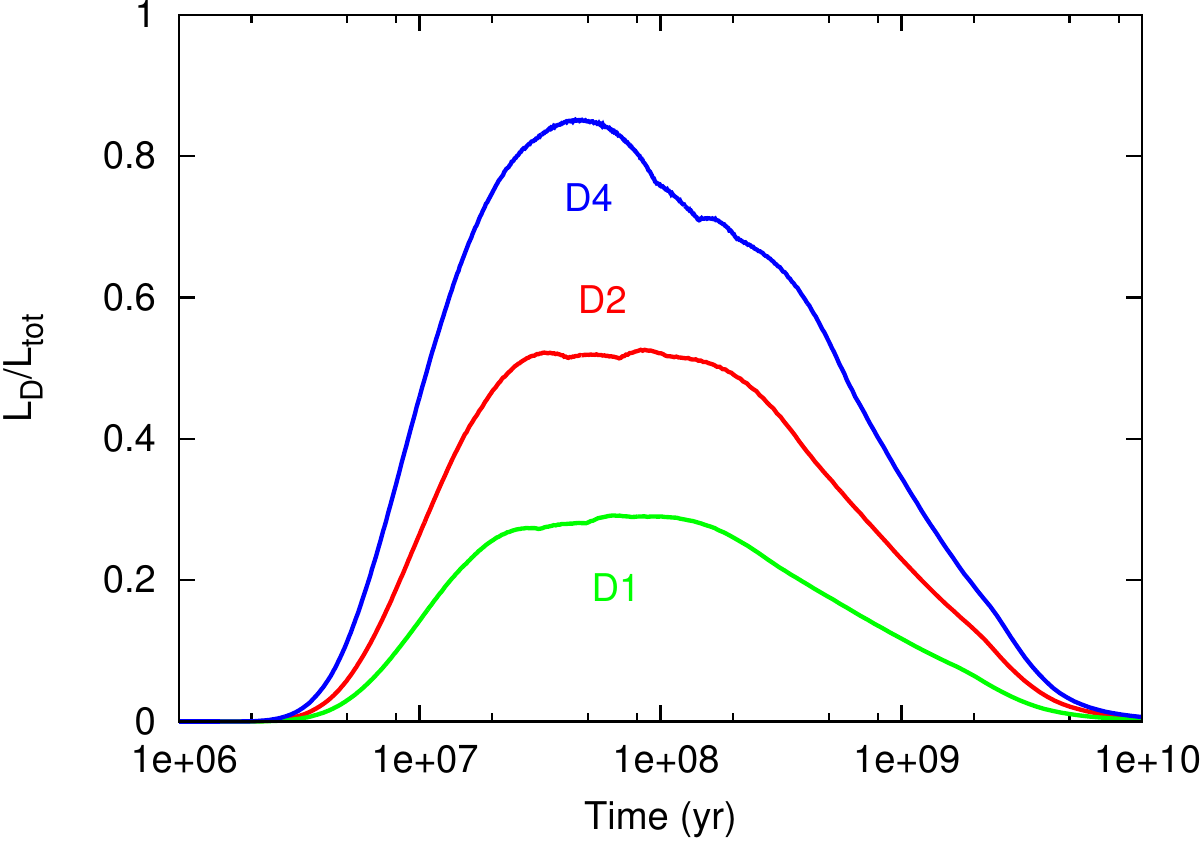}
     \end{minipage}\hfill
     \begin{minipage}{0.5\textwidth}
      \centering
       \includegraphics[width=0.95\textwidth]{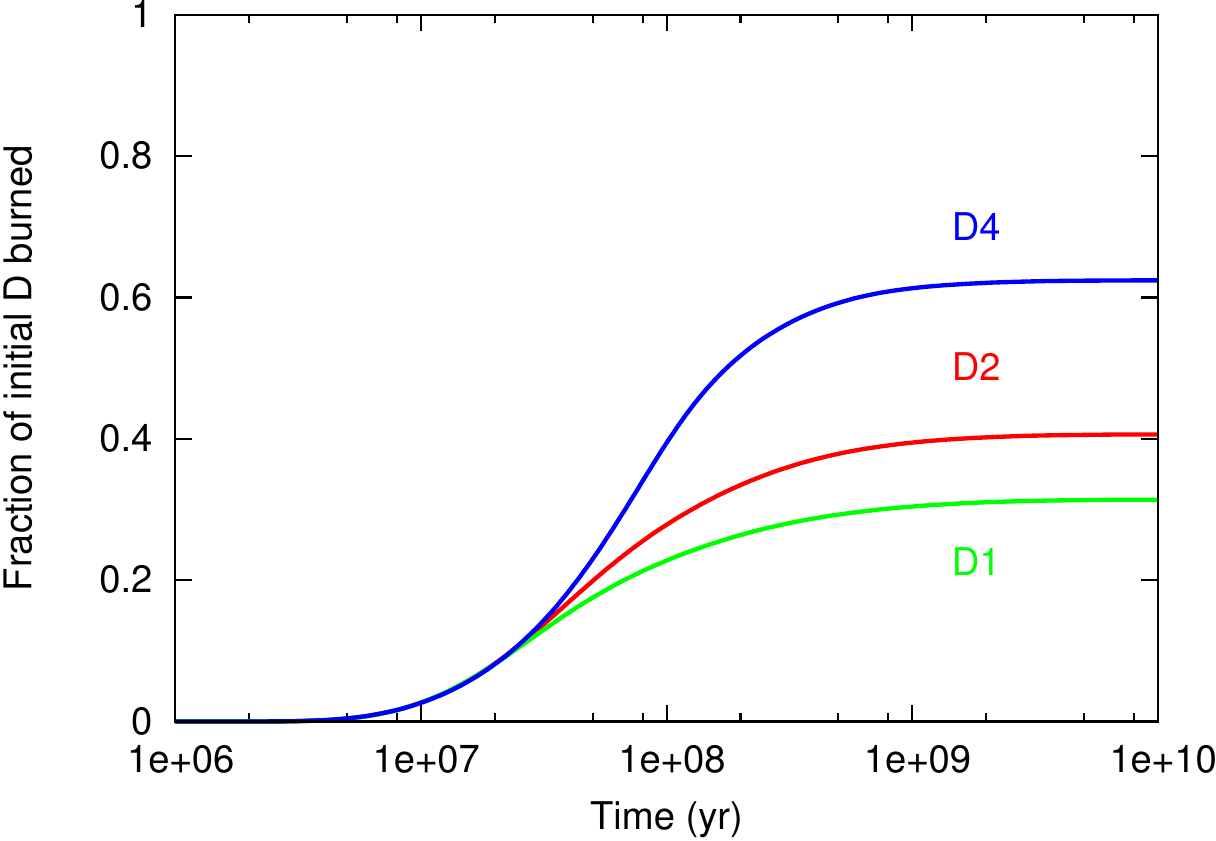}
     \end{minipage}
     \begin{minipage}{0.5\textwidth}
	      \centering
       \includegraphics[width=0.95\textwidth]{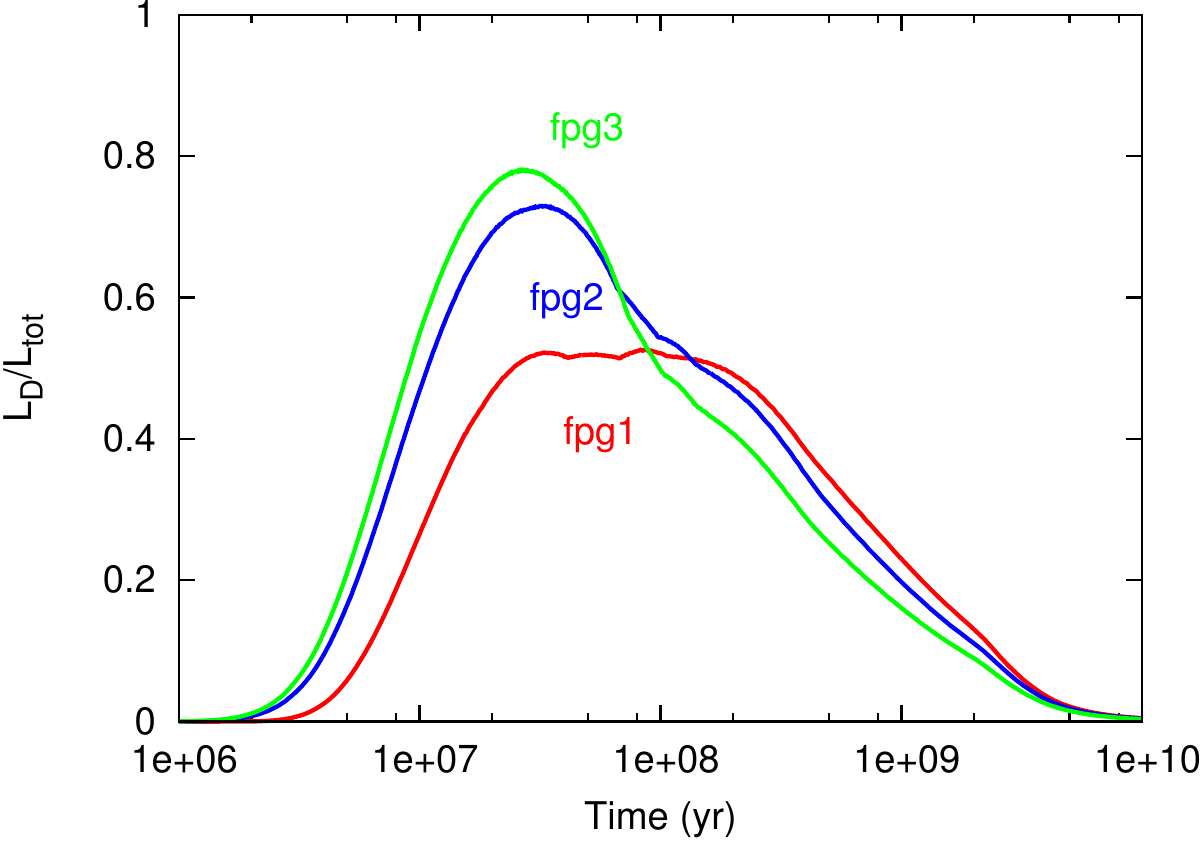}
     \end{minipage}\hfill
     \begin{minipage}{0.5\textwidth}
      \centering
       \includegraphics[width=0.95\textwidth]{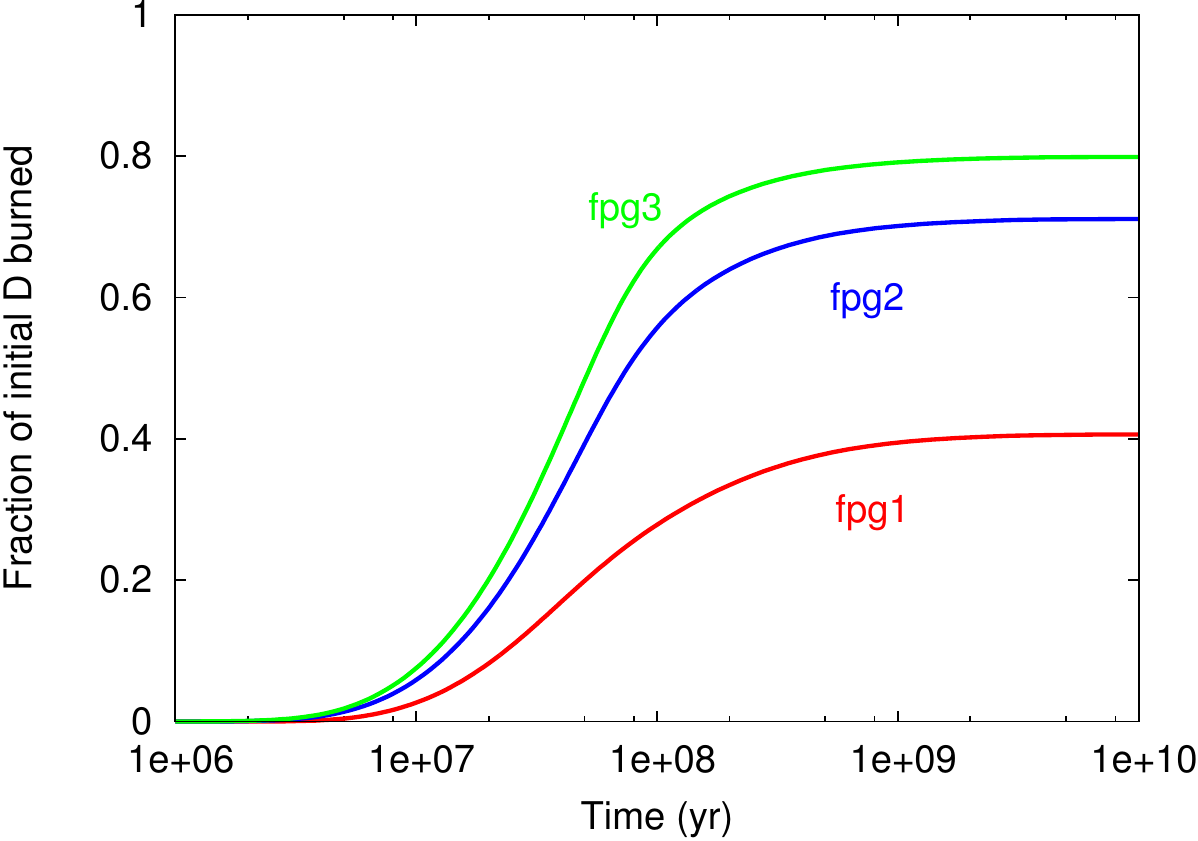}
     \end{minipage}
     \caption{Left column: The ratio $L_{\rm D}/L_{\rm tot}$. Right column:
              The depletion of deuterium. The first row shows our results for different helium
              abundances, the second row shows the influence of different deuterium abundances and the last row shows different dust-to-gas
              ratios (as defined in Tab. \ref{tab:compruns}). The mass of all objects is 13 $\mj$ (\textbf{hot start}). The run names corresponding to the individual lines are indicated in the plot.} 
      \label{fig:paramshot1}
\end{figure*}
In their paper, SBM11 tested how the deuterium burning process is influenced when parameters such as the
hydrogen abundance, the deuterium abundance and the metallicity are varied. We also carry out this parameter study to further test our own simulation, before
we proceed to the study of the cold start results of objects forming by core accretion in Section \ref{subsect:coldstart}. In the upper left and right panel of Fig.
\ref{fig:paramshot1} one can see the $L_{\rm D}/L_{\rm tot}$ evolution and the deuterium depletion for the runs with different helium abundances (Y22-Y30)
for a 13 $\mj$ object. \\ \\
The bumpy pattern in some $L_{\rm D}/L_{\rm tot}$ plots is due to the interpolation of the opacities in the tables of Freedman et al. (\cite{freedmanetal}) {and thus a numerical artifact}. \\ \\
Concerning the height of the $L_{\rm D}/L_{\rm tot}$ peaks we again find a very good agreement with the results of SBM11. We do, however,
also observe
that our objects burn 10-50 \% more deuterium as they sustain a high burning rate over a longer period of time.
Overall, we thus confirm the fact that a higher helium abundance yields a stronger deuterium burning process, due to the increased central
density of the objects (see Eq. \ref{equ:stahler})\footnote{In the terminology used in this paper, by saying ''central'' we always mean the innermost
layer of gas just above the solid core}. The results of the runs with different deuterium abundances (D1, D2 and D4) are shown in middle row of Fig.
\ref{fig:paramshot1}. Again, we find a very good agreement when comparing to the peak value of $L_{\rm D}/L_{\rm tot}$ of the Spiegel et al.
(\cite{spiegeletal2011}) results. Furthermore, we again find the already mentioned fact that our objects burn more deuterium than in the SBM11 case.
A possible reason for this could be the use of different opacities: Higher opacities enable the objects to keep high central temperatures over a longer period of time (the so-called blanket effect). Furthermore SBM11 use a proper atmospheric model while we use gray photospheric boundary conditions.  As the deuterium burning rate is depending on the temperature very strongly (see Eq. \ref{equ:stahler}) this can sustain the burning rate
over a longer period of time as well.
We conclude that we are in a good agreement with the results by SBM11 but out objects burn up to 50 \% more deuterium.
\begin{figure*}
\begin{minipage}{0.5\textwidth}
	      \centering
       \includegraphics[width=0.95\textwidth]{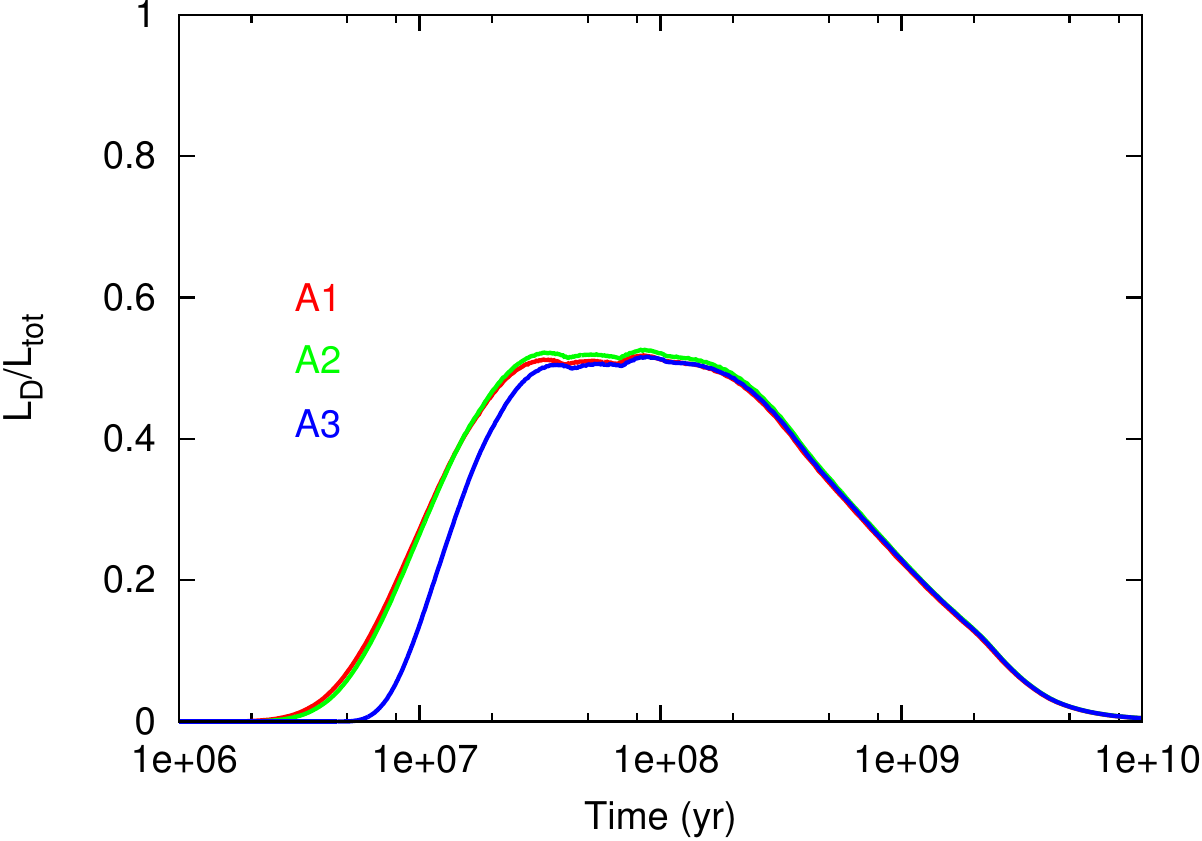}
     \end{minipage}\hfill
     \begin{minipage}{0.5\textwidth}
      \centering
       \includegraphics[width=0.95\textwidth]{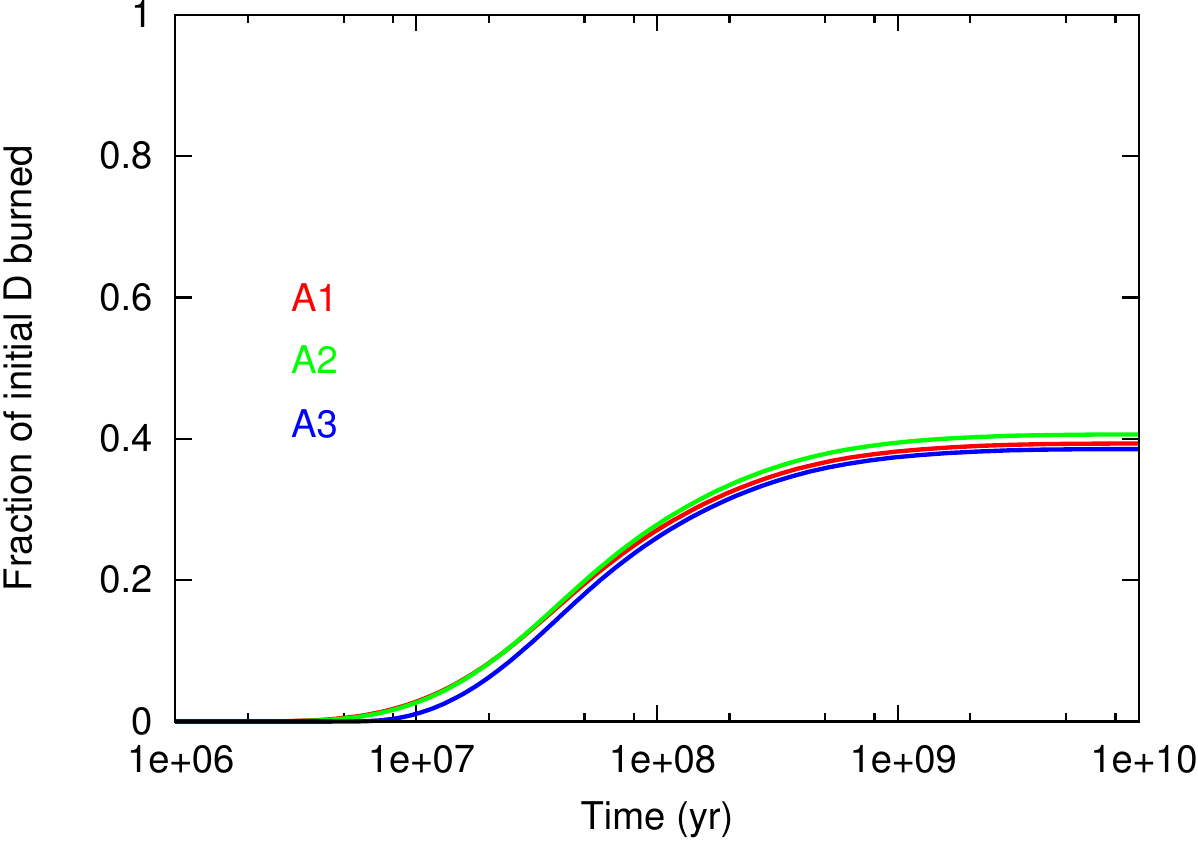}
     \end{minipage}
     \caption{Left panel: The ratio $L_{\rm D}/L_{\rm tot}$ as a function of time. Right panel:
              The depletion of deuterium for the runs with different maximum allowed mass accretion
              rates (as defined in Tab. \ref{tab:compruns}). The mass of the objects is 13 $\mj$ (\textbf{hot start}). The run names corresponding to the individual lines are indicated in the plot.} 
      \label{fig:mdothot}
\end{figure*}
Finally we consider the runs with varied dust-to-gas ratio (i.e. varied metallicity) and with varied maximum allowed gas accretion rates, which we
cannot compare to the SBM11 results, as they do not consider the formation process or the presence of solid cores. It is worthwhile
to study the impact of these parameters nonetheless. In the lower row of Fig. \ref{fig:paramshot1} one can see the results obtained for the $L_{\rm D}/L_{\rm tot}$ evolution
and for the deuterium depletion for runs with different dust to gas ratios fpg1, fpg2 and fpg3, corresponding to dust-to-gas ratios of 0.04, 0.09 and 0.12, respectively.
The considered dust-to-gas ratios have a significant influence on the deuterium burning behavior. The higher it is, the higher the nuclear burning
luminosity compared to the total luminosity and the more deuterium is burned. One can, in principle, imagine at least two reasons for this behavior. \\ \\
One reason is that a higher dust-to-gas ratio will lead to a higher solid surface density in the disk, which will in turn lead to a higher solid accretion
rate. A higher solid accretion rate also means a higher core luminosity, which might influence the objects structure as the core accretion luminosity is included into the
structure calculation. This is motivated by the assumption of our model that the solids impacting the object get destroyed and the debris then sinks to the core, following the so-called sinking approximation (Pollack et al. \cite{pollacketal}). Thus one would expect the central temperature to be higher, which has a large influence on deuterium burning, as the nuclear
energy generation rate is highly dependent on the temperature (see Eq. \ref{equ:stahler}). \\ \\
The other possible reason is due to the
fact that we assume that all solids accreted by the object eventually get incorporated in the core, leading to a higher core mass. A higher core mass will, even though the core is also growing
in size as it grows in mass, result in a higher gravitational acceleration on top of the core. In order to remain in hydrostatic equilibrium one thus needs
a higher pressure gradient, i. e. a higher temperature gradient, as the temperature gradient in a convective layer (which the central gas layer surely is) is
\beq
\frac{\partial ~ T}{\partial ~ r} = -\frac{1}{c_P}\frac{\partial ~ {\rm ln}\rho}{\partial ~ {\rm ln}T}\frac{Gm(r)}{r^2}
\label{equ:tempgradcore}
\eeq
which is proportional to the gravitational acceleration. This means that a higher temperature gradient might lead to a higher central temperature, if it dominates over the fact that the gas cannot reach as deep within the object as it could with a lighter core (as objects with a higher core mass are found to have a smaller total radius and a larger core radius). If it does, this would create a higher deuterium burning rate. \\ \\
We were able to identify that it is rather the increased core mass than the additional core accretion luminosity which affects the deuterium
burning process in the following way: In Fig. \ref{fig:mdothot}, one can see the $L_{\rm D}/L_{\rm tot}$ evolution and the deuterium depletion for
the runs with a different maximum gas accretion rate (A1-A3). As one can see, the object with the lowest maximum mass accretion rate reaches the
the peak value the latest, as it needs a longer time to form. However, all three runs have the same peak value of $L_{\rm D}/L_{\rm tot}$, and the same
subsequent evolution of $L_{\rm D}/L_{\rm tot}$ after having reached the peak value. One can also see that all three runs burn approximately the same amount
of deuterium. The remaining slight variations in the A1-A3 runs for the peak values of $L_{\rm D}/L_{\rm tot}$ and the deuterium depletion are mainly due to a simulation artifact: As we cannot switch off the accretion instantaneously (in order to remain numerically stable) it is difficult to exactly reach the desired total mass. The deuterium burning rate
depends on the objects density (see Eq. \ref{equ:stahler}) and thus one can see that a slightly different final mass will have an impact. \\ \\
In conclusion we see that the amount of deuterium
burned and the strength of deuterium burning is independent of the luminosity due to gas accretion, which is included in the internal luminosity in the hot start case (see Eq. \ref{equ:internallumhot}). We have furthermore found that the contribution of the solid accretion luminosity to the total accretion luminosity is very small in the phase
of runaway gas accretion. This makes it very unlikely that the different solid accretion luminosities in the runs with different dust-to-gas ratios (''fpg''-runs) have a strong impact on the deuterium burning process.
The luminosity due to gas accretion is much stronger, but apparently has very little effect on deuterium burning.
It rather seems to be the presence of the core itself which influences deuterium burning via Eq. \ref{equ:tempgradcore}. \\ \\
\begin{figure*}
\begin{minipage}{0.5\textwidth}
	      \centering
       \includegraphics[width=0.95\textwidth]{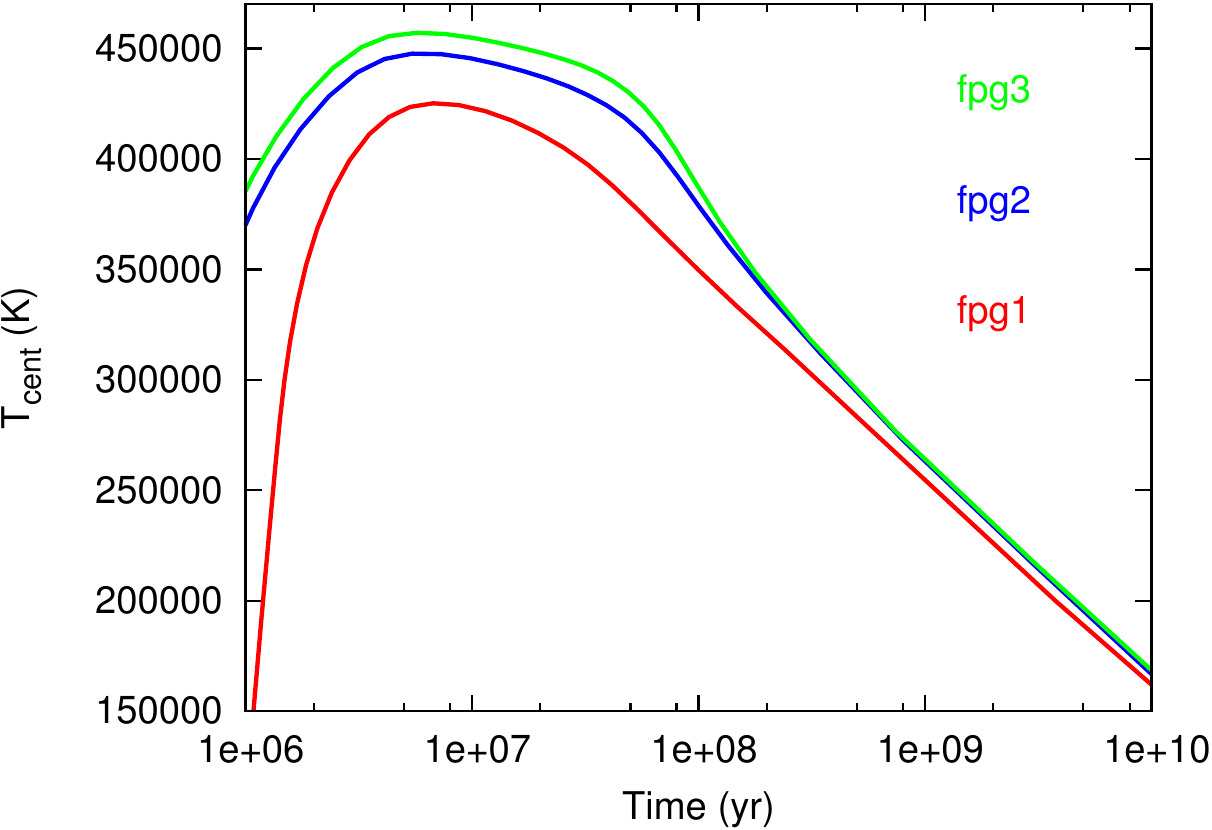}
     \end{minipage}\hfill
     \begin{minipage}{0.5\textwidth}
      \centering
       \includegraphics[width=0.95\textwidth]{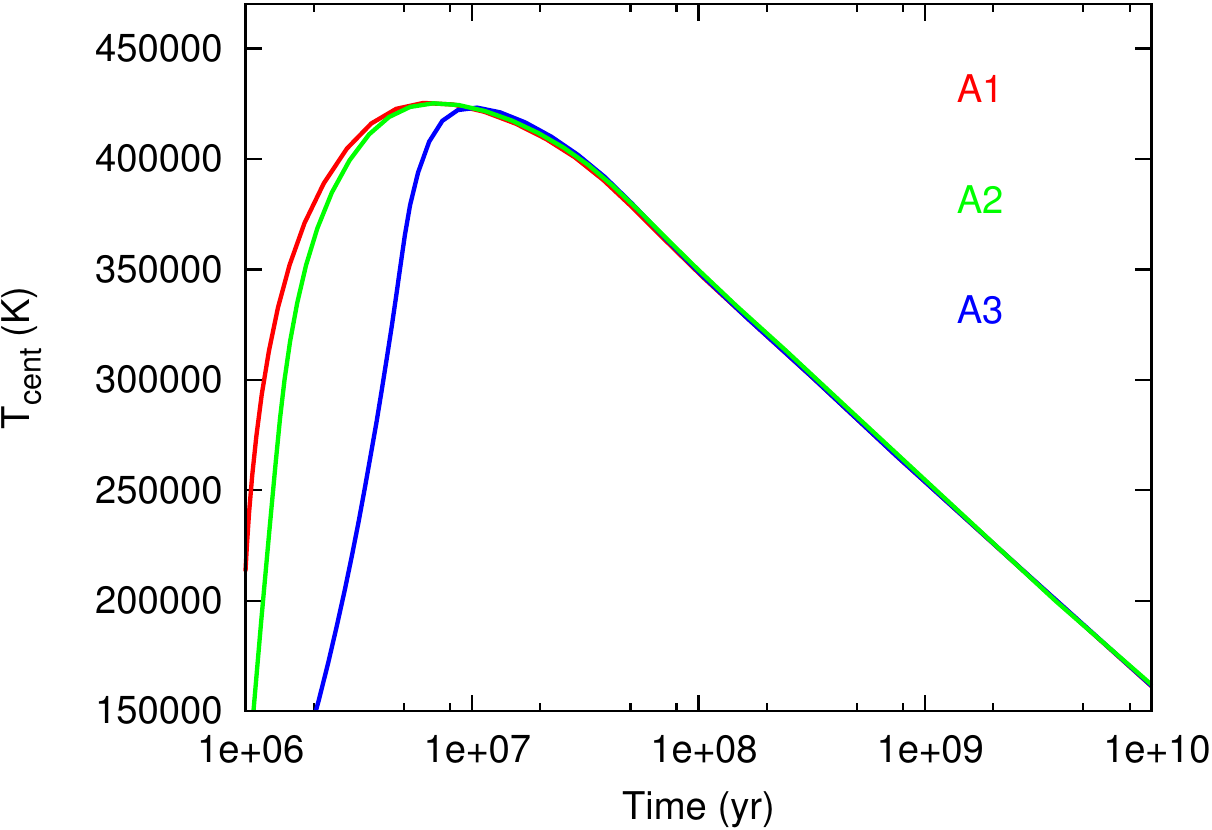}
     \end{minipage}
     \caption{Central temperature for the different dust-to-gas ratio runs (left plot) and for the different maximum allowed gas accretion runs (right plot) (\textbf{hot start}).} 
\label{fig:centtemphot}
\end{figure*}
In order to further show this, we plot in Fig. \ref{fig:centtemphot} the central temperature for the different dust-to-gas ratios and
for the different maximum gas accretion rates. One clearly sees that for the runs with different dust-to-gas ratios, the central temperature is always
higher for the runs with higher dust-to-gas ratio, which persists until the end of the simulation at $10^{10}$ years. The runs with different maximum
gas accretion rate, however, reach approximately the same central temperature at the end of formation and their subsequent central temperature
is the same. The core masses for the three runs ''fpg1'', ''fpg2'' and ''fpg3'' are 35.2 $\mearth$, 77.5 $\mearth$ and 101.5 $\mearth$, respectively. For the three runs ''A1'', ''A2'' and ''A3'' the core masses are 39.1 $\mearth$, 35.2 $\mearth$ and 31.2 $\mearth$.
\begin{table}
\caption{Parameter study for cold start simulations}\label{tab:minmassruns}
\begin{center}
\begin{tabular}{lc}
\hline\hline
Run name & Property \\ \hline
Y22c & $Y=0.22$ \\
Y24c & $Y=0.24$ \\
Y25c & $Y=0.25$ \\
Y26c & $Y=0.26$ \\
Y28c & $Y=0.28$ \\
Y30c & $Y=0.3$ \\ \hline
D1c & ${\rm [D/H]}=1\times 10^{-5}$ \\
D2c & ${\rm [D/H]}=2\times 10^{-5}$ \\
D2.5c & ${\rm [D/H]}=2.5\times 10^{-5}$ \\
D3c & ${\rm [D/H]}=3\times 10^{-5}$ \\
D4c & ${\rm [D/H]}=4\times 10^{-5}$ \\ \hline
fpg1c & Dust-to-gas ratio = 0.04 $\hat{=} \ {\rm [Fe/H]} = 0$\\
fpg2c & Dust-to-gas ratio = 0.06 $\hat{=} \ {\rm [Fe/H]} = 0.18$\\
fpg3c & Dust-to-gas ratio = 0.08 $\hat{=} \ {\rm [Fe/H]} = 0.3$\\
fpg4c & Dust-to-gas ratio = 0.1 $\hat{=} \ {\rm [Fe/H]} = 0.4$\\
fpg5c & Dust-to-gas ratio = 0.12 $\hat{=} \ {\rm [Fe/H]} = 0.48$\\ \hline
A1c & $\dot{M}_{\rm gas,max}$ = $10^{-1}$ $\mearth$/yr\\
A5-2c & $\dot{M}_{\rm gas,max}$ = $5\times10^{-2}$ $\mearth$/yr\\
A2c & $\dot{M}_{\rm gas,max}$ = $10^{-2}$ $\mearth$/yr\\
A5-3c & $\dot{M}_{\rm gas,max}$ = $5\times10^{-3}$ $\mearth$/yr\\
A3c & $\dot{M}_{\rm gas,max}$ = $10^{-3}$ $\mearth$/yr\\
\hline
\end{tabular}
\end{center}
\end{table}
\section{Cold start scenario}\label{subsect:coldstart}
In this main section, we consider the formation of objects forming via the core accretion scenario using the cold start assumption, which
means that all gravitational energy liberated at the accretion shock is radiated away, not contributing to the internal luminosity (see Eq. \ref{equ:internallumcold}). This is equivalent to the accretion of low entropy material. We stress again, however, that it is still an open question whether the total accretional luminosity is radiated away at the shock (Marley et al. \cite{marleyetal}, Stahler et al. \cite{stahleretal1980}). A cold start is the standart scenario for the formation via core accretion with a gradual accretion of matter. The typical timescale for the runaway gas accretion phase during the formation of a 13 $\mj$ object is approximately $4 \times 10^5$ years if one considers a $\dot{M}_{\rm max}$ of $10^{-2}$ $\mearth/{\rm yr}$ (almost the complete mass of the object is accreted in the runaway phase). This is quite long compared to the typical formation timescale of direct collapse models, which is approximately equal to the orbital timescale and of the order of $10^2$-$10^3$ years in the 
distances feasible for a direct collapse to occur ($a_{\rm direct}\ge 50 \ {\rm AU}$ for 10 $\mj$) (Janson et al. \cite{jansonetal}). It is, however, unclear if a clump at final mass forms in the direct collapse model or whether it would accrete gas as well.
For our fiducial model, we will use the parameters specified in Tab. \ref{tab:standart}. In Section \ref{subsubsec:overall}, we study the general
behavior of objects forming with a cold start. In Section \ref{sect:minmass} we investigate the minimum mass limit
for deuterium burning. In order to do so we vary the parameters of the model and carried out the runs given in Tab. \ref{tab:minmassruns}.
\begin{figure*}
     \begin{minipage}{0.5\textwidth}
	      \centering
       \includegraphics[width=0.95\textwidth]{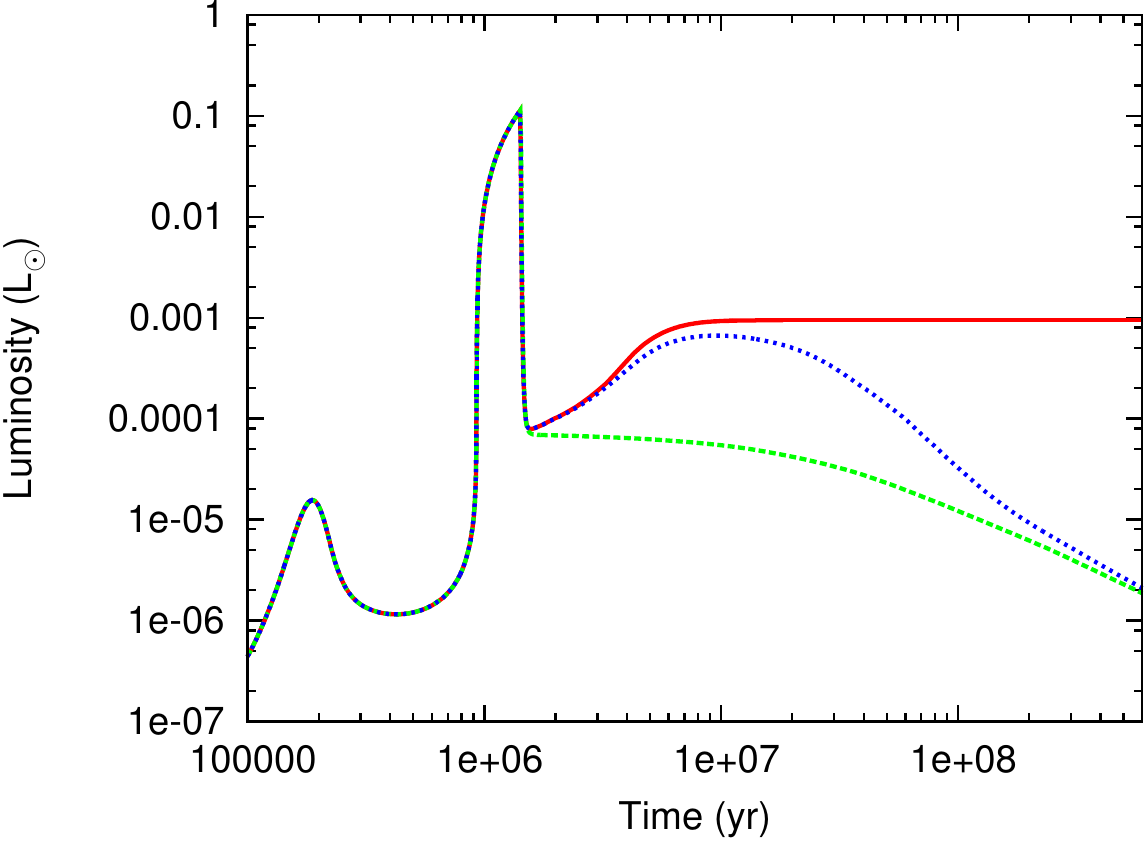}
     \end{minipage}\hfill
     \begin{minipage}{0.5\textwidth}
      \centering
       \includegraphics[width=0.95\textwidth]{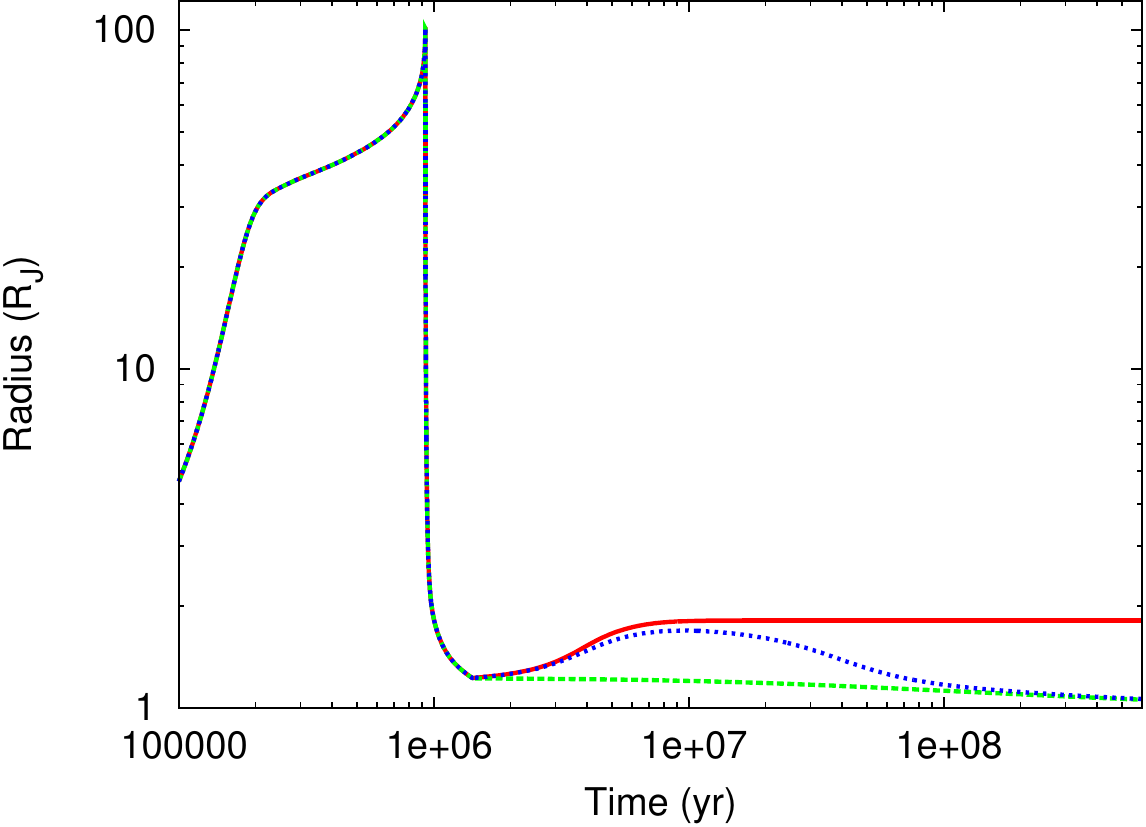}
     \end{minipage}
     \begin{minipage}{0.5\textwidth}
	      \centering
       \includegraphics[width=0.95\textwidth]{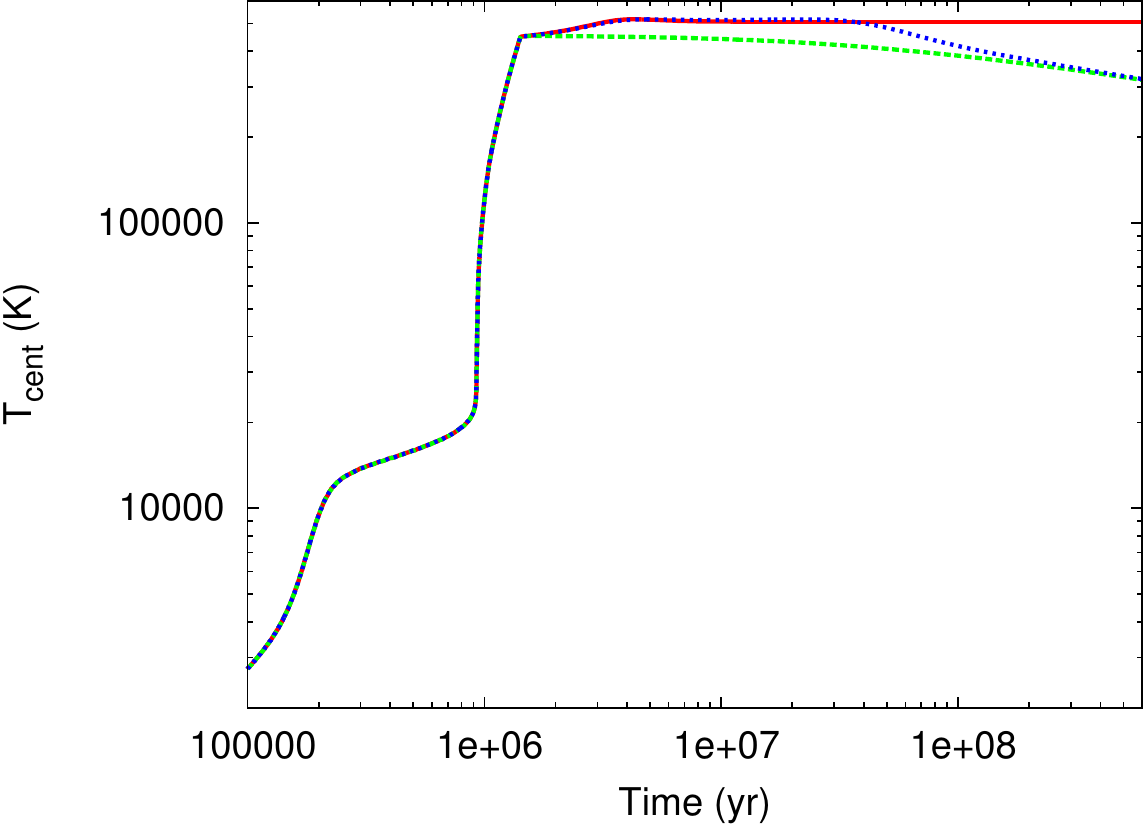}
     \end{minipage}\hfill
     \begin{minipage}{0.5\textwidth}
      \centering
       \includegraphics[width=0.95\textwidth]{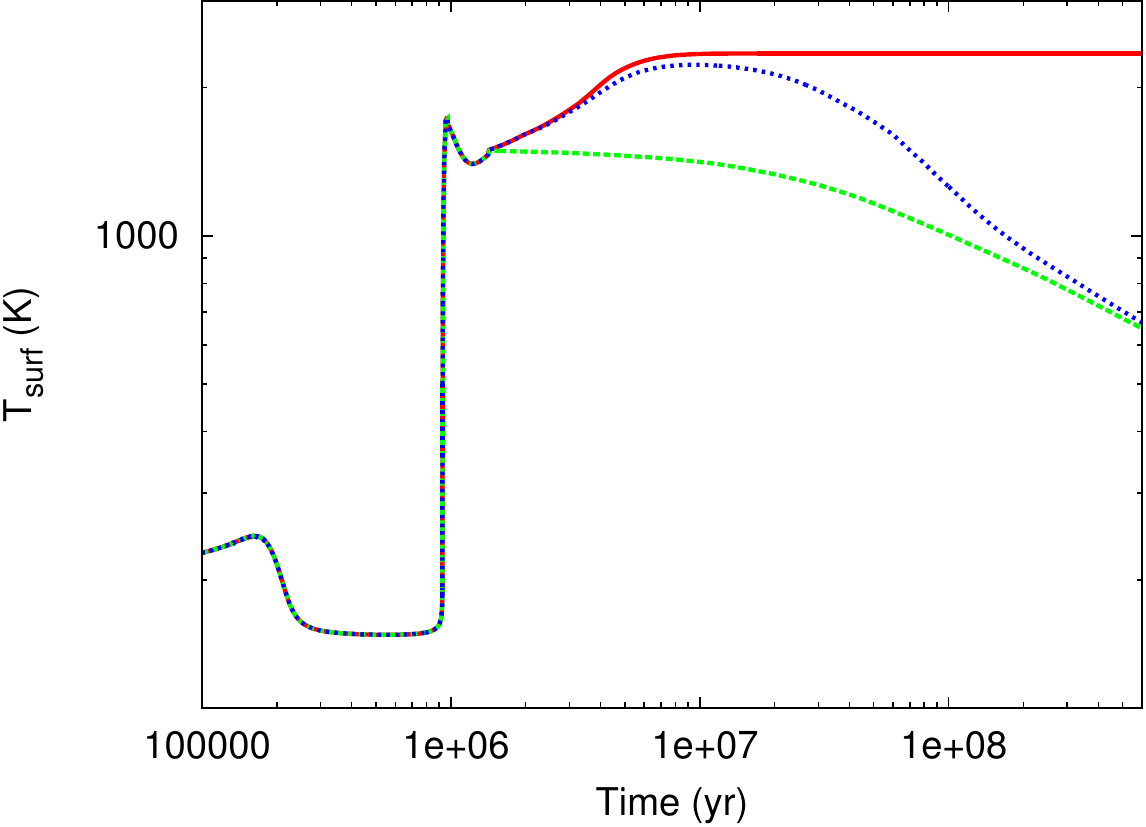}
     \end{minipage}
     \caption{Formation and subsequent evolution of a 16 $\mj$ object forming via core accretion with a radiatively efficient shock (\textbf{cold start}). The four panels show the temporal evolution of the total luminosity (internal $+$ shock) (upper left), the radius (upper right), the central temperature (lower left) and the surface temperature (lower right). The dotted blue line shows the results obtained for the fiducial model as
	      specified in table \ref{tab:standart}. For the solid red line, the deuterium abundance was kept constant, while for the dashed green line, deuterium burning
	      was disabled.}
\label{fig:coldstartgeneral}
\end{figure*}
\subsection{Overall behavior}\label{subsubsec:overall}
In order to understand the implications of deuterium burning on the evolution of objects forming via core accretion and a radiatively effective shock we first
describe the qualitative behavior studying the formation and subsequent evolution of a 16 $\mj$ object. In order to get an understanding for the changes
deuterium burning introduces to the internal structure we also consider the artificial cases where either deuterium burning is disabled, or where the
deuterium abundance is kept constant. Fig. \ref{fig:coldstartgeneral} shows the evolution of the object's luminosity, it's radius and it's
central and surface temperature for the three cases (deuterium burning, no deuterium burning, constant deuterium abundance).
Differences in the evolution can only be seen after the object's formation is completed, as marked by
the end of runaway gas accretion which produces the sharp decrease in the luminosity at approximately $1.3\times 10^6$ years. This is due to the disappearance of the accretion luminosity. It is clear, however, that the phases preceding the gas runaway
accretion cannot be important for deuterium burning, as the temperatures as well as the densities in the gaseous layers are far too low.
\begin{figure*}
\begin{center}
       \includegraphics[width=0.95\textwidth]{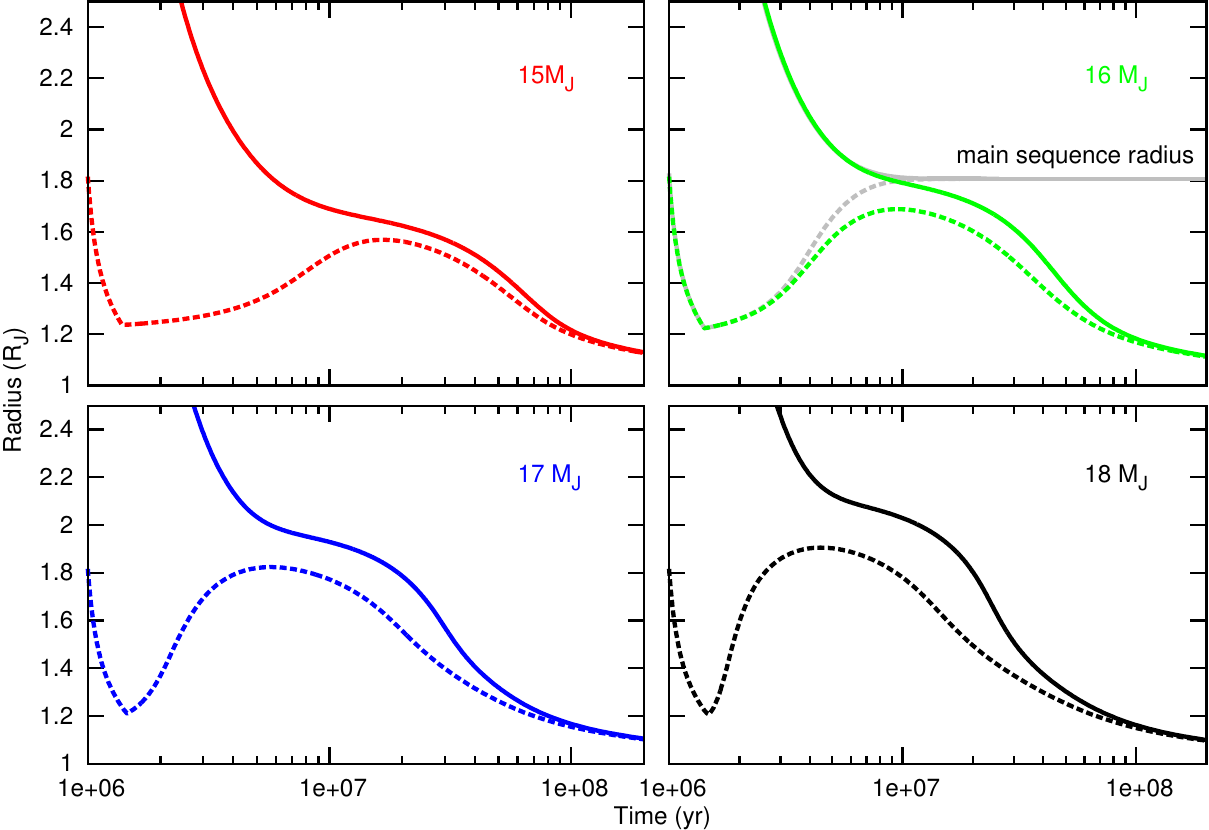}
     \caption{Radius as a function of time for objects forming via the hot start ansatz (solid line) or the cold start ansatz (dashed line).
	      The mass of the objects is indicated in each panel. In the upper right panel (16 $\mj$) the gray lines show the results obtained for
	      a constant deuterium abundance.}
      \label{fig:expandrad}
\end{center}
\end{figure*}
\subsubsection{Luminosity}
Concentrating on the panel with the luminosity in Fig. \ref{fig:coldstartgeneral}, we see that after the formation of the object, the deuterium burning will cause the
object to behave quite differently. We first note that deuterium burning produces an additional, rather flat peak in luminosity at approximately $10^7$
years where the luminosity is an order of magnitude higher than in the case with deuterium burning disabled. We also see that for objects where the
deuterium abundance is kept constant, the objects will reach a (hypothetical) deuterium burning main sequence, where the object's luminosity is generated by deuterium
burning only. This is what one would naively expect when keeping the deuterium abundance constant.
\subsubsection{Radius}
Comparing to the hot start scenario (see Fig. \ref{fig:overall}), a clearly different behavior can be seen for the evolution of the radius. Instead of stabilizing the objects against further
contraction as in the hot start case, the onset of deuterium burning is so strong that the object is inflated from 1.2 to approximately 1.7
$R_{\rm J}$ for the nominal case. This suggests that a large part of the deuterium burning luminosity is converted into gravitational potential energy
in that phase, which corresponds to a $L_{\rm D}/L_{\rm int}$ ratio (much) larger than 1 (we will discuss this later). Thus, instead of contracting and cooling at high
temperatures and gradually burning deuterium already at quite large radii as in the hot start case, a cold start yields a more abrupt start of deuterium burning. This is
due to the effect that the mass is accreted at low entropy onto an object with high density, due to the collapse from the attached to the detached phase.
Thus the objects have then much higher burning rates as the objects are already very dense when they reach the temperature needed for deuterium burning (see Eq.
\ref{equ:stahler}). The expansion process will be discussed in more detail below.
\subsubsection{Temperatures}
Finally, one can see that deuterium burning also produces a flat peak in the central and in the surface temperature when compared to the
case without deuterium burning. The case where the deuterium abundance was set constant yields a main sequence-like behavior again.
\subsubsection{Expansion}
\label{subsubsec:expansion}
In order to understand the expansion
{we} compare the radii to which the objects expand in the
cold start scenario with the radii at which the objects get partially stabilized against contraction in the hot start case. A comparison like
this seems sensible as it is well known that nuclear fusion processes in stars, such as hydrogen burning, have a thermostatic behavior (also known as the pressure temperature thermostat or stellar thermostat) (see, e.g. Palla et al. \cite{pallaetal}). This can also be seen
very clearly in the runs with constant deuterium abundance, as the objects, driven by deuterium burning, attain a stable state in the hypothetical deuterium main sequence.
Given the thermostatic nature of deuterium burning, we expect the cold start objects to expand to approximately the same radii found for
the hot start objects in the phase where they get partially stabilized. More precisely, as the objects in the hot start case are unable to completely stabilize
themselves against further contraction (as $L_{\rm D}/L_{\rm tot}$ is always smaller than 1) we expect that the cold start objects
will not be able to expand to the same radii as the ''quasi-stabilization''-radii of the hot start objects, but to a somewhat smaller radius.
The reason for $L_{\rm D}/L_{\rm tot}<1$ in the hot start case, as well as for the somewhat smaller expansion radius in the
cold start case is the decreasing deuterium abundance inside the objects.
When looking at Fig. \ref{fig:expandrad} we can see this behavior: The cold start objects expand to radii somewhat smaller than the
radii at which the hot start objects are partially stabilized. When this radius increases for the hot start objects (i.e. for higher masses) it also does so for the cold start
objects. For the 16 $\mj$ object in Fig. \ref{fig:expandrad}, we have also plotted the radius
evolution for hot and cold start with deuterium abundances held artificially constant. As one can see, they both get stabilized at the
same radius (which we call ''main sequence radius'' in the plot). The hot start object contracts to this radius from above, whereas the cold start object is at a smaller radius
when deuterium fusion ignites and must thus expand to the main sequence radius. One can observe that the main sequence radius approximately coincides with the ''quasi-stabilization''
radius of the hot start object. Thus, the consideration of the in principle unphysical (artificially held constant deuterium abundance)
main sequence radius and the thermostatic nature of nuclear fusion processes help to understand not only the stabilization behavior of the hot
start objects, but also the expansion behavior of the cold start objects.
\subsubsection{Evolutionary sequences for objects formed by core accretion, assuming a cold start}
Similar to Fig. \ref{fig:overall} for the hot start case we also show the temporal evolution of the luminosity, the surface temperature, the radius, the gravitational acceleration at the surface, the specific entropy and the mass averaged degeneracy parameter $\Theta_e$ for the cold start case in Fig. \ref{fig:overallcold}. Except for $\Theta_e$, the evolution is shown starting at $10^6$ years, i.e. shortly before the final mass is reached, and therefore shows the final stages of the object's formation in the runaway accretion phase. The time axis of $\Theta_e$ starts at $10^5$ years in order to see that the envelope indeed was fully non-degenerate in the beginning. \\ \\
The luminosity shown in the plot is the total luminosity, i.e. the sum of the internal luminosity and the accretion luminosity, as this is the luminosity which would be seen by an observer. The sharp drop in luminosity (from approximately 0.1 $\lsun$ for the 23 $\mj$ object) marks the end of runaway gas accretion and thus the end of the formation process. One can see that for the lower masses up to 13 $\mj$ there is no additional bump in the luminosity. Instead it decreases monotonously. For the objects of higher masses, however, the luminosity does not decrease monotonously but has an additional bump due to deuterium burning. For almost all masses shown in the plot the maximum luminosity due to deuterium burning is reached far after the end of formation. For the 23 $\mj$ object this is not the case: The luminosity does not increase strongly anymore after the end of formation and indeed one finds that the 23 $\mj$ object burns almost all of it's deuterium during the formation phase (this is discussed in more 
detail in Section \ref{sect:accretionimp}). One can also observe that the higher the mass of the object the higher is the bump in luminosity.\\ \\
A similar behavior is found for the surface temperature: The objects of masses up to 13 $\mj$ do not exhibit an additional bump due to deuterium burning. All objects with masses above, however, do. The peak value of the increased surface temperature is increasing with mass. Directly after formation the 23 $\mj$ object is over 1000 K hotter than the 13 $\mj$ object. As well as in the luminosity case the peak value of the deuterium burning induced temperature bump is reached later for the lower mass objects. The time difference between the 23 $\mj$ object and the 14 $\mj$ object is approximately $5 \times 10^7$ years. \\ \\
Everything which was said about the luminosity and the surface temperature also applies for the radius, the entropy and the degeneracy factor $\Theta_e$. Deuterium burning  introduces an additional bump in the temporal evolution of these quantities for masses above 13 $\mj$. The peak value of these bumps will be reached later for the less massive objects and the peak value will be smaller. \\ \\
For the surface gravity it also holds that no deuterium burning caused effect can be seen for masses up to 13 $\mj$. For the more massive objects deuterium burning produces a dip, rather than a bump, as the gravitational acceleration is proportional to $R^{-2}$, where $R$ is the radius of the object. \\ \\
Finally it is important to note that the fact that no additional bump or dip is seen for masses up to 13 $\mj$ does not necessarily mean that these objects do not burn deuterium at all. As we will discuss in the following sections, objects with masses around 13 $\mj$ already burn significant amounts of their deuterium. The effect of deuterium burning does not introduce an additional bump or dip but it slows down the monotonous decrease (or increase for ${\rm log}\left(g\right)$) of the discussed quantities.
\begin{figure*}
\begin{minipage}{0.5\textwidth}
	      \centering
       \includegraphics[width=0.95\textwidth]{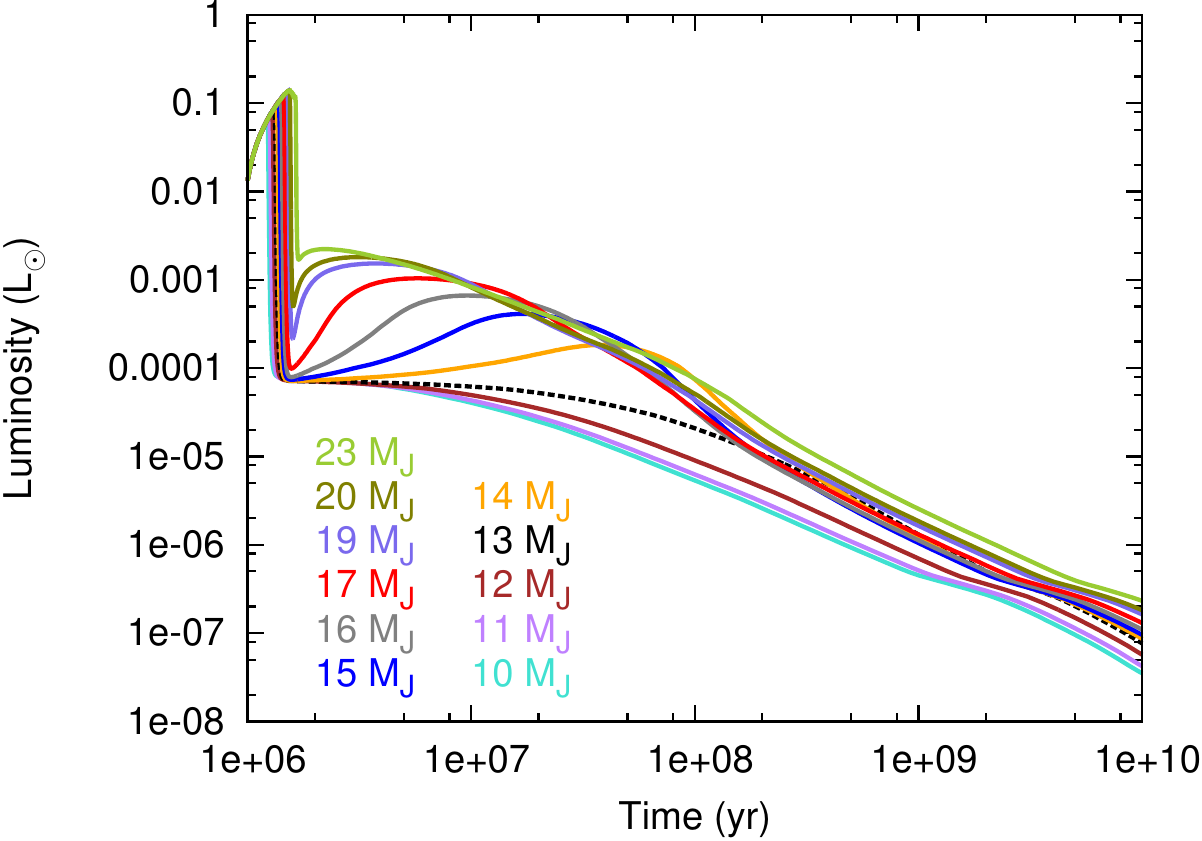}
     \end{minipage}\hfill
     \begin{minipage}{0.5\textwidth}
      \centering
       \includegraphics[width=0.95\textwidth]{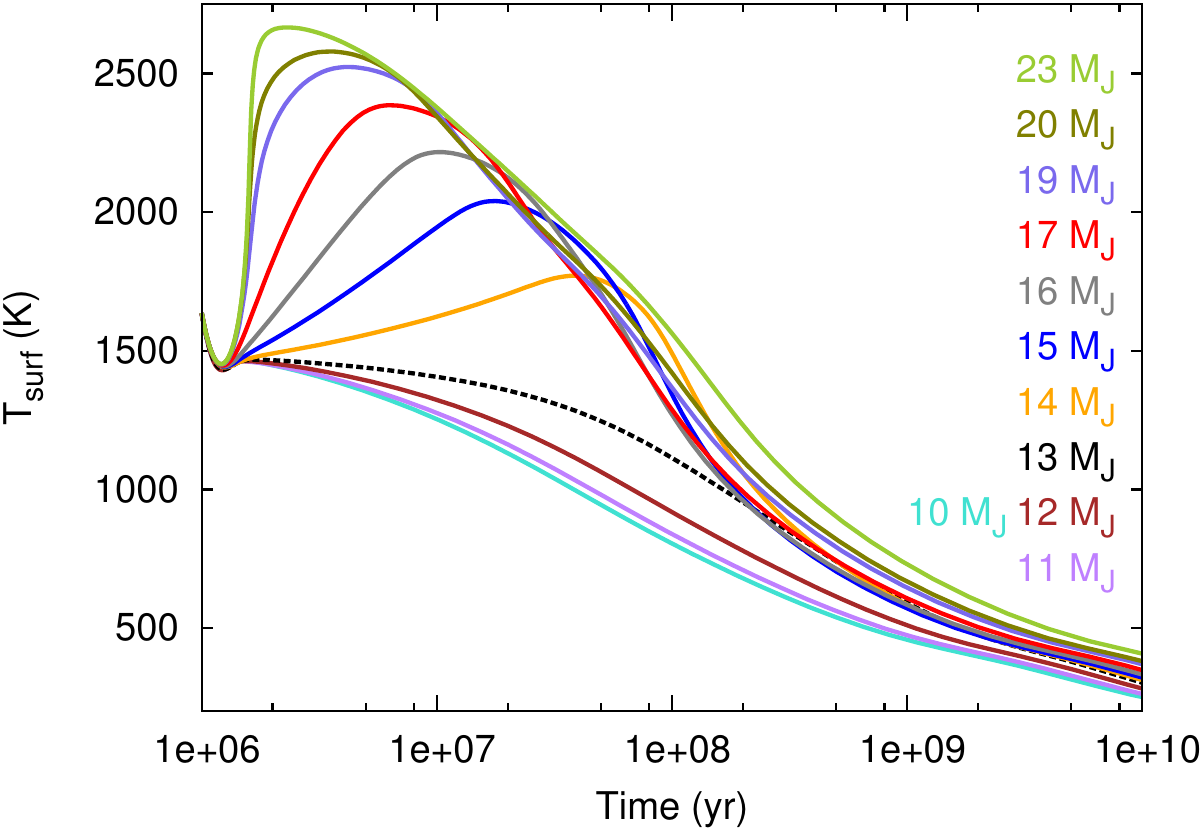}
     \end{minipage}
     \begin{minipage}{0.5\textwidth}
	      \centering
       \includegraphics[width=0.95\textwidth]{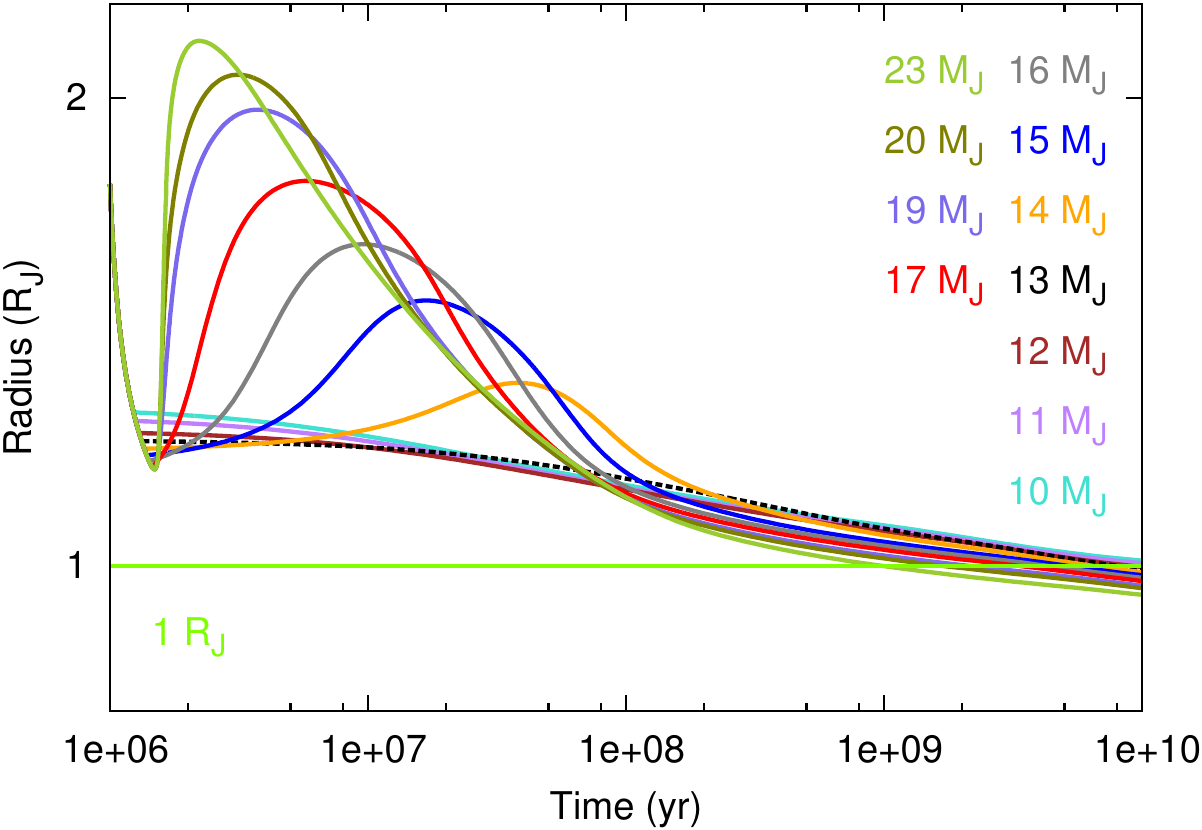}
     \end{minipage}\hfill
     \begin{minipage}{0.5\textwidth}
      \centering
       \includegraphics[width=0.95\textwidth]{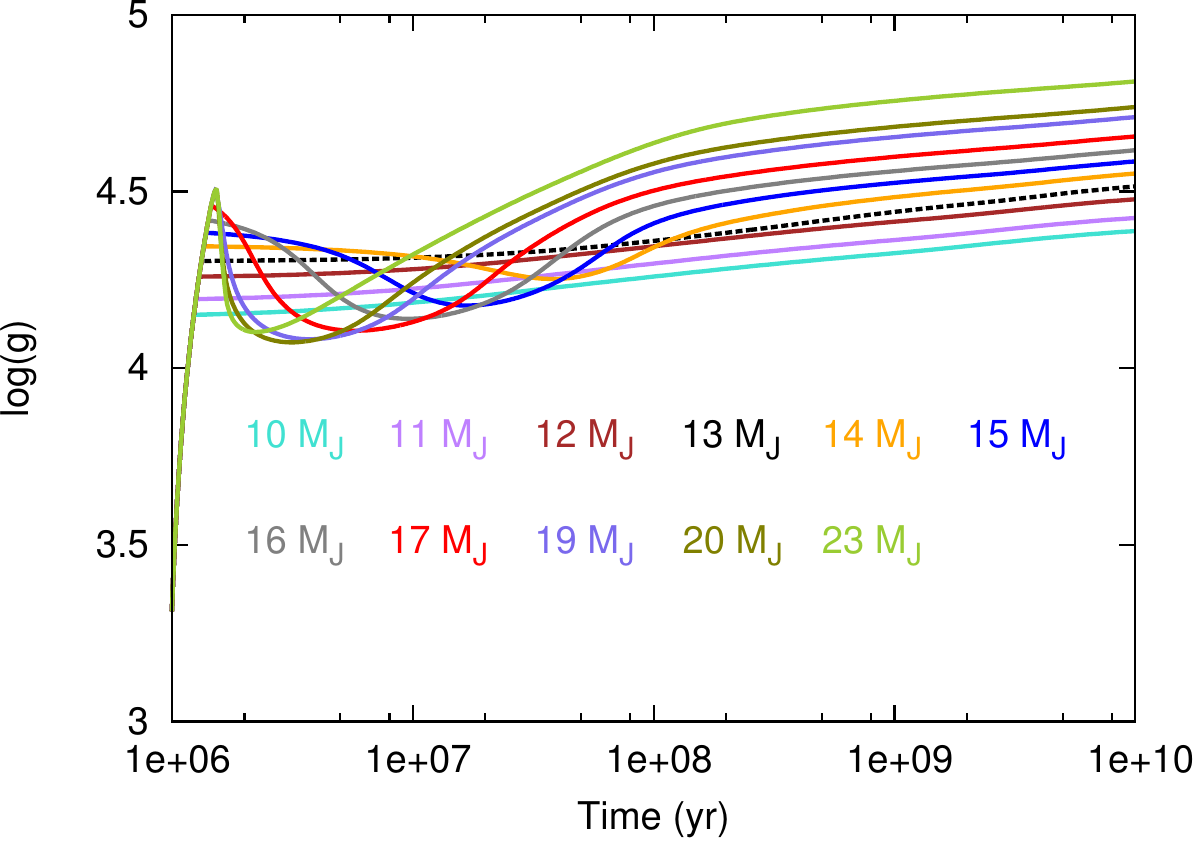}
     \end{minipage}
     \begin{minipage}{0.5\textwidth}
	      \centering
       \includegraphics[width=0.95\textwidth]{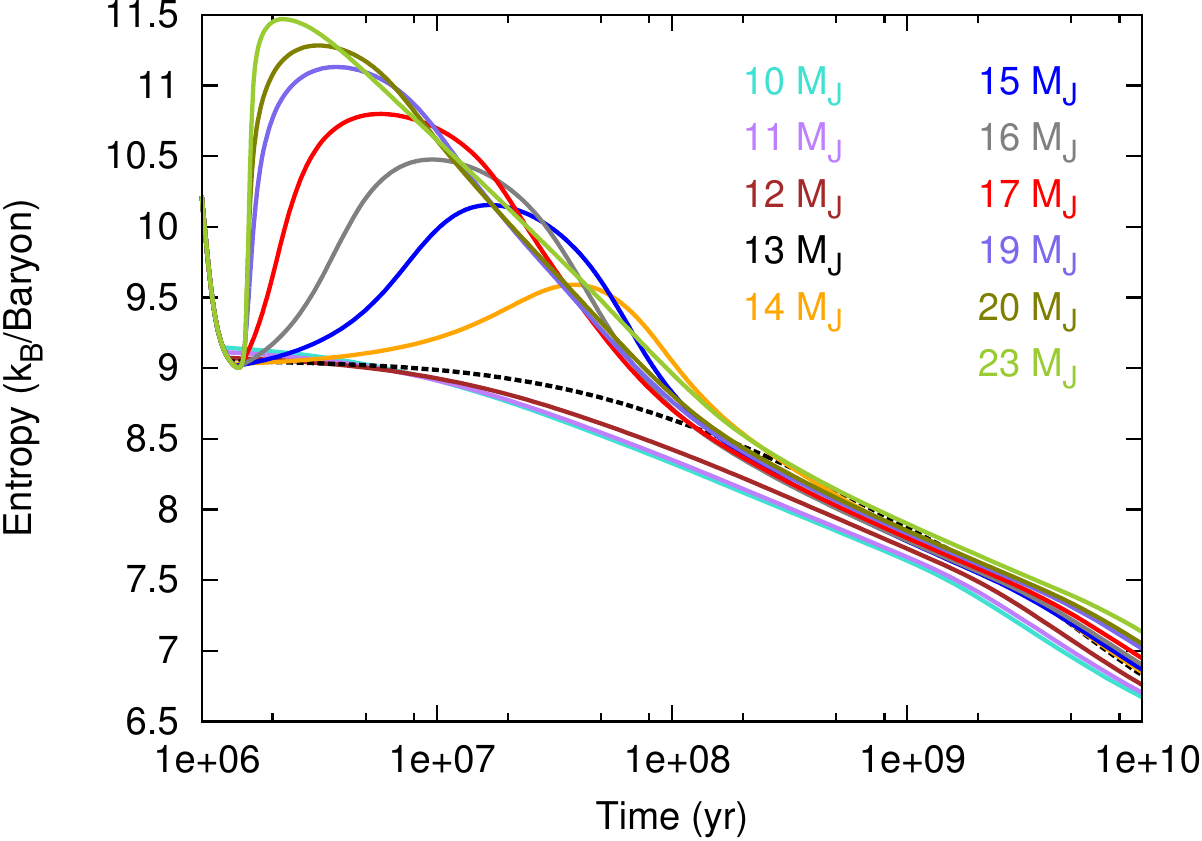}
     \end{minipage}\hfill
     \begin{minipage}{0.5\textwidth}
      \centering
       \includegraphics[width=0.95\textwidth]{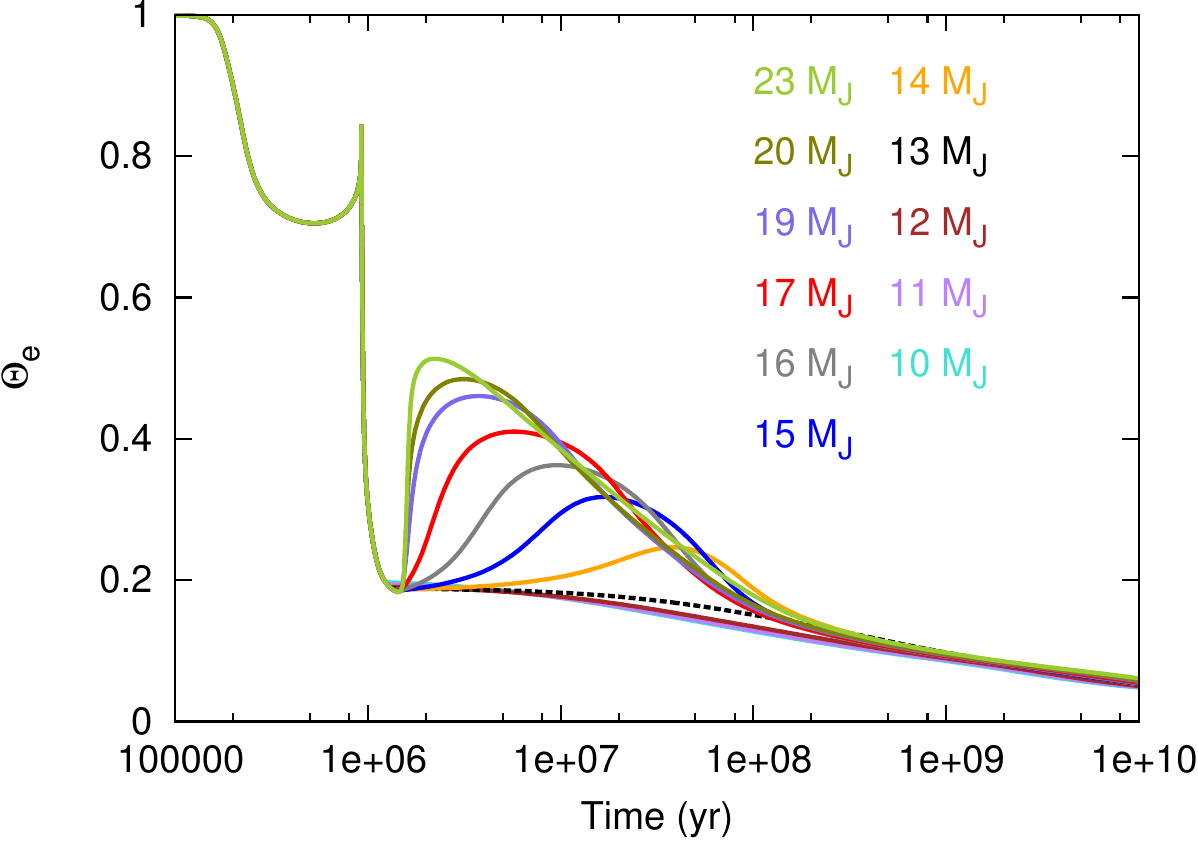}
     \end{minipage}
     \caption{Temporal evolution (for \textbf{cold start} objects) of the luminosity as seen by an observer, i.e. including the accretion luminosity (first row, left), the surface temperature (first row, right), the radius (second row, left), the gravitational acceleration at the surface (second row, right), the specific entropy above the core (third row, left) and the degeneracy related factor $\Theta_e$ (third row, right) for objects with masses varying from 10 $\mj$
              to 23 $\mj$ using our fiducial model as defined in Tab. \ref{tab:standart}. The black dashed line in all plots corresponds to the 13 $\mj$ object. The horizontal green line in the plot of the radius indicates the position of 1 $R_{\rm J}$. The masses corresponding to the lines are indicated in the plot.
	      }
     \label{fig:overallcold}
\end{figure*}
\subsubsection{Implications of the gas runaway accretion phase}
\label{sect:accretionimp}
\begin{figure}
     \includegraphics[width=1.\columnwidth]{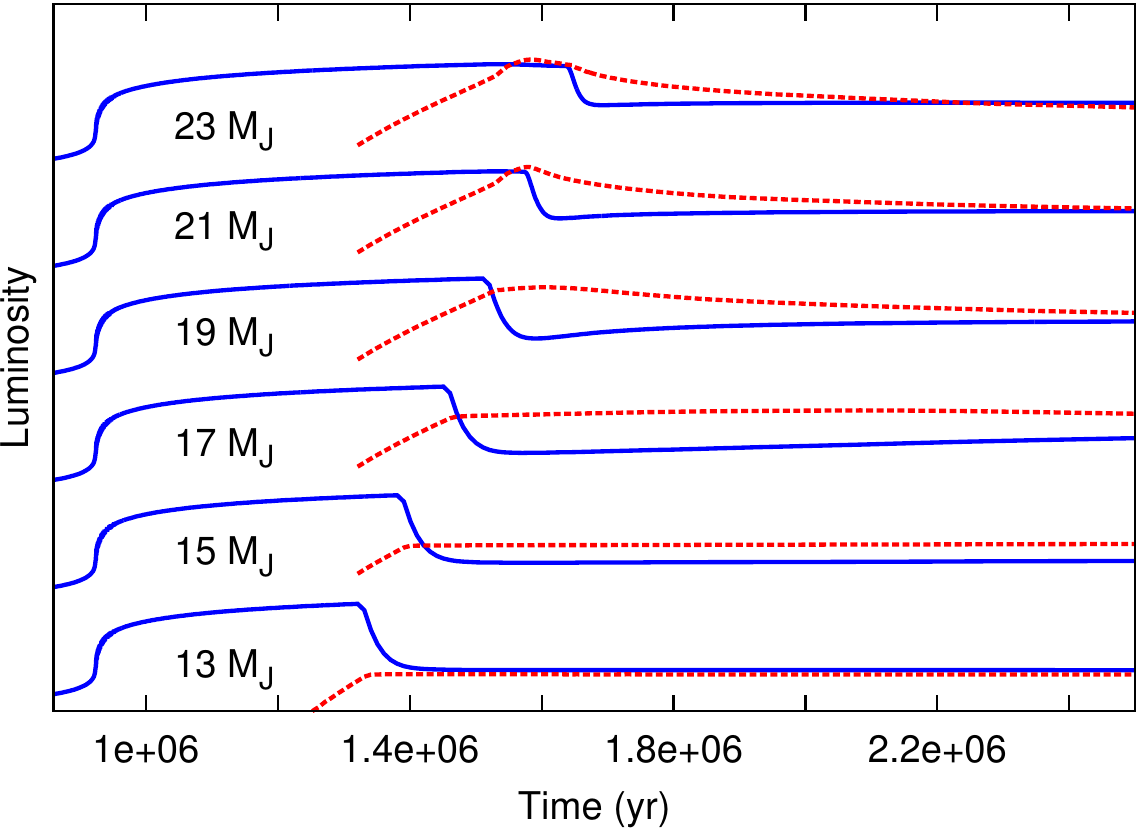}
     \caption{Evolution of the luminosities of \textbf{cold start} objects with masses as indicated in the plot. The blue solid lines show the total
	      luminosities of the objects (i.e. the luminosity an observer would see), whereas the dashed red lines denote the
	      luminosity due to deuterium burning only. The luminosities for the different masses have been separated
	      in the y-axis by an arbitrary offset for better visibility. Up to approximately $1.3\times 10^6$ years,
	      where the gas runaway accretion ends for the 13 $\mj$ object, all objects have the same luminosity.}
\label{fig:manydeulum}
\end{figure}
We now concentrate on the behavior of deuterium burning and how it is influenced by the mode of formation of the
objects, which is a core accretion formation with a radiatively efficient shock (cold start). As we have already seen, a major difference comes from
the fact that cold start objects start deuterium burning at a much smaller radius than in the hot start case. Then
they expand approximately to the radius they attain in a hypothetical deuterium main sequence, due to the thermostatic nature
of deuterium burning. In Fig. \ref{fig:manydeulum} one sees the evolution of the total luminosity $L_{\rm tot}$
and of the luminosity due to deuterium burning ($L_{\rm D}$) for objects of different masses. When studying the plot
one can notice that the end of gas runaway accretion (marked by the steep decrease in the total luminosity of the objects)
causes a bend in $L_{\rm D}$ for all masses, which is clearer for the lower masses. For the lower masses up to 17 $\mj$, the growth of $L_{\rm D}$,
which is strongly increasing in the gas runaway phase, is brought to a halt and remains approximately constant. For the heavier masses, $L_{\rm D}$
will decrease after the end of the gas runaway accretion. Furthermore we observe that for the higher masses
($\approx$ 23 $\mj$),  $L_{\rm D}$ reaches a peak still during the phase of gas runaway accretion, then decreases slightly before
showing the bend at the end of gas runaway {accretion}, after which it decreases even faster. This behavior can be explained with the equations \ref{equ:stahler} and \ref{equ:centtempapprox}. After the object has collapsed at
the transition from the attached to the detached phase, it is not yet massive enough to burn deuterium. Furthermore, due to
the cold start assumption the (major) part of the planets mass is then added onto the planet at low entropy, meaning that it
grows in mass while contracting even further.
As the central temperature is proportional to $M/R$ (Eq. \ref{equ:centtempapprox}) and the deuterium
burning rate is depends on the temperature very strongly and linearly on the density (Eq. \ref{equ:stahler}), this process of adding mass
on the
{object} will cause the observed increase of $L_{\rm D}$ in the phase of runaway mass accretion.
The sharp bend in $L_{\rm D}$ at the end of gas runaway accretion nicely shows the dependence of $L_{\rm D}$ on the process of runaway accretion (as the object has stopped mass accretion
$L_D$ ceases it's strong increase). For the objects of heavier masses, which exhibit a peak of deuterium burning inside the phase of gas
runaway accretion, another effect becomes important. As one can see in Eq. \ref{equ:stahler}, the fusion rate of
deuterium burning is proportional to the deuterium abundance. Indeed we find that those objects burn most of their deuterium
already during their formation phase (as higher masses mean a higher central temperature and thus a higher deuterium burning rate). Thus, at some point, $L_{\rm D}$ will not increase further even though the central
temperature is still rising, as the effect of the decreasing deuterium abundance becomes stronger. The reason for the
following modest decrease of $L_{\rm D}$ in the runaway phase is that
the still increasing central temperature (due to Eq. \ref{equ:centtempapprox}) will partially counterbalance the impact
of the decreasing deuterium abundance. \\ \\
Finally, as already said above, the objects expand due to deuterium burning. Thus $L_{\rm D}$ must overshoot the intrinsic luminosity (to which it contributes) in the expansion phase. 
\subsection{Sensitivity of the minimum mass limit of deuterium burning}\label{sect:minmass}
Similar to the paper of SBM11, we will investigate the implication of certain quantities
on the process of deuterium burning, namely in terms of how they affect the minimum mass limit for deuterium fusion. In order to do so, we assume
a core accretion formation, use the cold start assumption and consider the runs as specified in Tab. \ref{tab:minmassruns}. As SBM11
pointed out, no exact border can be drawn between deuterium burning and non-deuterium burning objects, as the amount of burned deuterium varies
smoothly as a function of mass instead of exhibiting a sharp transition. In this paper, if an object decreases it's initial deuterium abundance by more than 50 \%, we will call it for simplicity a deuterium burning object and objects below this border will be called non-deuterium burning objects. The mass at which this transition happens
will be called $M_{50}$. One should
however bear in mind that this choice of the position of the border is partially arbitrary, even though 50 \% seems to be a sensible choice, as it is in the middle of the two cases of (almost) no deuterium burning and complete deuterium combustion. This illustrates the problem which arises if one persists
on the rigorous distinction of planets and Brown Dwarfs in the mass regime considered henceforth. \\ \\
\begin{figure*}
     \begin{minipage}{0.5\textwidth}
	      \centering
       \includegraphics[width=0.95\textwidth]{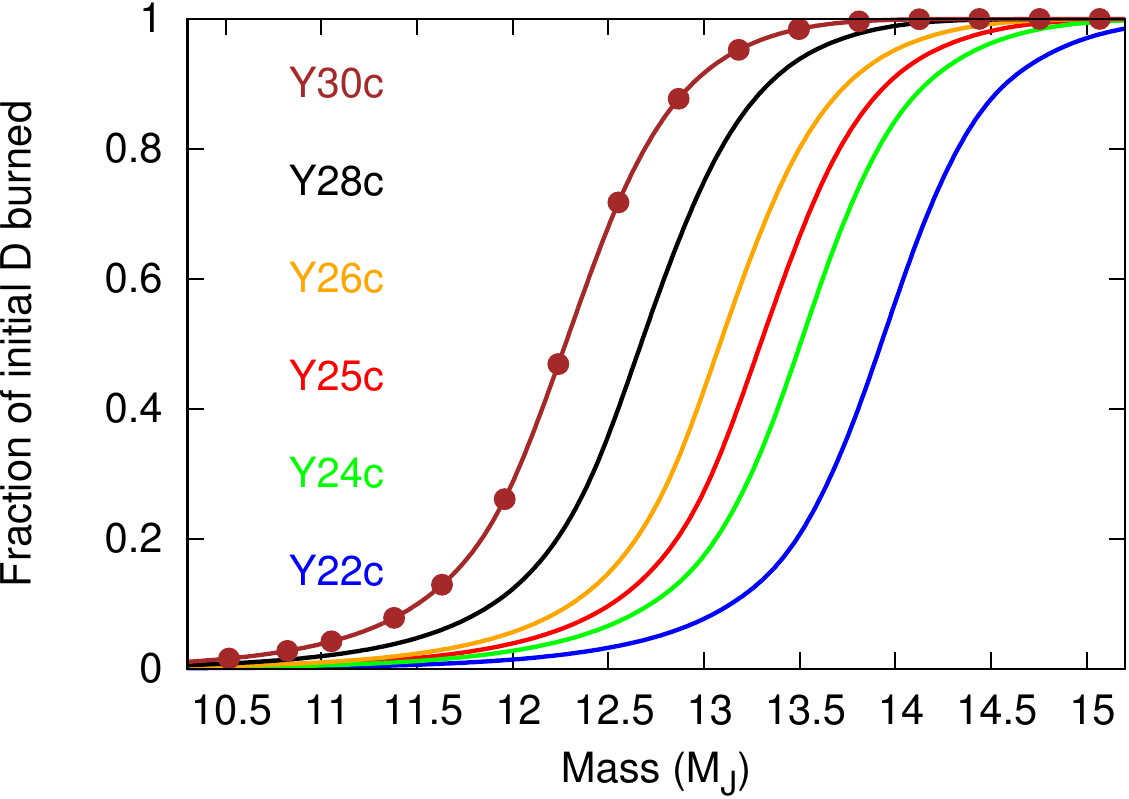}
     \end{minipage}\hfill
     \begin{minipage}{0.5\textwidth}
      \centering
       \includegraphics[width=0.95\textwidth]{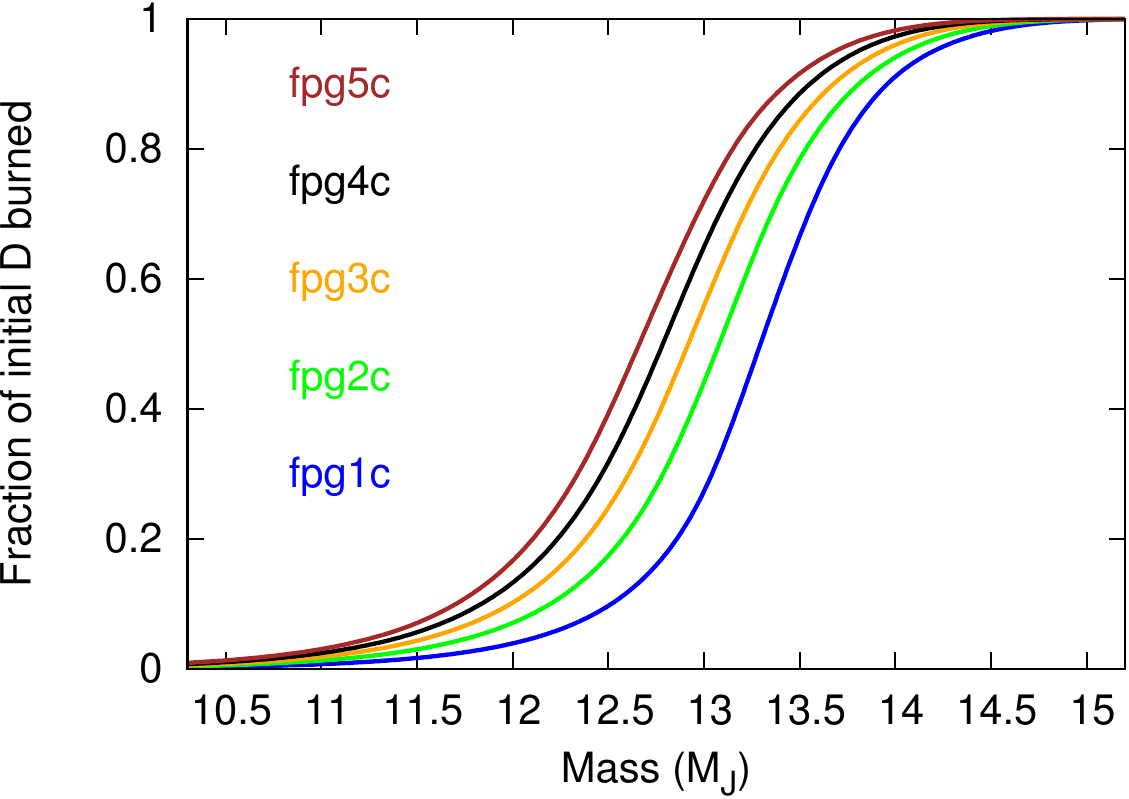}
     \end{minipage}
     \begin{minipage}{0.5\textwidth}
	      \centering
       \includegraphics[width=0.95\textwidth]{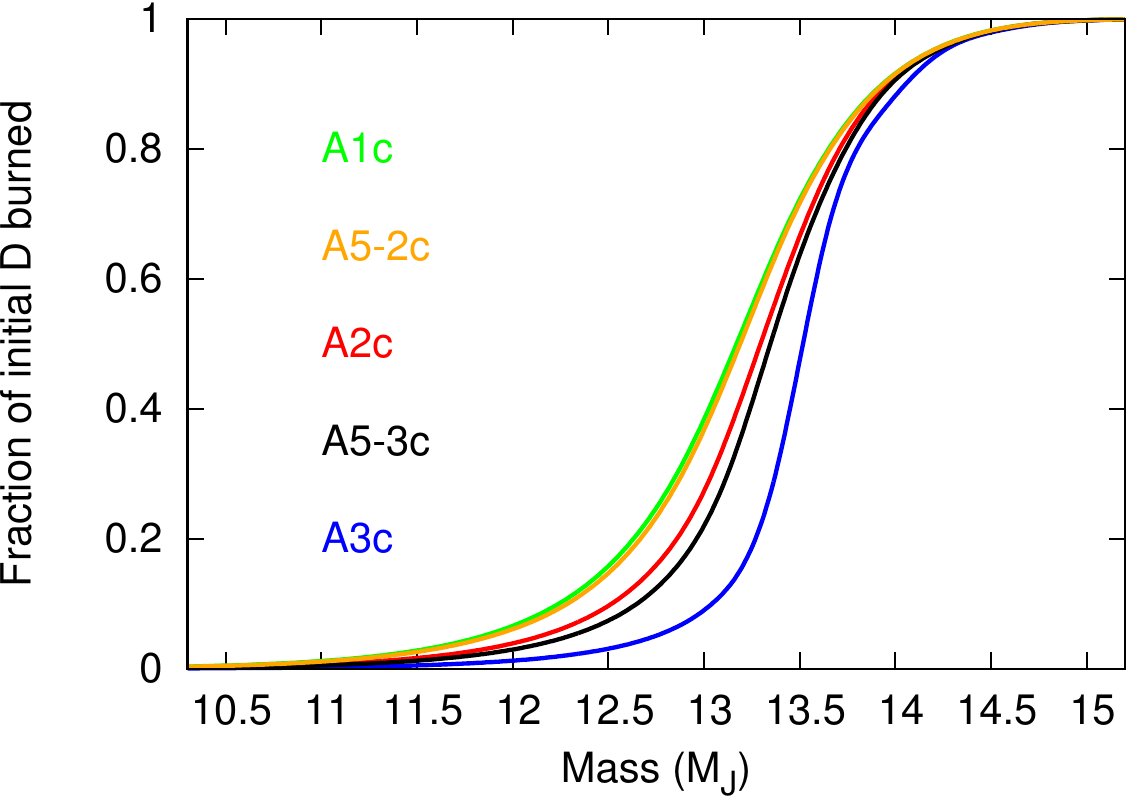}
     \end{minipage}\hfill
     \begin{minipage}{0.5\textwidth}
      \centering
       \includegraphics[width=0.95\textwidth]{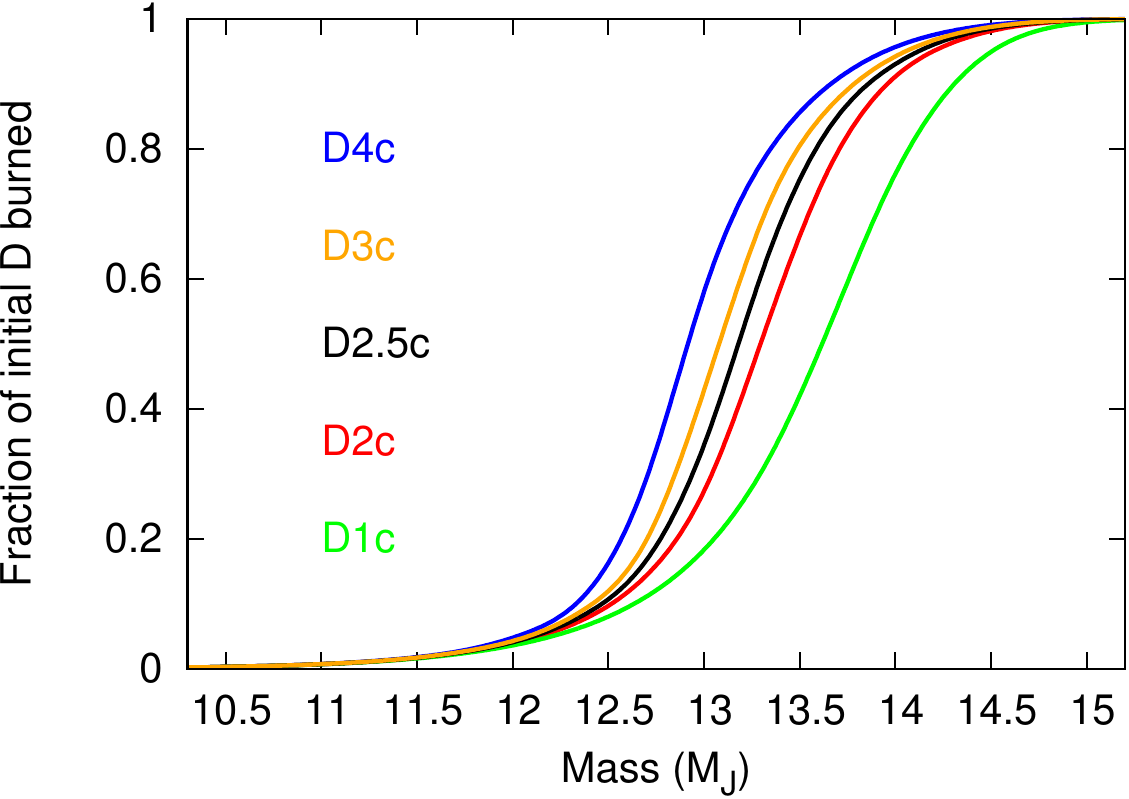}
     \end{minipage}
     \caption{Fraction of the initial deuterium burned as a function of mass until the end of the simulation ($10^{10}$ years) for \textbf{cold start} objects.
	      Top left: Results for different helium fractions. Top right: Results with different dust-to-gas ratios (or metallicities). Bottom left: Results for different maximum gas accretion rates. Bottom right: Results for different deuterium abundances. The uppermost run names in the plots correspond to the leftmost curves, the next lower run names correspond to the second leftmost curves etc. For the Y30 run in the upper left panel the positions of the individual mass runs are indicated with filled brown circles. For the sake of clarity we omitted plotting the individual mass run positions for all other runs. Instead a spline function was plotted through the individual mass run positions.}
\label{fig:masseslimburn}
\end{figure*}
In Fig. \ref{fig:masseslimburn} one can see the results we obtain for the runs in Tab. \ref{tab:minmassruns}. The fraction of the initial deuterium which
has been burned until the end of the simulation ($10^{10}$ years) is plotted against the total mass of the objects. As one can see, the transition from
objects which burn no deuterium to objects which burn all their deuterium has a width of approximately 5 $\mj$ and takes place roughly between 10 and 15 $\mj$. \\ \\
As one can see in Fig. \ref{fig:masseslimburn}, the largest impact on changing $M_{50}$ has the variation of the helium abundance.
It must be noted, however, that a helium abundance as low as 0.22 is rather improbable, as the helium fraction due to the primordial nucleosynthesis was approximately
0.25 (Spergel et al. \cite{spergeletal}) and has, since then, increased in time due to hydrogen fusion in stars. The top and bottom right panel of Fig \ref{fig:masseslimburn} show that a variation of the metallicity, as well as the
variation of the deuterium abundance, will shift the place of the transition region between deuterium burning and non deuterium burning objects. \\ \\
When considering the maximum mass accretion rate $\dot{M}_{\rm max}$ (bottom left panel) we find that, opposite to the results obtained for the hot start case, $\dot{M}_{\rm max}$ now also has an impact on the process of deuterium burning. The reason for this is {the following:} The objects collapse at the transition from the attached to the detached phase (as can be seen at approximately $10^6$ years in the upper right plot in Fig. \ref{fig:coldstartgeneral} showing the radius evolution). After this collapse the objects will contract further while they accrete most of their mass during the runaway accretion process. This is in contrast to the hot start case, where the objects expand again during the runaway accretion. It is clear that the higher the maximum allowed mass accretion rate, the shorter {is} the time the object will have to contract during
the runaway accretion process. The consequence of this is the following: At higher $\dot{M}_{\rm max}$, the gas accretes onto a larger object, which means that the free fall velocity of the gas at the moment it reaches the surface of the object will be smaller, as the free fall velocity for the infalling matter is
\beq
v_{\rm ff} = \sqrt{\frac{2GM}{R}}
\eeq
$M$ denotes the mass of the object, $R$ is the radius of the object and $G$ the gravitational constant. The lower free fall velocity would result in a smaller accretion luminosity {if $\dot{M}_{\rm max}$ was constant} (c.f. Bodenheimer et al. \cite{bodenheimeretal2000})
\beq
L_{\rm acc} = \frac{1}{2}\dot{M}_{\rm max}v_{\rm ff}^2 = \frac{GM\dot{M}_{\rm max}}{R} .
\eeq
The lower {accretion luminosity per accreted mass unit} means that less liberated gravitational potential energy is radiated away at the shock, or in other words, that matter of higher entropy is incorporated into the object. {This can be understood by the fact that although the object forms according to a cold start the material is incorporated into the object at larger radii. Thus it must have a larger specific entropy, just as the initial radius of hot start simulations is increasing with the chosen initial entropy for fully adiabatic objects.}
\begin{figure}[htb]
       \includegraphics[width=1.\columnwidth]{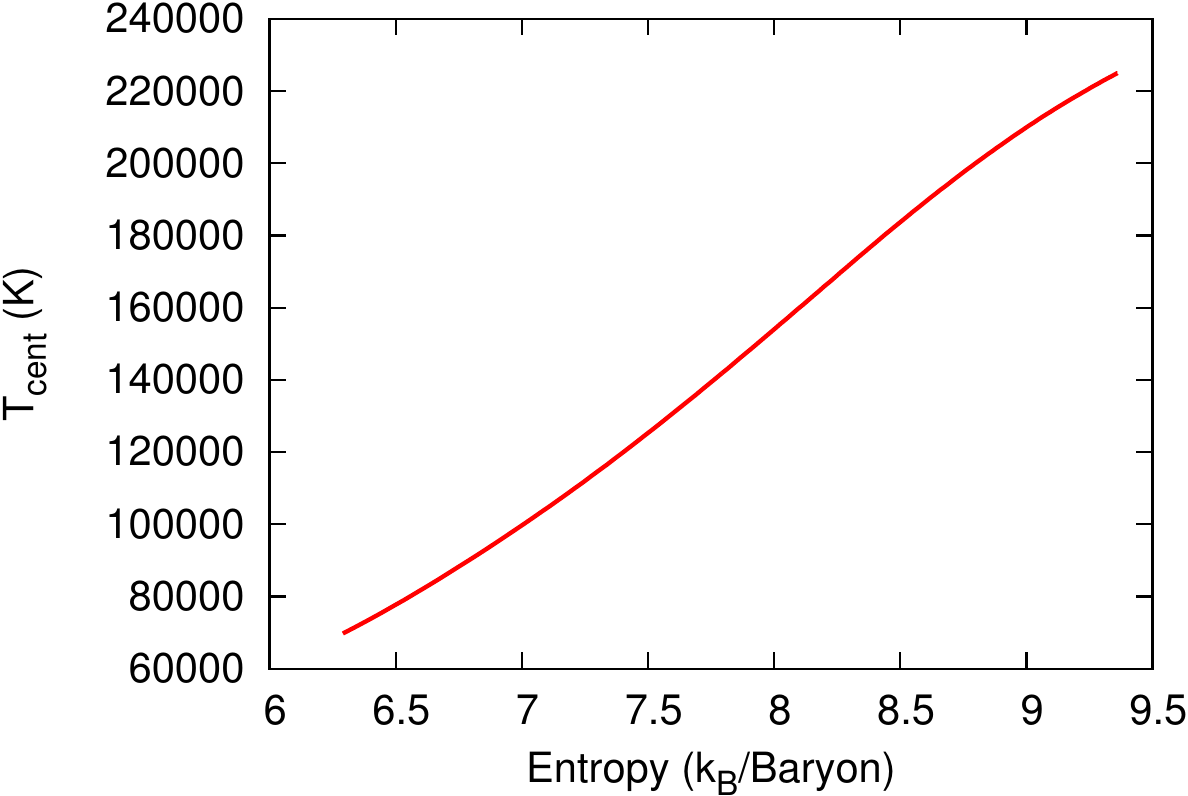}
     \caption{Central temperature of a 7 $\mj$ planet as a function of it's specific entropy showing the complete post-formation evolutionary phase.}
     \label{fig:entrotcent}
\end{figure}
{We find that} the accretion of matter with a higher entropy means that the object will have a higher central temperature, leading to higher deuterium burning rates. {The effect of a higher central temperature due to a higher specific entropy might surprise in view of the classical result obtained for the central temperature of a self-gravitating sphere of ideal gas (see Eq. \ref{equ:centtempapprox}): A higher specific entropy in fully convective (i.e. adiabatic) objects generally implies a larger radius (see also Spiegel \& Burrows \cite{spiegeletal2012}). This, in turn, should cause a lower central temperature (see Eq. \ref{equ:centtempapprox}). In the cases considered here, however, partial degeneracy already plays an important role. The fact that the degeneracy becomes stronger for smaller radii as well as the fact that fully degenrate objects cool as they shrink (e.g. white dwarfs) underlines that a larger radii due to a higher specific entropy must not generally imply a smaller central temperature. Also the presence of the core can cause the envelope to cool. Both arguments are outlined analytically in the appendix \ref{appendix:A} and \ref{appendix:B}. In Fig. \ref{fig:entrotcent} one can see see how the central temperature decreases as the entropy decreases in a 7 $\mj$ planet in the post-formation evolutionary phase.}
\subsubsection{Detailed study of the behavior of $M_{50}$}
Fig. \ref{fig:masslimburn} shows how $M_{50}$ varies due to changes in the helium abundance $Y$, the maximum allowed runaway gas accretion rate $\dot{M}_{\max}$,
the deuterium abundance [D/H] and the metallicity [Fe/H] (corresponding to different dust-to-gas ratios).
\begin{figure*}
     \begin{minipage}{0.5\textwidth}
	      \centering
       \includegraphics[width=0.95\textwidth]{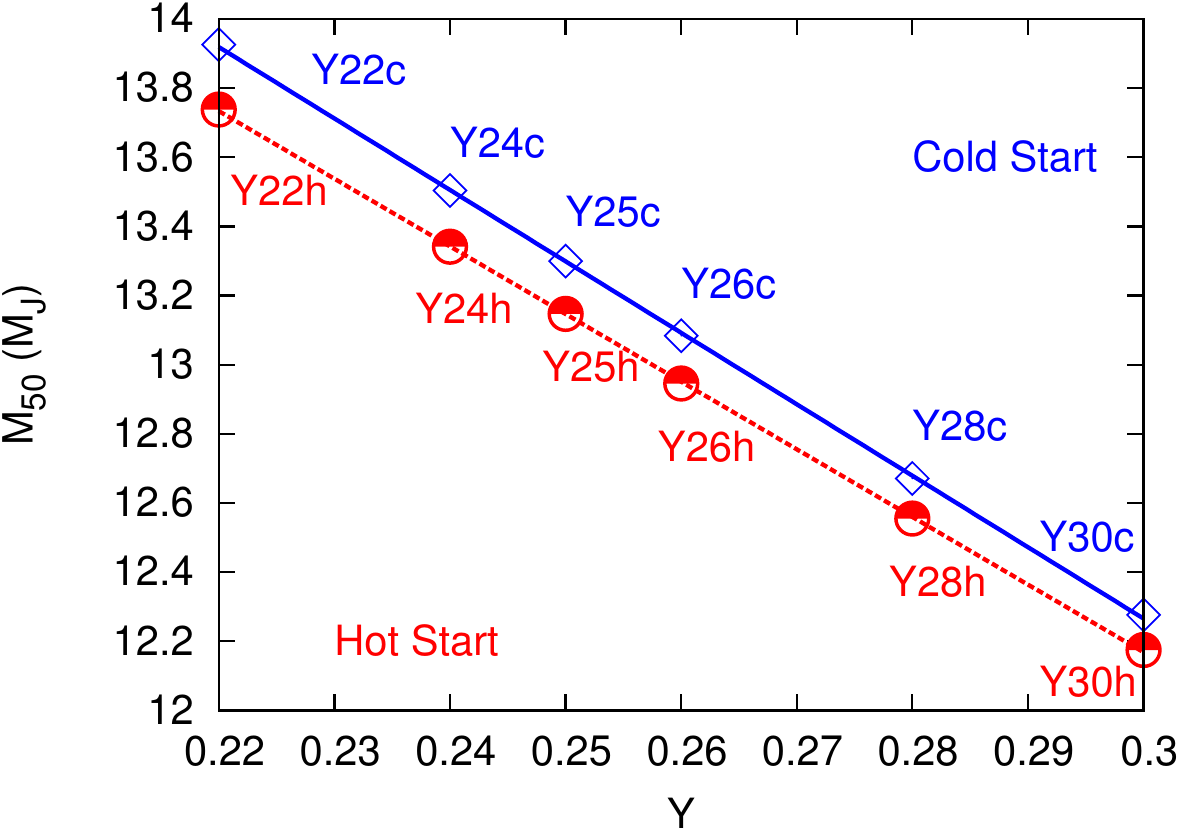}
     \end{minipage}\hfill
     \begin{minipage}{0.5\textwidth}
      \centering
       \includegraphics[width=0.95\textwidth]{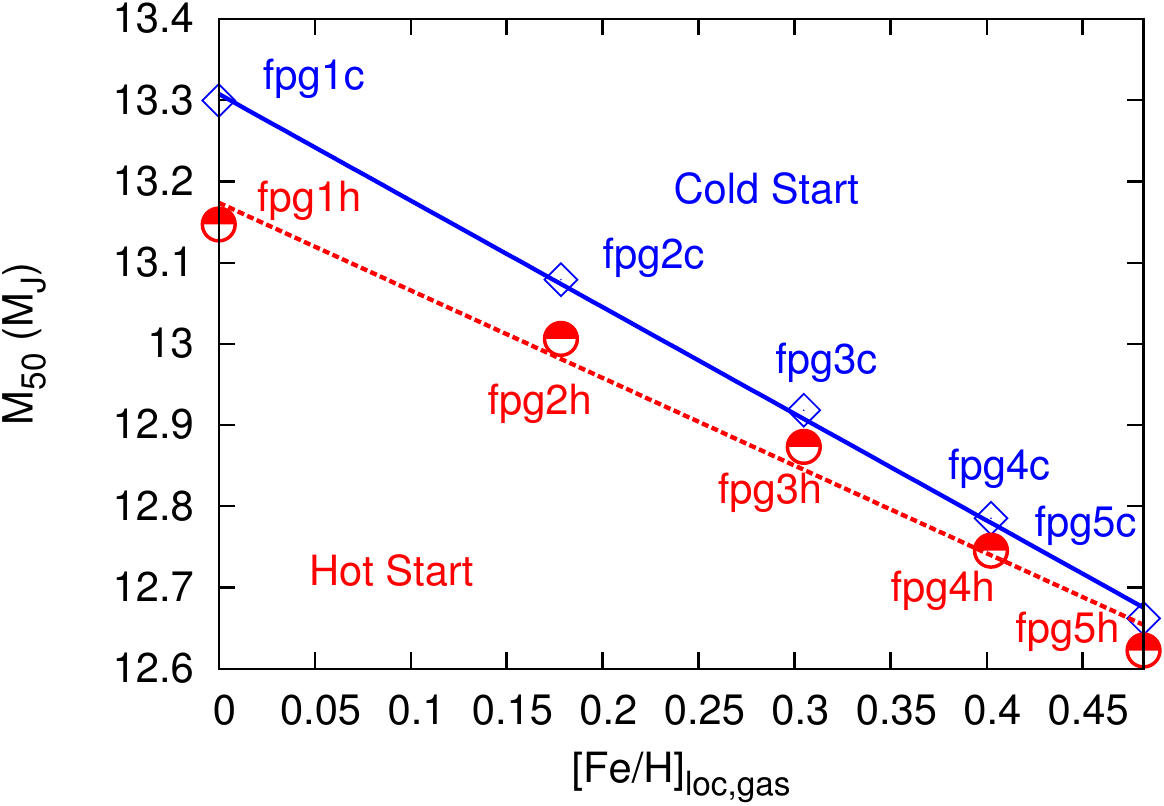}
     \end{minipage}
     \begin{minipage}{0.5\textwidth}
	      \centering
       \includegraphics[width=0.95\textwidth]{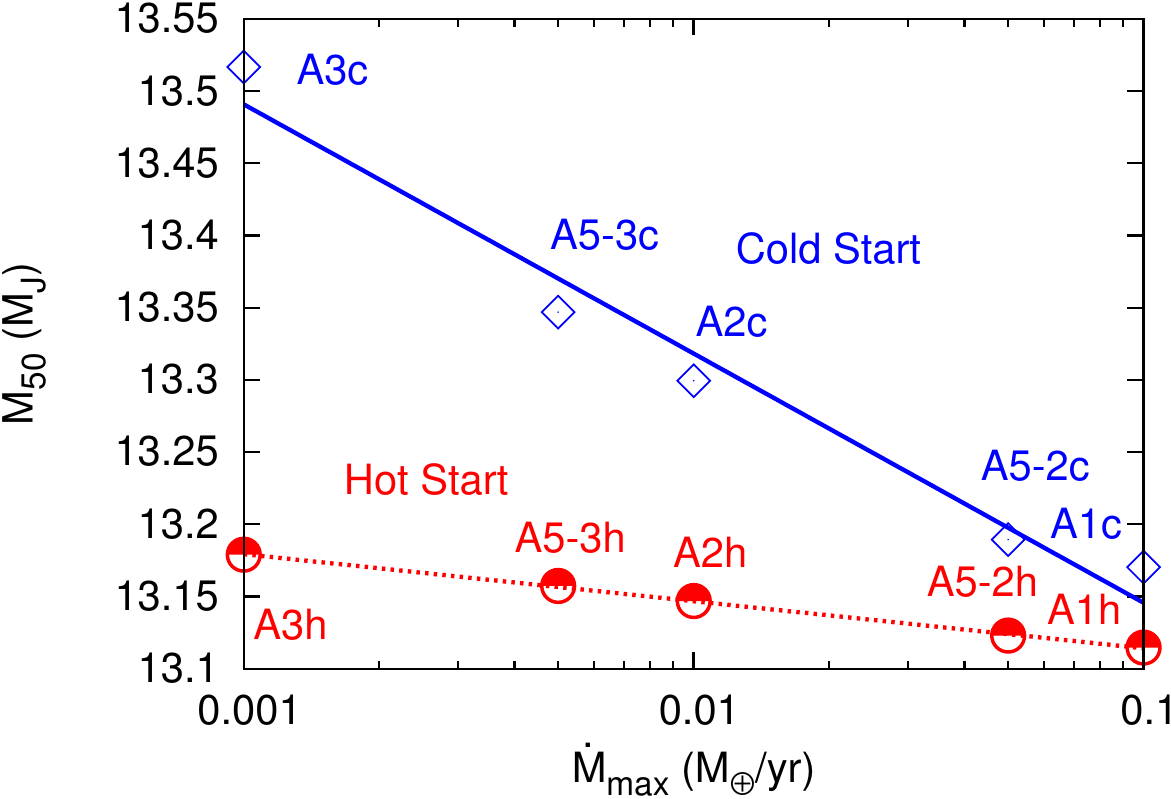}
     \end{minipage}\hfill
     \begin{minipage}{0.5\textwidth}
      \centering
       \includegraphics[width=0.95\textwidth]{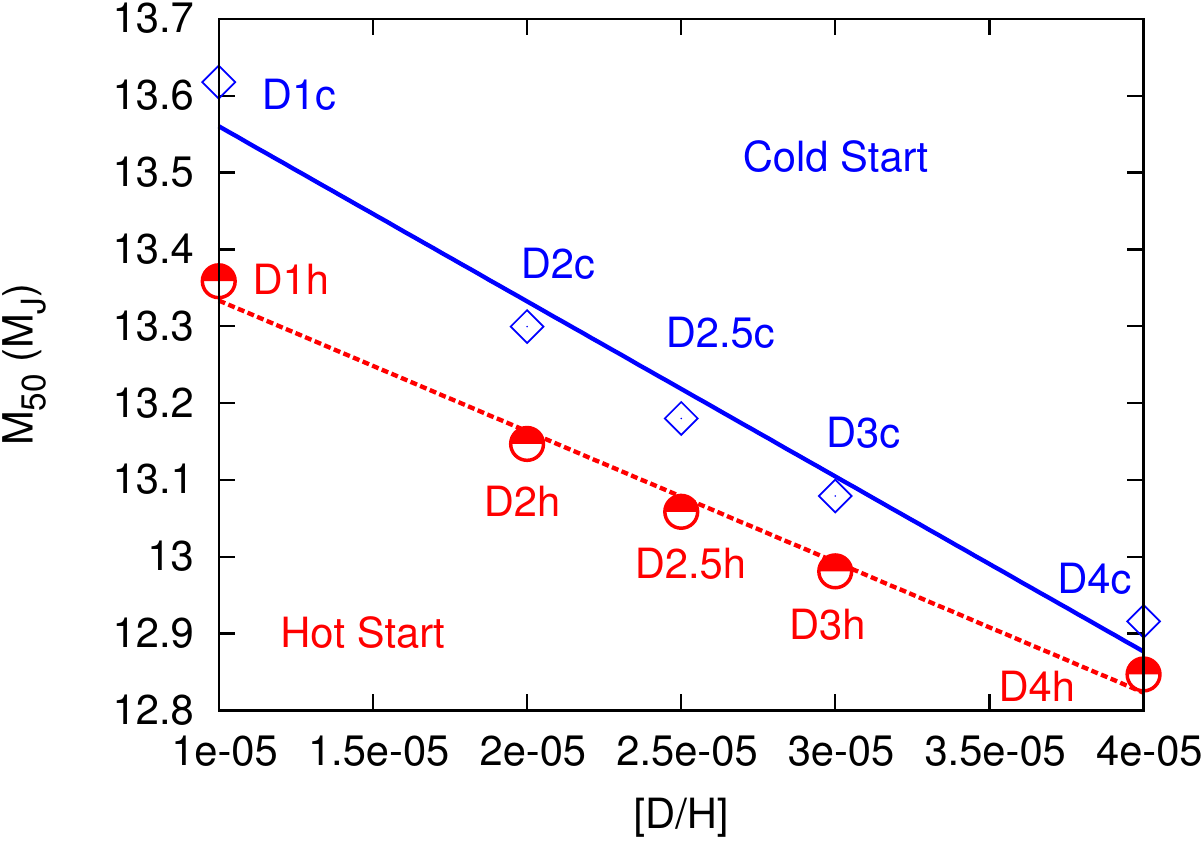}
     \end{minipage}
     \caption{Masses at which 50 \% of the deuterium has been burned by $10^{10}$ years. Upper row, left: varied helium fraction. Right: varied metallicities. Lower row, left: varied maximum gas accretion rates. Right: varied deuterium abundances.
	      The red dashed line shows the fitted functions for the hot start objects (lower line in all plots), while the fits for the cold start objects are plotted in a blue solid line (upper line in all plots). The blue diamonds show the positions of $M_{50}$ in the cold start runs, while the red semi-filled circles show the positions of $M_{50}$
	      for the hot start runs.}
\label{fig:masslimburn}
\end{figure*}
\begin{figure*}
     \begin{minipage}{0.5\textwidth}
	      \centering
       \includegraphics[width=0.95\textwidth]{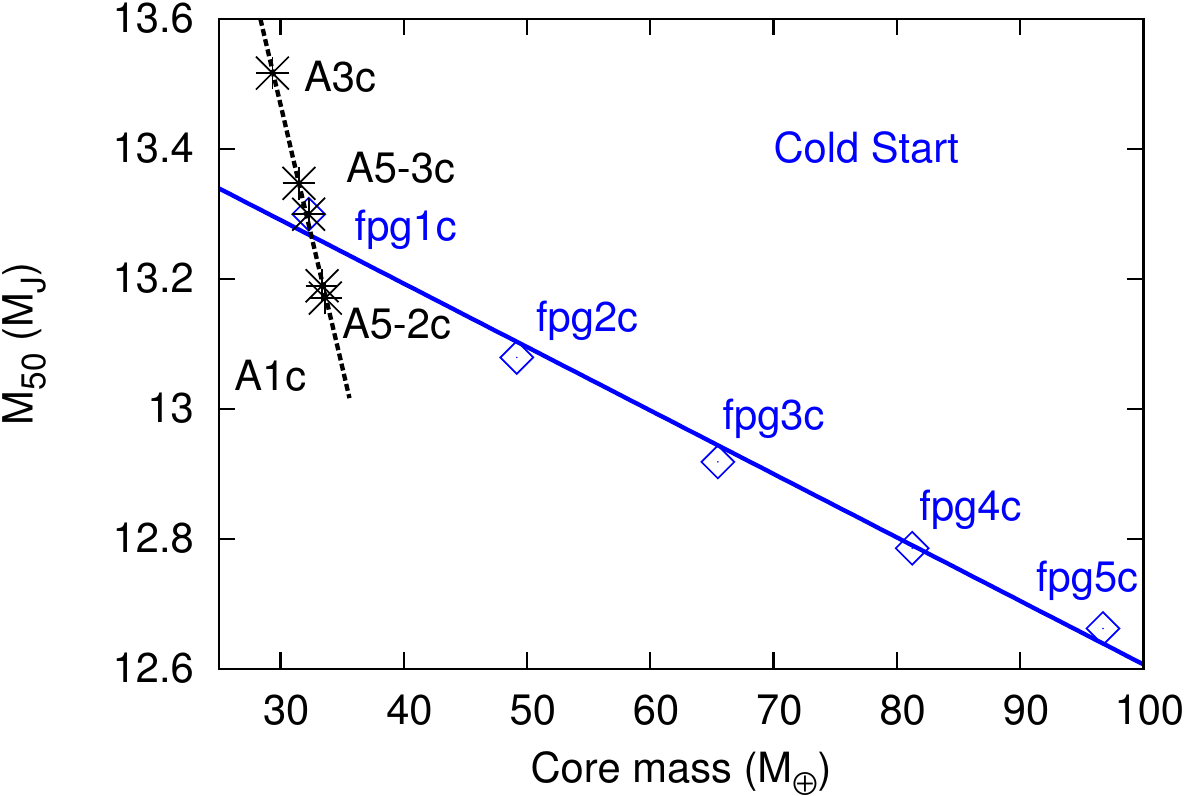}
     \end{minipage}\hfill
     \begin{minipage}{0.5\textwidth}
      \centering
       \includegraphics[width=0.95\textwidth]{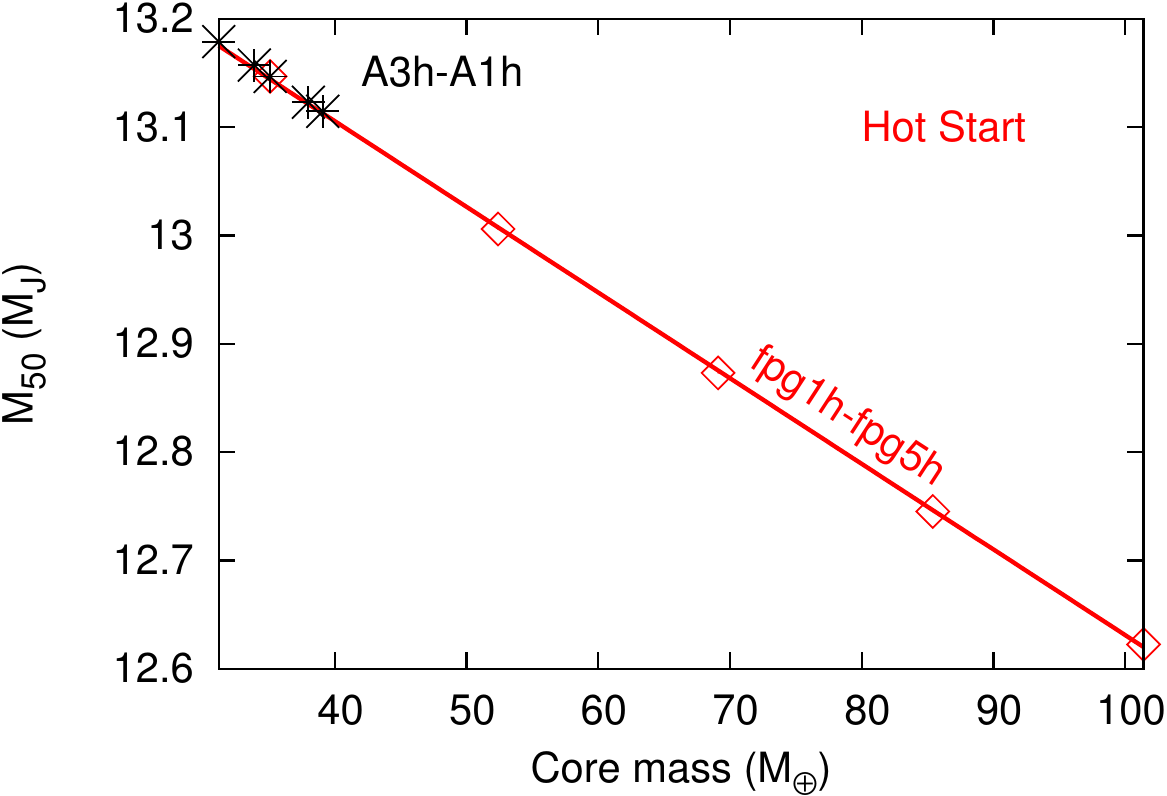}
     \end{minipage}
     \caption{$M_{50}$ as a function of the core mass for different dust-to-gas ratios and different $\dot{M}_{\rm max}$. The left panel shows the results for a cold start, while the the right panel is for a hot start. The black stars correspond to the runs with different $\dot{M}_{\rm max}$, while the diamonds correspond to the runs with different dust-to-gas ratios.
	      }
\label{fig:mdotmaxeffect}
\end{figure*}
The results obtained for cold start objects are plotted in blue, while the results obtained for hot start objects are plotted in red (we varied the fiducial model
for the hot start objects in the same way as we varied it for the cold start objects given in Tab. \ref{tab:minmassruns}, but suffixed the run names with an ''h''
instead of a ''c'').
For $Y$, [D/H] and [Fe/H] the results were fitted to a line using the least square method:
\begin{align}
\nonumber M_{50}/\mj & = m_Y \cdot Y + b_Y \\
M_{50}/\mj & = m_{\rm [D/H]} \cdot {\rm [D/H]} + b_{\rm [D/H]}\\
\nonumber  M_{50}/\mj & = m_{\rm [Fe/H]} \cdot {\rm [Fe/H]} + b_{\rm [Fe/H]}
\end{align}
For $\dot{M}_{\rm max}$ (note the logarithmic x-axis in Fig. \ref{fig:masslimburn} for $\dot{M}_{\rm max}$) the results were fitted to the function
\beq
M_{50}/\mj = {\rm log}_{10}\( c_{\dot{M}} \left(\frac{\dot{M}_{\rm max}}{1 \ \mearth/{\rm yr}}\right)^a \)
\eeq
This means that in a first approximation $M_{50}$ depends linearly on the helium abundance $Y$, the deuterium number fraction [D/H] and the metallicity [Fe/H]. For $\dot{M}_{\rm max}$ and the dust-to-gas ratio (see Eq. \ref{equ:metall}), however, there is a logarithmic dependence. \\ \\
The upper right panel of Fig. \ref{fig:masseslimburn} shows the metallicity dependence of $M_{50}$. However, instead of labeling the x-axis with ''[Fe/H]'' it was labeled with ''${\rm [Fe/H]_{loc,gas}}$''. The intention of this is the following: In our model the metallicity sets the dust-to-gas ratio via Eq. \ref{equ:metall}, which is only valid at approximately 5.2 AU (see discussion of Equ. \ref{equ:metall}). The dust-to-gas ratio (fpg), in turn, sets the local initial solid surface density ($\Sigma_{\rm s,0}={\rm fpg}\cdot\Sigma_0$) which determines the final core mass. Thus it is important to keep in mind that the metallicity as it is given here serves as a proxy for the core mass of the object and only in our fiducial model at a distance of 5.2 AU it is correct to infer the $M_{50}$ value from the metallicity. \\ \\
The results for the fitting parameters can be found in Tab. \ref{tab:fittingparams}. \\ \\
\begin{table}
\caption{Fitting parameters for $M_{50}$}\label{tab:fittingparams}
\begin{center}
\begin{tabular}{lcc}
\hline\hline
Parameter & cold start & hot start \\ \hline
$m_Y$ & $-20.66$ & $-19.56$ \\
$b_Y$ & 18.46 & 18.04 \\ \hline
$m_{\rm [D/H]}$ & $-2.28 \times 10^4$ & $-1.7 \times 10^4$ \\
$b_{\rm [D/H]}$ & 13.79 & 13.5 \\ \hline
$m_{\rm [Fe/H]}$ & $-1.31$ & $-1.08$\\
$b_{\rm [Fe/H]}$ & $13.31$ & $13.17$\\ \hline
$a$ & $-0.17$ & $-0.03$ \\
$c_{\dot{M}}$ & $9.33\times 10^{12}$ & $1.21\times 10^{13}$ \\
\hline
\end{tabular}
\end{center}
\end{table}
Looking at the helium dependency of $M_{50}$, one finds that the slope of both fitted lines (i.e. for hot and cold start) agrees quite well and approximately coincides with the values from
SBM11, who found $m_Y = -20$. Under variation of [D/H] and [Fe/H], the cold start objects react somewhat stronger than the hot start objects, resulting in a steeper slope. The slope for the hot start objects with different deuterium abundances [D/H], again coincides quite well with the results of SBM11 who found $m_{\rm [D/H]} = -1.6\times 10^4$. \\ \\
The dependence of $M_{50}$ SBM11 found for variations of the metallicity [Fe/H] is different. This is, however, not surprising, as the metallicity is determining only the opacity in their results, whereas it determines only the core mass in our results. It is clear that it is not expected that those two different effects should have the same consequences for
deuterium burning. \\ \\
The most obvious difference between cold and hot start objects is found in the $M_{50}$ sensitivity to different $\dot{M}_{\rm max}$ values.
As described in Section \ref{subsubsec:paramstudy}, the deuterium burning process (in terms of how much deuterium is burned) is rather insensitive to $\dot{M}_{\rm max}$ for the hot start case (an explanation for the remaining sensitivity will be given below). In the cold start case, however, the objects are much more sensitive to changes in $\dot{M}_{\rm max}$. Here, $M_{50}$ shows a variation that is about 3 times larger. In absolute terms, however, the variation is very small: An increase of $\dot{M}_{\rm max}$ by a factor of 100 (which should cover the range of $\dot{M}_{\rm max}$ expected to occur in a self-gravitationally stable accretion disk) only changes $M_{50}$ by $\sim \frac{0.35}{13.3} \hat{=} 2.6 $ \%. The reason for this was already
given above: As objects with higher $\dot{M}_{\rm max}$ accrete a larger part of their mass at larger radii than
the objects with lower $\dot{M}_{\rm max}$ {(because due to the high accretion rate the mass is added at times where the objects have not yet contracted as much as in the low $\dot{M}_{\rm max}$ cases)}, they start their post-formation evolution with a higher specific entropy and have to release a higher amount of remaining gravitational potential energy via contraction (contributing to $L_{\rm int}$). {As outlined in Sect. \ref{sect:minmass}}  this makes them hotter and leads to a higher burning rate. As there is no entropy reducing shock in the hot start scenario, or, in other words, as the entire released gravitational potential energy must be processed out of the object by $L_{\rm int}$, this effect is not observed for hot start objects. \\ \\
Different maximum gas accretion rates will also lead to slightly different core masses in our simulations. This is a natural consequence of the fact that the object's capture radius for planetesimals, and thus it's core accretion rate, depends on the object's structure (see Mordasini et al. \cite{mordasiniklahr2011}), and thus on $\dot{M}_{\rm max}$. We therefore want to check whether the dependence of $M_{50}$ on $\dot{M}_{\rm max}$ is not simply a consequence of the different associated core masses.
However, for the cold start objects, a varying core mass for varying $\dot{M}_{\rm max}$ can be ruled out as the (main) reason for the sensitivity of the deuterium burning process to $\dot{M}_{\rm max}$: \\ \\
In the transition region between no deuterium burned at all and complete deuterium fusion, all objects of one given $\dot{M}_{\rm max}$ or dust-to-gas ratio, but of different total mass, have nearly the same core mass. Thus we are able to plot $M_{50}$ against the core mass for the different dust-to-gas ratios and for the different $\dot{M}_{\rm max}$. $M_{50}$ as a function of the core mass for hot and cold start simulations is shown in Fig. \ref{fig:mdotmaxeffect}. If the varying amount of burned deuterium in the cold start case for different $\dot{M}_{\rm max}$ was caused by different core masses,
then the $\dot{M}_{\rm max}$ simulations should lie on the same line as the dust-to-gas ratio simulations (the fpg runs). Fig. \ref{fig:mdotmaxeffect} shows that this is not the case, which implies that indeed the accretion
at larger radii is important for objects with larger $\dot{M}_{\rm max}$. For the hot start objects, in contrast, we find that the dependence of $M_{50}$ on $\dot{M}_{\rm max}$ can be explained by different core masses alone, as the $\dot{M}_{\rm max}$ runs in Fig. \ref{fig:mdotmaxeffect} lie on the same line as the dust-to-gas ratio runs. It is thus not the case that a different
$\dot{M}_{\rm max}$ directly influences the deuterium burning process in the hot start case, but it is a second order effect that different $\dot{M}_{\rm max}$ result in slightly different core masses.
In order to get the deuterium burning mass limit $M_{50}$ as a function of the core mass, we fit lines through the cold start runs with varying dust-to-gas ratio and varying $\dot{M}_{\rm max}$, as well as through the hot start dust-to-gas ratio runs.
\begin{align}
\nonumber M_{50}/\mj & = m_{\rm dust,cold} \cdot M_{\rm core}/\mearth + b_{\rm dust,cold} \\
M_{50}/\mj & = m_{\dot{M},{\rm cold}} \cdot M_{\rm core}/\mearth + b_{\dot{M},{\rm cold}}\\
\nonumber  M_{50}/\mj & = m_{\rm dust,hot} \cdot M_{\rm core}/\mearth + b_{\rm dust,hot}
\end{align}
This yielded the fitting parameters given in Tab. \ref{tab:fittingcore}. \\ \\
\begin{table}
\caption{Fitting parameters for $M_{50}$ as a function of the core mass}\label{tab:fittingcore}
\begin{center}
\begin{tabular}{lc}
\hline\hline
Parameter & Value \\ \hline
$m_{\rm dust,cold}$ & $-9.77 \times 10^{-3}$\\
$b_{\rm dust,cold}$ & $13.58$\\ \hline
$m_{\dot{M},{\rm cold}}$ & $-8.07 \times 10^{-2}$\\
$b_{\dot{M},{\rm cold}}$ & $15.89$\\ \hline
$m_{\rm dust,hot}$ &  $-7.91 \times 10^{-3}$\\
$b_{\rm dust,hot}$ &  $13.42$\\
\hline
\end{tabular}
\end{center}
\end{table}
We find that $M_{50}$ is approximately equally sensitive to the core masses for both the cold and the hot start scenario: In the core mass range from $\sim 30 \ \mearth$ to $\sim 100 \ \mearth$, $M_{50}$ decreases by $\sim 0.6 \ \mj$. The y-intercept of the fitted lines furthermore indicates that for objects without a core $M_{50}$ is 13.58 $\mj$ for a cold start and 13.42 $\mj$ for a hot start in our fiducial model. However, this result should be treated cautiously, as an object formed via the core accretion scenario harbors a solid core at all times during its formation and subsequent evolution in our model (no core erosion) and thus we cannot further test this result. The second reason is that the mass limit $M_{50}$ for a core mass of 0 $\mearth$ is an extrapolation of our data. \\ \\
\section{Distinguishability of cold and hot start objects}
\label{sec:distinguish}
As both cold and hot start objects will eventually forget their initial conditions their evolutionary sequences will converge after a certain time, making an observational identification of the accretion mode (i.e. cold or hot start) difficult. The timescale after which the luminosity of a 10 $\mj$ cold start object converges to the luminosity of a 10 $\mj$ hot start object is in the order of $10^8$ to $10^9$ years (Marley et al. \cite{marleyetal}). This timescale will increase with mass, which can be understood by looking at the specific entropies of the objects. For hot start objects the specific entropy directly after formation is increasing with mass, whereas the specific entropy of cold start objects is decreasing with mass. As the objects are nearly fully convective the specific entropy can be used as a proxy for the thermodynamic state of the objects. Thus one finds that the difference between the thermodynamic states of cold and hot start objects after formation increases with mass and thus shifts the 
convergence to 
later times. In Section \ref{subsubsec:expansion} we have shown that deuterium burning objects of the same mass would attain the same thermodynamic state if one holds their deuterium abundance artificially constant (the hypothetical ''deuterium main sequence''), independent from whether they formed hot or cold. With non-constant deuterium abundance we found that hot and cold start objects will try to attain the ''main-sequence''-state, even though they are unable to reach it fully. As both hot and cold start objects tend to attain the same ''main-sequence'' state it is clear that their thermodynamic properties will start to converge. After {having burned up all their deuterium} they will then continue the normal evolution, governed by contraction and cooling.
This tendency of both hot and cold start objects to attain the same state due to deuterium burning should thus have an influence on the convergence timescale between cold and hot start objects. This is indeed what we find. \\ \\
First we define the timescale $\tau_{L}$ as the timescale after which the ratio $L_{\rm hot}/L_{\rm cold}$, where $L_{\rm hot}$ and $L_{\rm cold}$ stand for the total luminosity of a hot and cold start object, has reached a value of 1.1 as a function of mass. We considered two cases: For one run deuterium burning was disabled artificially, while for the other run our normal model was used, including deuterium burning. In our fiducial model, we found that for masses between $14 \ \mj$ to $21 \ \mj$ $\tau_L$ increases monotonously as a function of mass for disabled deuterium burning, going from $4 \times 10^8$ years to $1.1 \times 10^9$ years. For deuterium burning objects it attains a constant value of approximately $2\times 10^8$ years in the same mass range. We thus find that deuterium burning shortens the convergence timescale of massive cold and hot start objects. Nevertheless, one should keep in mind that the exact numerical values of $\tau_L$ depend on the assumptions (e.g. the structure of the accretion shock) and initial conditions used for a model. However, the qualitative behavior should stay the same.
\section{Summary}\label{sect:summary}
We have studied deuterium burning in objects with masses between 10 to 30 $\mj$ that form via the core accretion mechanism. Such high masses could be reached if gap formation does not act as a limit for gas accretion (Kley \& Dirksen \cite{kleyetal}), which might be supported by the observational finding that the planetary mass function seems relatively continuous up to a mass of $\sim 25 \ \mj$ (Sahlmann et al. \cite{sahlmann}). Our simulations self-consistently link the formation and the evolution phase. We have considered two limiting cases for the property of the accretion shock occurring during the gas runaway accretion phase. First, that it is completely radiatively inefficient, so that no liberated gravitational potential energy is radiated away. Instead, high entropy material is incorporated into the object, leading to a so-called hot start. We find that the properties of such objects forming gradually with the core accretion but without radiative losses at the shock are the same as found in the 
classical hot start simulations where one starts with an already fully formed object with some specified high entropy, usually thought to have formed via direct collapse. This shows that 
the structure of the shock is as decisive to determine the thermodynamic properties of an object as it's formation mechanism. Second, we assumed that the shock is fully radiatively efficient so that all potential energy liberated at the shock is radiated away. This leads to low entropy, so-called cold start objects. We have then studied the following {topics}:
\begin{itemize}
 \item \textbf{Comparison with Burrows et al. (\cite{burrowsetal1997}) and Spiegel et al. (\cite{spiegeletal2011})}:
As seen in Section \ref{sect:hotstartcomp}, we were able to cover the results of Spiegel et al. (\cite{spiegeletal2011}) for deuterium burning in classical hot start objects, {who started} at the final mass at an arbitrary entropy without considering the formation. The radii at which the objects get partially stabilized against contraction as well as the peak values of $L_{\rm D}/L_{\rm tot}$ and the temperatures agree very well. The only difference is that our objects in the transition zone between non deuterium burning objects and deuterium burning objects seem to be able to maintain a higher burning rate over a longer period of time, enabling them to burn more deuterium in total. This is probably due to different opacities (leading to different cooling tracks) and the high temperature dependence of deuterium burning. Also, looking at the sensitivity of the minimum mass limit of deuterium burning in hot start objects, we find approximately the same slopes for $M_{50}$ (which is the mass at which 50 \% of 
the deuterium has been burned) as 
Spiegel et al. (\cite{spiegeletal2011}) when considering variations of the helium fraction $Y$ and the deuterium abundance [D/H] (see Section \ref{sect:minmass}). In the hot start case we find a mass limit of $M_{50} = 13.15 \ \mj$ for our fiducial model ($Y=0.25$, ${\rm [D/H]} = 2 \times 10^{-5}$).
 \item \textbf{Behavior of deuterium burning in hot start objects}:
In addition to the fact that we could reconfirm the results of Spiegel et al. (\cite{spiegeletal2011}) we were furthermore able to investigate the sensitivity of deuterium
burning to quantities linked to the underlying core accretion formation process. We found that deuterium burning in hot start objects is virtually independent of
the maximum allowed gas accretion rate $\dot{M}_{\rm max}$ and that a higher core mass (determined by the dust-to-gas ratio, i.e. the metallicity of the protoplanetary disk) yields higher central
temperatures which will increase the deuterium burning rate. Using our fiducial model, we find that $M_{50}=13.15 \mj$ for a core mass of $35 \ \mearth$ and $M_{50}=12.62 \mj$ for a core mass of $101 \ \mearth$. Extrapolation to $M_{\rm core}=0$ yields $M_{50} = 13.42 \ \mj$.
 \item \textbf{Behavior of deuterium burning in cold start objects}:
As cold start objects accrete most of their mass at small radii during the runaway phase, they will be below their (hypothetical) deuterium main sequence radius when their central density and temperature become high enough to burn deuterium. The thermostatic nature of deuterium fusion will then cause the objects to expand. Due to the concurrent decrease
of deuterium they will not, however, be able to fully reach the hypothetical deuterium main sequence radius. This leads to a re-inflation of the planets to about 1.4 $R_J$ and 2.3 $R_J$ for for a 14 $\mj$ and 23 $\mj$ object, respectively. Objects with masses higher than approximately 23 $\mj$ burn almost all their deuterium during the formation phase (for $\dot{M}_{\rm max}=10^{-2} \ \mearth/{\rm yr}$), as higher masses lead to higher central temperatures and thus higher deuterium burning rates. Objects with masses $\le 20 \ \mj$ burn most of their deuterium after the formation with $\dot{M}_{\rm max}=10^{-2} \ \mearth/{\rm yr}$ is complete. 
 \item \textbf{Sensitivity of the deuterium burning mass limit}:
The sensitivity of the minimum mass limit for deuterium burning for cold start objects, $M_{50}$, is comparable to the hot start objects when considering variations in the
helium mass fraction Y, the deuterium abundance [D/H] and the core mass. The dependence is, however, somewhat steeper for cold start objects. For an increase of Y = 0.22 to Y = 0.3, $M_{50}$ decreases from 13.9 $\mj$ to 12.3 $\mj$. For an increase of [D/H] from $10^{-5}$ to $4 \times 10^{-5}$, $M_{50}$ decreases from 13.6 $\mj$ to 12.9 $\mj$. When varying the maximum allowed gas accretion rate $\dot{M}_{\rm max}$, the deuterium burning in cold start objects reacts stronger than in the hot start case. The reason for
this is that the cold start objects with higher $\dot{M}_{\rm max}$ will accrete mass in the runaway phase at larger radii than those with smaller $\dot{M}_{\rm max}$.
Accretion onto a larger object means that the accretion shock is weaker, leading to less radiative losses. The object therefore has a higher specific entropy and, {quite counterintuitively}, burns deuterium more intensively {(see Sect. \ref{sect:minmass} and appendix \ref{appendix:A} and \ref{appendix:B})}. Similar as for the hot start, we find (using our fiducial model) that $M_{50}=13.3 \mj$ for a core mass of $32 \ \mearth$ and $M_{50}=12.66 \mj$ for a core mass of $97 \ \mearth$. Extrapolation to $M_{\rm core}=0$ yields $M_{50} = 13.58 \ \mj$.
\item \textbf{Evolution of the luminosity:} The evolution of the luminosity (as well as the radius and effective temperature) of hot start objects is characterized by a monotonic decrease in time, once the formation process is over. Cold start objects in contrast begin after formation with a luminosity one to two orders of magnitude lower, but then get brighter, due to deuterium  burning. The luminosity of hot and cold start objects more massive than $\sim 15 \ \mj$ converges after $\sim200 $ Myrs, due to deuterium burning. 
\end{itemize}
In summary, we find that the transition between deuterium burning and non-deuterium burning objects lies at approximately 13 $\mj$ also for objects forming via the core accretion scenario (see, e.g., Fig. \ref{fig:masslimburn}), independent of whether they might have formed with a hot or a cold start. The exact behavior of deuterium burning, however, depends on the initial conditions such as the helium mass fraction, the deuterium abundance and the metallicity. Furthermore the exact thermodynamics of the accretion shock (cold start or hot start) and the maximum allowed mass accretion rate during the runaway formation phase play a role. Nevertheless the transition from no deuterium burning at all to complete fusion of all deuterium nuclei always goes from $\sim 10 \ \mj$ to $\sim 15 \ \mj$ for a realistic range of parameters (see Fig. \ref{fig:masseslimburn}).
\section{Conclusion}\label{sect:conclusion}
Deuterium burning so far has been studied in detail only in objects which have not formed via the core accretion scenario (although Baraffe et al. \cite{baraffeetal} found that deuterium burning in objects harboring a solid core is possible).  It is, however, important to consider deuterium burning
as well in objects with a core which have formed via the core accretion scenario, as it cannot be excluded that objects forming in a protoplanetary disk via core accretion can grow beyond the required mass of $\sim 13 \ \mj$ (Kley \& Dirksen \cite{kleyetal}). This scenario is particularly interesting as objects that form
via the core accretion scenario in a disk are usually named ''planets''. This contradicts the planet definition on deuterium burning alone, leading to an interesting class of deuterium burning planets (Baraffe et al. \cite{baraffeetal}).
Therefore one should consider the possibility of objects with masses in the Brown Dwarf domain that have a solid core, at least initially, as the exact fate of the core under the extreme conditions in objects as heavy as considered here is uncertain and our results are in the limit that the core is not eroded. \\ \\ Considering a core accretion formation of the objects in the hot start scenario, we find, when comparing to the results of Spiegel et al. (\cite{spiegeletal2011}) and Burrows et al. (\cite{burrowsetal1997}), that such objects with a core burn their deuterium in approximately the same way as the objects without a core. The sensitivity of the burning process to changes in the parameters like $Y$ and [D/H] is approximately also the same (neglecting those parameters which are characteristic to the core accretion scenario). \\ \\
It is important to note, however, that if an object forms via core accretion and a low initial entropy (cold start) this will considerably alter the way in which the deuterium is burned and how the objects react to it. The biggest difference between the two formation modes is that cold start objects will expand to radii somewhat smaller than the hypothetical deuterium main sequence radius (which would be attained with a constant deuterium abundance). The most massive cold start object considered here (23 $\mj$) reinflates to a radius larger than 2 $R_J$, coming from a radius of just over 1 $R_J$. For hot start objects no reinflation occurs. Future direct imaging observations of young objects, if possible coupled with dynamical mass estimates may allow to observationally test these findings, and to constrain the formation mode of objects at the transition between planets and Brown Dwarfs.\\ \\
In addition to the results of Spiegel et al. (\cite{spiegeletal2011}), who found that deuterium burning is depending on the metallicity of the objects via
the opacity, it should be kept in mind that the metallicity also determines the core mass (if the local gas surface density in the protoplanetary disk is known). A higher core mass leads to a higher burning rate. Thus this should add up to the opacity related deuterium burning enhancement effect of the metallicity. \\ \\
Finally we saw that the consideration of a core accretion scenario also yields that the border between deuterium burning and non deuterium burning objects remains at approximately 13 $\mj$. To model the exact deuterium burning behavior, however, it is important to know certain characteristics of a system, such as the helium abundance, the solid surface density, the deuterium abundance and an upper limit for the gas runaway accretion rate.  \\ \\
Numerical data showing the formation and subsequent evolution of cold and hot start objects in our fiducial model can be found at  \href{http://www.mpia.de/homes/mordasini/Site7.html}{http://www.mpia.de/homes/mordasini/Site7.html}.

\acknowledgements{We thank Hubert Klahr, Yann Alibert and Kai-Martin Dittkrist for helpful discussions. We especially thank David Spiegel and Adam Burrows for dedicated comparison calculations concerning the expansion phase of cold start objects. Finally, we thank our referee Gilles Chabrier for questions which helped us to gain a deeper understanding of the governing physical mechanisms. Christoph Mordasini acknowledges the support as an Alexander von Humboldt fellow.}

\begin{appendix}
\section{Envelope cooling due to the core}
\label{appendix:A}
\begin{figure}
       \includegraphics[width=1.\columnwidth]{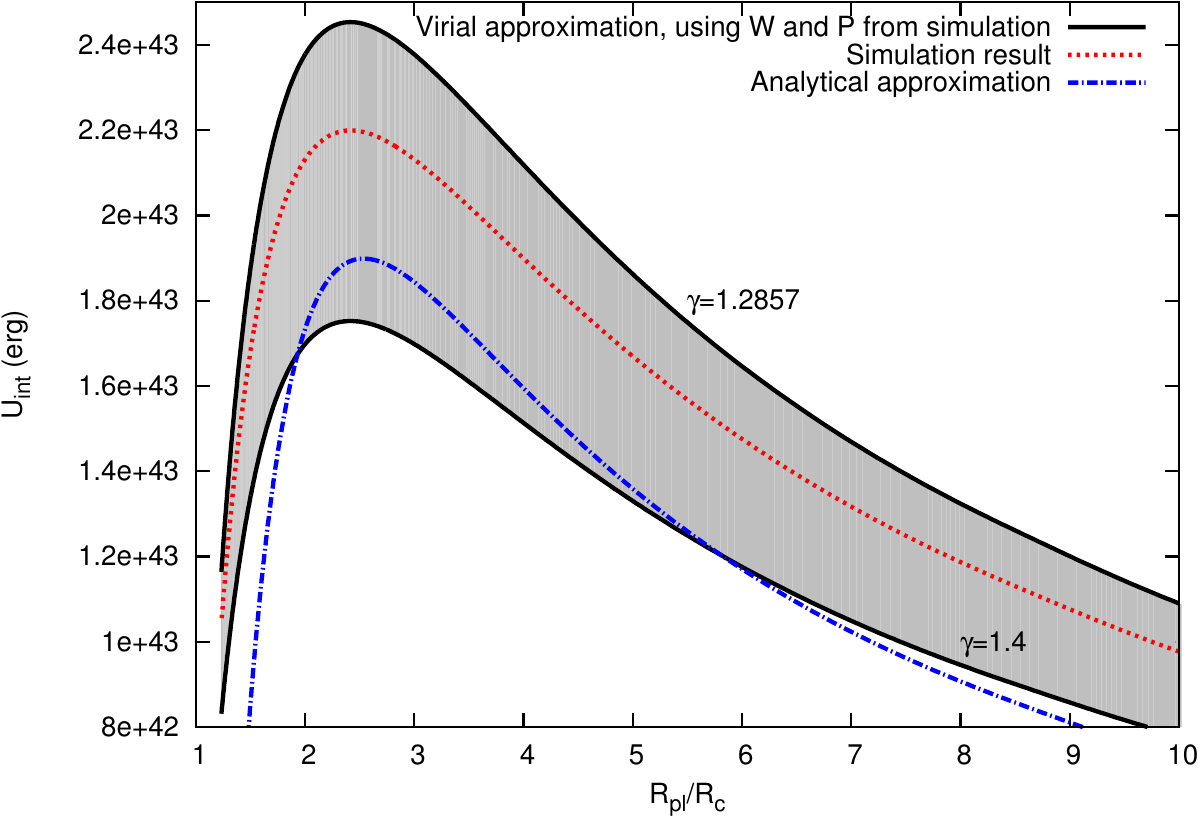}
     \caption{Internal energy of the envelope obtained from the simulation (red dotted line), the internal energy obtained by using the potential energy and the pressure from the simulation in Eq. \ref{equ:approxuint} for the two limiting $\gamma$ (black solid line) and the internal energy obtained by using the fully analytical approximation from Eq. \ref{equ:approxuintfinal} and $\gamma=1.2857$ (blue dot-dashed line). The gray area shows where one would expect the internal energy (red dotted line) to be.}
     \label{fig:idealuint}
\end{figure}
The goal of this approximation is to show that even an ideal gas in the envelope will eventually cool in the presence of a solid core. From the usual analysis without a core one obtains that $T_{\rm cent} \propto 1/R$ (see Eq. \ref{equ:centtempapprox}). In order to show this we make a virial analysis. Assuming virial equilibrium of the envelope we can start with the equation of hydrostatic equilibrium
\beq
0 = - \frac{1}{\rho}\vec{\nabla}P-\vec{\nabla}\phi \ \ ,
\eeq
where $P$ is the pressure, $\rho$ the density and $\phi$ the gravitational potential. We assume the gas in the envelope to be ideal. Furthermore we assume that the core density is constant. In this idealized analysis, one writes
\beq
0 = - \int_{\rm env}\vec{r}\cdot\vec{\nabla}P\frac{{\rm d}m}{\rho}-\int_{\rm env}\vec{r}\cdot\vec{\nabla}\phi {\rm d}m \ \ .
\label{equ:appendix2}
\eeq
We integrate over the envelope only as we are interested in the envelope only. The core is merely treated as an external factor; this also means that the second integral on the RHS of the above equation contains the potential energy of the envelope in the gravitational field of the core, but not the potential energy of the core itself. Using ${\rm d}m/\rho = {\rm d}V$ the first integral on the RHS of Eq. \ref{equ:appendix2} transforms to
\beq
- \int_{\rm env}\vec{r}\cdot\vec{\nabla}P\frac{{\rm d}m}{\rho} = \underbrace{- \oint_{\rm env} P \vec{r} {\rm d}\vec{S}}_{-3P_{\rm s}V_{\rm pl}+3P_{\rm c}V_{\rm c}} + 3(\gamma-1)\underbrace{\int_{\rm env}\rho u {\rm d}V}_{U_{\rm int}} \ \ ,
\eeq
where ${\rm d}\vec{S}$ is an infinitesimal vectorial surface element of the envelope,  $\gamma$ is the adiabatic index and $u$ is the specific energy density of the gas. $P_{\rm s}$ is the surface pressure which is negligible in our problem, $V_{\rm pl}$ is the object's volume, $P_{\rm c}$ is the gas pressure on top of the core, $V_{\rm c}$ is the core volume  and $U_{\rm int}$ denotes the internal energy of the envelope. Furthermore one can show that the second term on the RHS of Eq. \ref{equ:appendix2} is equal to the gravitational potential energy $W$ of the envelope with
\beq
W = W_{\rm env-core} + W_{\rm env-env} .
\eeq
where $W_{\rm env-core}$ is the potential energy of the envelope in the gravitational field of the core and $W_{\rm env-env}$ is the potential energy of the envelope due to its own gravitational field. 
Thus we finally have
\beq
U_{\rm int} = -\frac{1}{3(\gamma-1)}\left(W_{\rm env-env}+W_{\rm env-core} +3P_{\rm c}V_{\rm c}\right) \ \ ,
\label{equ:approxuint}
\eeq
We furthermore found that $W_{\rm env-env}$ is reasonably well described by
\beq
W_{\rm env-env} \approx - \frac{3}{5}\frac{GM_{\rm env}^2}{R_{\rm pl}} \ \ .
\label{equ:approxwenvenv}
\eeq
$W_{\rm env-core}$ can be estimated by
\beq
W_{\rm env-core} \approx - 2\frac{GM_{\rm env}M_{\rm c}}{(R_{\rm pl}+R_{\rm c})} \ \ .
\eeq
To approximate the core pressure we use the equation of hydrostatic equilibrium again and neglect all terms linear in the core mass. One then finds
\beq
\frac{P_{\rm s}-P_{\rm c}}{R_{\rm pl}-R_{\rm c}} \approx - \frac{GM_{\rm env}}{\left(R_{\rm pl}/2+R_{\rm c}/2\right)^2}\frac{M_{\rm env}}{4/3\pi \left(R_{\rm pl}^3-R_{\rm c}^3\right)} \ \ .
\label{equ:approxpress}
\eeq
Again using that $P_{\rm s}$ is negligible one finally gets by using Eq. \ref{equ:approxuint}, \ref{equ:approxwenvenv} and \ref{equ:approxpress}: 
\beq\emph{}
U_{\rm int} = \frac{GM_{\rm env}}{\gamma-1}\left(\frac{1}{5}\frac{M_{\rm env}}{R_{\rm pl}}+\frac{2}{3}\frac{M_{\rm c}}{R_{\rm pl}+R_{\rm c}}-4M_{\rm env}\frac{(R_{\rm pl}-R_{\rm c})}{(R_{\rm pl}+R_{\rm c})^2}\frac{R_{\rm c}^3}{(R_{\rm pl}^3-R_{\rm c}^3)}\right) \ \ .
\label{equ:approxuintfinal}
\eeq
With this approach one finds that the envelope starts to cool if the $R_{\rm pl}$ becomes smaller than $2.83 \ R_{\rm c}$ if $M_{\rm c}\rightarrow0$ and somewhat later for non negligible core masses. The physical reason for this cooling process is that the excess energy is transferred to the core which has to withstand an evergrowing external pressure. 
In order to test this approximation we carried out a simulation forming a 1 $\mj$ planet, using a constant core density as well as an ideal $\rm H_2$ EOS for the envelope. $\gamma$ then varies between 1.4 near the surface and 1.2857 at the center. The core mass was 35 $\mearth$. In Fig. \ref{fig:idealuint} one can see the internal energy of the envelope obtained from the simulation, the internal energy obtained by using the potential energy and the pressure from the simulation in Eq. \ref{equ:approxuint} and the internal energy obtained by using the fully analytical approximation from Eq. \ref{equ:approxuintfinal} and $\gamma=1.2857$. One sees that the analytical approximations recover the shape of the internal energy curve obtained from the simulation quite well. Furthermore the simulation result is bracketed by the two results obtained from Eq. \ref{equ:approxuint} using $\gamma = 1.4$ and $\gamma = 1.2857$.
\section{Envelope cooling due to the core and the electron degeneracy}
\begin{figure*}[htb!]
     \begin{minipage}{0.5\textwidth}
	      \centering
       \includegraphics[width=0.95\textwidth]{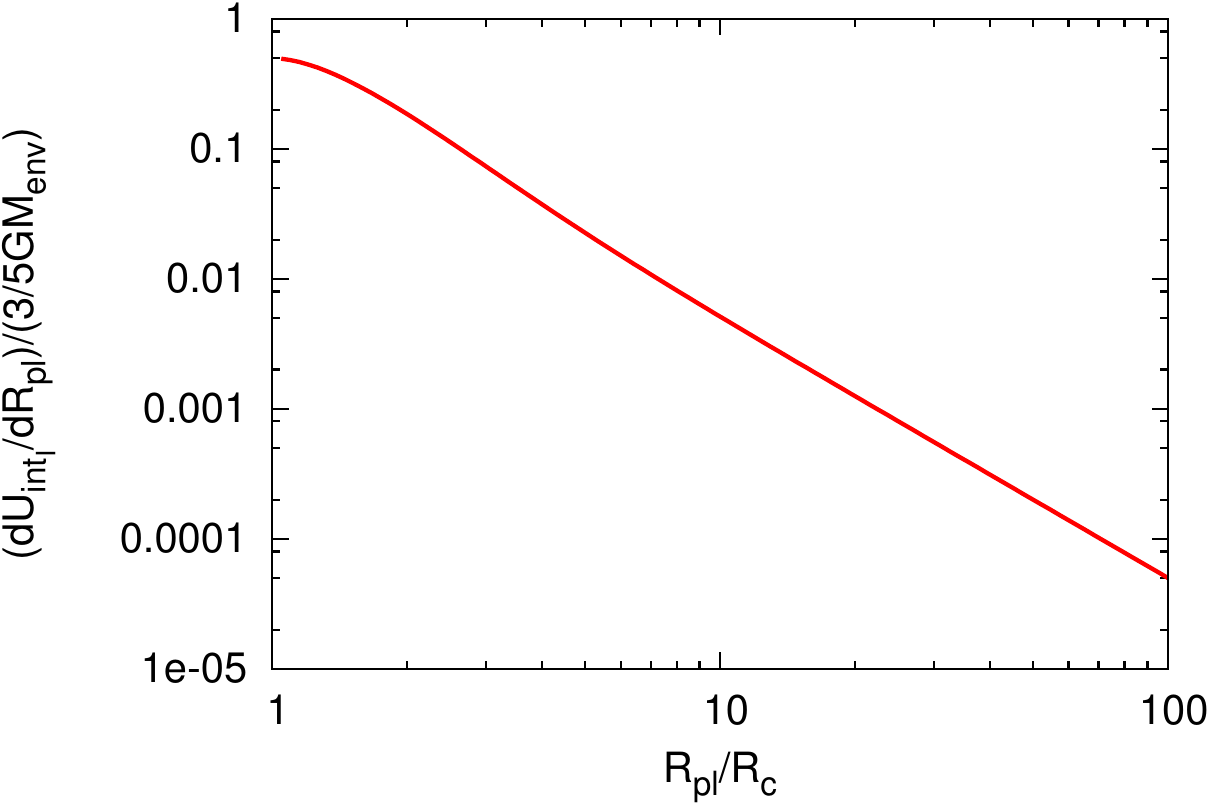}
     \end{minipage}\hfill
     \begin{minipage}{0.5\textwidth}
      \centering
       \includegraphics[width=0.95\textwidth]{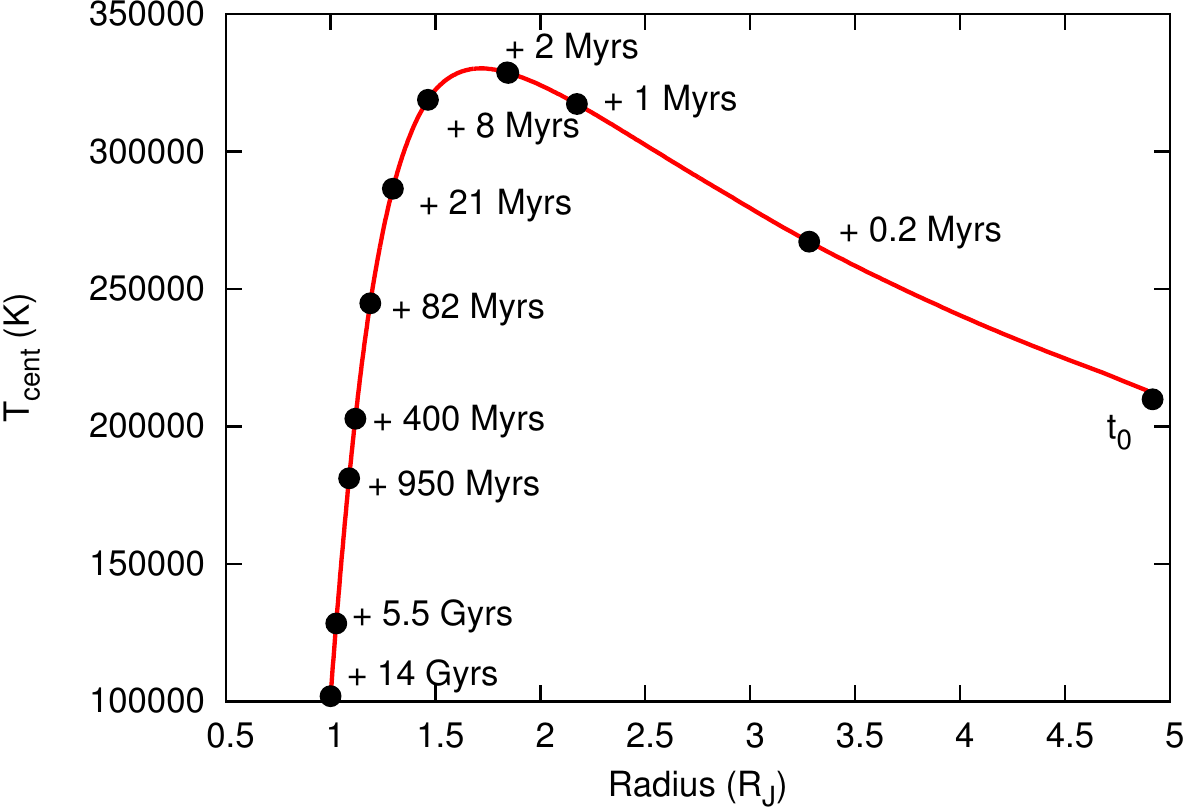}
     \end{minipage}
     \caption{Left panel: Analytical approximation of the derivative of the internal energy of the ions with respect to the object's radius plotted against $R_{\rm pl}/R_{\rm c}$. Right panel: Central temperature (i.e. the temperature in the envelope above the core) of a 10 $\mj$ hot start planet as a function of $R_{\rm pl}$. $t_0$ denotes the time where the formation of the object is finished.}
\label{fig:ionscooling}
\end{figure*}

\label{appendix:B}
It is well known that the ions in fully degenerate objects, such as White Dwarfs, will cool as they contract. The excess energy is put into the degenerate electrons. In the objects considered in this paper we have partially degenerate envelopes. However, in order to make an analytical analysis simpler, we will consider a fully degenerate envelope. In combination with the effect that the presence of the core can already cool an ideal gas envelope we expect a cooling behaviour for the degenerate envelope as well. We use the simplification that the internal energy of  the electrons is approximately
\beq
U_{\rm int_e} \approx N_eE_F \ \ ,
\eeq
where $E_F$ is the Fermi energy and $N_e$ the number of electrons.
Next we use that $E_F \propto p_F^2$ and $p_F^2\propto n_e^{2/3}$, where $p_F$ is the Fermi momentum and $n_e$ the electron number density. As $n_e \propto \left( R_{\rm pl}^{3} - R_{\rm c}^{3}  \right)^{-1}$ one finally obtains that
\beq
\frac{{\rm d \ ln}\ U_{\rm int_e}}{{\rm d \ ln}\ R_{\rm pl}} = -2 \frac{R_{\rm pl}^3}{R_{\rm pl}^3-R_{\rm c}^3} \approx -2\ \ .
\label{equ:loguinte}
\eeq
Here we made the approximation that $R_{\rm pl}^{3} \approx R_{\rm pl}^{3} - R_{\rm c}^{3}$, which is justified in the case considered here as degenerate envelopes do not contract as much as the ideal ones do. Furthermore it holds for the potential energy of the envelope
\beq
\frac{{\rm d \ ln}\ W_{\rm env-env}}{{\rm d \ ln}\ R_{\rm pl}} = -1 \ \ ,
\eeq
which yields, together with Eq. \ref{equ:loguinte}:
\beq
{\rm d}U_{\rm int_e} = \frac{2U_{\rm int_e}}{W_{\rm env-env}}{\rm d}W_{\rm env-env} \ \ .
\label{equ:duinte}
\eeq
We use that the ions can be considered as an ideal gas. Then it holds for both the ions and the degenerate electrons that $P = 2/3 \rho u$. One finds again from a virial analysis that one can write then
\beq
U_{\rm int} = -\frac{1}{2}\left(W_{\rm env-env}+3P_{\rm c}V_{\rm c}\right) \ \ .
\label{equ:unitdegen}
\eeq
Here we neglected $W_{\rm env-core}$ this time since for the objects we are ultimately interested in, it holds that $M_{\rm env}\gg M_{\rm c}$.
As we consider a fully degenerate envelope the internal energy of the electrons should be much larger than the internal energy of the ions:
\beq
U_{\rm int_e} \gg U_{\rm int_I} \ \ .
\eeq
Thus one finds from Eq. \ref{equ:unitdegen}:
\beq
U_{\rm int_e} = -\frac{1}{2}\left(W_{\rm env-env}+3P_{\rm c}V_{\rm c}\right) \ \ .
\eeq
From Eq. \ref{equ:approxwenvenv} and Eq. \ref{equ:approxpress} we know that
\beq
3P_{\rm c}V_{\rm c} = - W_{\rm env-env} K_R(R_{\rm pl},R_{\rm c}) \ \ ,
\label{equ:appdpv}
\eeq
with
\beq
K_R(R_{\rm pl},R_{\rm c}) = 20 \  R_{\rm pl}\frac{R_{\rm pl}-R_{\rm c}}{(R_{\rm pl}+R_{\rm c})^2}\frac{R_{\rm c}^3}{R_{\rm pl}^3-R_{\rm c}^3} \ \ .
\eeq
For the change of the internal energy ${\rm d}U_{\rm int} = {\rm d}U_{\rm int_e} + {\rm d}U_{\rm int_I}$ we now find
\beq
{\rm d}U_{\rm int_e} + {\rm d}U_{\rm int_I} = -\frac{1}{2} {\rm d} W_{\rm env-env} - \frac{1}{2} {\rm d} (3P_{\rm c}V_{\rm c}) \ \ .
\eeq
Using Eq. \ref{equ:duinte} and Eq. \ref{equ:appdpv} then yields
\beq
{\rm d}U_{\rm int_I}=\left(\frac{1}{2}-K_R\right) {\rm d} W_{\rm env-env}+\frac{1}{2}{\rm d} (W_{\rm env-env}K_R)  .
\eeq
As the only free parameter for a given object is $R_{\rm pl}$ (i.e. $M_{\rm env}$ and $R_{\rm c}$ are fixed) one finds the above equation to be equal to
\beq
{\rm d}U_{\rm int_I}=\left[\left(\frac{1}{2}-K_R\right) \frac{{\rm d} W_{\rm env-env}}{{\rm d} R_{\rm pl}}+\frac{1}{2}\frac{{\rm d} (W_{\rm env-env}K_R)}{{\rm d} R_{\rm pl}}\right]{\rm d} R_{\rm pl} .
\eeq
For the cooling of the ions one must have
\beq
\left(\frac{1}{2}-K_R\right) \frac{{\rm d} W_{\rm env-env}}{{\rm d} R_{\rm pl}}+\frac{1}{2}\frac{{\rm d} (W_{\rm env-env}K_R)}{{\rm d} R_{\rm pl}} > 0
\eeq
and therefore
\beq
\frac{{\rm d}\ U_{\rm int_I}}{{\rm d} \ R_{\rm pl}} > 0 \ \ ,
\eeq
which means that the internal energy of the ions decreases as the objects shrink.
In the left panel of Fig. \ref{fig:ionscooling} one can see that this is indeed the case, i.e. we expect the degenerate envelope around the solid core to cool.
For a partly degenerate envelope however, if the objects radius is much larger than the cooling radius derived for ideal gas ($2.83 \ R_{\rm c}$), the object should heat up during contraction. The reason for this is that the degeneracy in the envelope is not yet strong enough to result in ion cooling. In the right panel of Fig. \ref{fig:ionscooling} one sees the central temperature (i.e. the temperature in the envelope above the core) of a 10 $\mj$ hot start planet as a function of the radius for the whole post-formation phase (starting at time $t_0$). The envelope EOS is the EOS of Saumon et al. (\cite{saumonetal1995}), and the core density was not kept constant artificially. One can see that the envelope heats up until it has reached a radius of approximately 1.7 $R_J$ and then cools down. Non deuterium burning cold start objects will always cool in the post formation phase as they are already strongly degenerate. 
\end{appendix}


\begin{thebibliography}{}

\bibitem[2004]{alibertmordasini2004} 
Alibert, Y., Mordasini, C., \& Benz, W.\ 2004, \aap, 417, L25 

\bibitem[1999]{anguloetal}
Angulo, C., Arnould, M., Rayet, M. et al. \ 1999, \npa, 656, 3

\bibitem[2008]{baraffeetal}
Baraffe, I., Chabrier, G., \& Barman, T. \ 2008, \aap, 482, 315

\bibitem[2000]{bodenheimeretal2000}
Bodenheimer, P. H., Hubickyj, O., \& Lissauer, J. J. \ 2000, Icarus, 143, 2

\bibitem[2007]{bossetal} 
Boss, A.~P., Butler, R.~P., Hubbard, W.~B. et al. \ 2007, Transactions of the International Astronomical Union, Series A, 26, 183

\bibitem[2001]{burrowsetal2001}
Burrows, A., Hubbard, W.~B., Lunine, J.~I., \& Liebert, J. \ 2001, \rmp, 73, 719

\bibitem[1997]{burrowsetal1997}
Burrows, A., Marley, M., Hubbard, W.~B. et al. \ 1997, \apj, 491, 856

\bibitem[2000]{chabrieretal2000}
Chabrier, G., \& Baraffe, I. \ 2000, \araa, 38, 337

\bibitem[2010]{dangeloetal}
D'Angelo, G., Durisen, R.~H., \& Lissauer, J.~J. \ 2010, Giant Planet Formation, in Exoplanets, ed. S. Seager (Univ. of Arizona Pr., Tucson), 319

\bibitem[1973]{dewittetal}
Dewitt, H.~E., Graboske, H.~C., \& Cooper, M.~S. \ 1973, \apj, 181, 439

\bibitem[2005]{fischeretal}
Fischer, D.~A. \& Valenti, J. \ 2005, \apj, 622, 1102

\bibitem[1967]{fowleretal}
Fowler, W.~A.,  Caughlan, G.~R., \& Zimmerman, B.~A. \ 1967 \araa, 5, 525

\bibitem[2008]{freedmanetal}
Freedman, R.~S., Marley, M.~S., \& Lodders, K. \ 2008, \apjs, 174, 504

\bibitem[1973]{graboskeetal}
Graboske, H.~C., Dewitt, H.~E., Grossman, A.~S., \& Cooper, M.~S. \ 1973, \apj, 181, 457

\bibitem[2004]{guillotetal2003}
Guillot, T., Stevenson, D.~J., Hubbard, W.~B., \& Saumon, D. \ 2004, The interior of planets, in Jupiter.~The Planet, Satellites and Magnetosphere, ed. Bagenal, F., Dowling, T.~E., \& McKinnon, W.~B. (Cambridge Univ. Pr., Cambridge), 35

\bibitem[2011]{jansonetal}
Janson, M., Bonavita, M., Klahr, H., Lafreniere, D., Jayawardhana, R.,  Zinnecker, H. \ 2011, \apj, 736, 89

\bibitem[1990]{kippenhahnetal}
Kippenhahn, R., \& Weigert, A. \ 1990, Stellar Structure and Evolution, (Springer Verlag, Berlin, Heidelberg, New York)

\bibitem[2006]{kleyetal}
Kley, W., \& Dirksen, G. \ 2006, \aap, 447, 369 

\bibitem[2012]{leconte}
Leconte, J., \& Chabrier, G. \ 2012, \aap, 540, A20 

\bibitem[2003]{lodders}
Lodders, K. \ 2003, \apj, 591, 1220

\bibitem[1999]{lubowetal}
Lubow, S.~H., Seibert, M., \& Artymowicz, P. \ 1999, \apj, 526, 1001

\bibitem[2007]{marleyetal}
Marley, M.~S., Fortney, J.~J., Hubickyj, O., Bodenheimer, P., \&  Lissauer, J.~J.\ 2007, \apj, 655, 541

\bibitem[1984]{mercersmith}
Mercer-Smith, J.~A., Cameron, A.~G.~W., \& Epstein, R.~I. \ 1984, \apj, 279, 363

\bibitem[2011]{mordasinietal2011}
Mordasini, C., Alibert, Y., Klahr, H., \& Benz, W. \ 2011, EPJ Web of Conferences, 11, 4001

\bibitem[2012]{mordasiniklahr2011}
Mordasini, C.,  Alibert, Y., Dittkrist, K.-M.,  Klahr, H., Henning, T., \& Benz, W. \ 2012, \aap, in review

\bibitem[2002]{pallaetal}
Palla, F., Zinnecker, H., Maeder, A., \& Meynet, G. \ 2002, Physics of Star Formation in Galaxies, ed. Palla, F., Zinnecker, H., Maeder, A., \& Meynet, G. (Springer Verlag, Berlin, Heidelberg, New York)

\bibitem[1996]{pollacketal}
Pollack, J.~B., Hubickyj, O., Bodenheimer, P., Lissauer, J.~J., Podolak, M., \& Greenzweig, Y.\ 1996, Icarus, 124, 62

\bibitem[2010]{prodanovicetal}
Prodanovi{\'c}, T., Steigman, G., \& Fields, B.~D. \ 2010, \mnras, 406, 1108

\bibitem[2004]{rozyczkaetal}
R{\'o}zyczka, M., Kornet, K., Bodenheimer, P., \& Stepinski, T. \ 2004, Rev. Mex. Astron. Astrof. Conf. Ser., 22, 91

\bibitem[2011]{sahlmann}
Sahlmann, J.,S{\'e}gransan, D.,Queloz, D., \& Udry, S. \ 2011, IAU Symposium, ed. Sozzetti, A., Lattanzi, M.~G., \& Boss, A.~P.

\bibitem[2003]{santosetal}
Santos, N.~C., Israelian, G., Mayor, M., Rebolo, R., \& Udry, S. \ 2003, \aap, 398, 363

\bibitem[1995]{saumonetal1995}
Saumon, D., Chabrier, G., \& van Horn, H.~M. \ 1995, \apjs, 99, 713

\bibitem[1996]{saumonetal1996}
Saumon, D., Hubbard, W.~B., Burrows, A., Guillot, T., Lunine, J.~I., \& Chabrier, G. \ 1996, \apj, 460, 993

\bibitem[2010]{segransanetal2010}
S{\'e}gransan, D., Udry, S., Mayor, M. et al. \ 2010, \aap, 511, A45

\bibitem[2007]{spergeletal}
Spergel, D.~N., Bean, R., Dor{\'e}, O. et al. \ 2007, \apjs, 170, 377

\bibitem[2011]{spiegeletal2011}
Spiegel, D.~S.,Burrows, \& A.,Milsom, J.~A. \ 2011, \apj, 727, 57

\bibitem[2012]{spiegeletal2012}
Spiegel, D.~S., \& Burrows, A. \ 2012, \apj, 745, 174

\bibitem[1988]{stahler1988}
Stahler, S.~W. \ 1988, \apj, 332, 804

\bibitem[1980]{stahleretal1980}
Stahler, S.~W., Shu, F.~H., \& Taam, R.~E. \ 1980, \apj, 241, 637

\bibitem[2012]{wilsonetal}
Wilson, H.~F., \&  Militzer, B. \ 2012, \apj, 745, 54

\end{thebibliography}
\end{document}